\documentclass[a4paper,11pt]{article}

\usepackage{jheppub} 

\usepackage[T1]{fontenc} 

\usepackage{amsmath}
\usepackage{amsthm}
\usepackage{physics}
\usepackage[compat=1.1.0]{tikz-feynman}
\usepackage[T1]{fontenc} 
\usepackage{tensor}
\usepackage{graphicx}
\usepackage[export]{adjustbox}
\usepackage{slashed}
\usepackage{stackrel,amssymb}
\usepackage{tikz}
\usepackage{mathtools}
\usepackage{comment}
\newcommand{\ee}{\mathrm{e}}
\newcommand{\ii}{\mathrm{i}}
\newcommand{\vvev}[1]{{\left\langle #1 \right\rangle}}
\makeatletter
\newcommand*{\letterdef@}{}
\newcommand*{\letterdef}[3]{%
	\def\letterdef@##1{\expandafter\newcommand\csname #1\endcsname{#2{##1}}}%
	\@tfor\@tempa :=#3\do{\expandafter\letterdef@\expandafter{\@tempa}}}
\makeatother
\letterdef{c#1} {\mathcal}{ABCDEFGHIJKLMNOPQRSTUVWXYZ} 
\letterdef{rm#1}{\mathrm} {dDeimM} 

\title{Remarks on BPS Wilson loops in non-conformal ${\cal N}=2$ gauge theories and localization}

\author[a]{M. Bill\`o,}
\author[b]{L. Griguolo,}
\author[b]{A. Testa}

\affiliation[a]{Universit\`a di Torino, Dipartimento di Fisica and INFN, Sezione di Torino,\\
	Via P. Giuria 1, I-10125 Torino, Italy}
\affiliation[b]{Dipartimento SMFI, Universit\`a di Parma and INFN, Gruppo Collegato di Parma,
	\\ Viale G.P. Usberti 7/A, 43100 Parma, Italy}

\emailAdd{marco.billo@unito.it}
\emailAdd{luca.griguolo@unipr.it}
\emailAdd{alessandro.testa@unipr.it
}

\abstract{We consider 1/2 BPS supersymmetric circular Wilson loops in four-dimensional $\mathcal{N}=2$ $\mathrm{SU}(N)$ SYM theories with massless matter content and non-vanishing $\beta$-function. Following Pestun's approach, we can use supersymmetric localization  on the sphere $\mathbb{S}^4$ to map these observables into a matrix model, provided that the one-loop determinants are consistently regularized.
Employing a suitable procedure, we construct the regularized matrix model for these theories and show that, at order $g^4$, the predictions for the $1/2$ BPS Wilson loop match standard perturbative renormalization based on the direct evaluation of Feynman diagrams on $\mathbb{S}^4$. Despite conformal symmetry begin broken at the quantum level, we also demonstrate that the matrix model approaches perfectly captures the expression of the renormalized observable in flat space at this perturbative order.
Moreover, we revisit in detail the difference theory approach, showing that when the  $\beta$-function is non-vanishing, this method does not account for evanescent terms which are made finite by the renormalization procedure and participate to the corrections at order $g^6$.
}

\begin{document}
\maketitle
\flushbottom

\section{Introduction}
\label{sec:intro}
Supersymmetric localization is a powerful mathematical technique which allows to study super-Yang-Mills (SYM) theories with extended supersymmetry on different spacetimes at the perturbative and non-perturbative level.
In four dimensions, localization applied to $\mathcal{N}=2$ SYM theories compactified\footnote{This can be extended to squashed spheres \cite{Hama:2012bg}.} on $\mathbb{S}^4$  leads to the reduction of the path-integral to a matrix model \cite{Pestun:2007rz,Pestun:2016zxk}. This reduction yields an exact expression for the partition function at finite volume, enabling the investigation of various protected observables, including supersymmetric Wilson/t' Hooft loops and correlation functions of chiral/anti-chiral operators.

In ${\cal N}=4$ SYM theories the analytical expression for the circular 1/2 BPS Wilson loop, originally conjectured in \cite{Erickson:2000af,Drukker:2000rr}, was obtained by means of supersymmetric localization on the four-sphere in 
\cite{Pestun:2007rz}. In this case, the localized partition function is described by a Gaussian matrix model. Subsequently, it was demonstrated that less supersymmetric Wilson loops of general shape \cite{Drukker:2007qr} and families of BPS local operators \cite{Giombi:2009ds}  localize on matrix models associated with Yang-Mills theory in two-dimensions \cite{Pestun:2009nn,Bassetto:1998sr,Giombi:2009ms}. Perturbation theory reproduces the localization results in all these scenarios \cite{Bassetto:2008yf,Bassetto:2009rt,Bonini:2014vta,Rodriguez-Gomez:2016cem}, confirming the consistency of the matrix model with standard Feynman diagram techniques. 

In the context of ${\cal N}=2$ SYM theories, 
the matrix model generated by localization exhibits a more intricate structure compared to the Gaussian one associated with ${\cal N}=4$ SYM. In particular, this complexity arises from interaction terms of both perturbative and non-perturbative origin. The predictions resulting from localization for several protected observables, such as supersymmetric Wilson loops \cite{Andree:2010na,Billo:2019fbi,Fiol:2020bhf,Fiol:2020ojn,Galvagno:2021bbj}, chiral operators \cite{Billo:2017glv,Billo:2018oog,Galvagno:2020cgq,Beccaria:2020hgy,Fiol:2021icm} and  Bremsstrahlung functions \cite{Correa:2012at,Bonini:2015fng,Fiol:2015spa,Mitev:2015oty,Gomez:2018usu},  have been successfully tested against standard perturbative techniques in $\mathcal{N}=2$ superconformal theories defined on flat Euclidean space. The observed agreement with the matrix model approach is expected since in these models the four-sphere remains conformally equivalent to the flat space at the quantum level.  

Conversely, in $\cN=2$ theories where conformal symmetry is broken, either at the classical level through an explicit mass scale or at  the quantum level through a non-zero $\beta$ function, 
the matrix model is expected to encode the perturbative expansions only on the sphere. Therefore, understanding which 
information, if any, about flat-space perturbation theory is provided by  localization in non-conformal theories is an interesting issue and is the primary focus of our present work. Let us provide some more details. 

Localization has been extensively employed to study various properties of massive $\mathcal{N}=2$ SYM theories, such as $\mathcal{N}=2^*$. This massive deformation of  ${\cal N}=4$ SYM preserves $\mathcal{N}=2$ supersymmetry but explicitly breaks conformal symmetry. Over the years, this set-up has been investigated  in a beautiful series of papers \cite{Buchel:2013id,Russo:2013qaa,Russo:2013kea,Chen-Lin:2014dvz,Russo:2019lgq}, mainly focusing on its behaviour at large-$N$ and for strong 't Hooft coupling. The studies have unveiled a complicated phase-structure and carefully examined the decompactification limit as well as the holographic regimes. Throughout these investigations, the expected consistency of the localization results with perturbation theory on the sphere 
was assumed. 
A direct check in this direction can be found in \cite{Belitsky:2020hzs},  where the authors demonstrated that the perturbative evaluation of the expectation value of $1/2$ BPS supersymmetric Wilson loops on $\mathbb{S}^4$  precisely reproduces the localization predictions but significantly differs from the result in flat space. Therefore, when conformal invariance is explicitly broken by a mass scale, the matrix model provides limited information about the flat space predictions. 

In this work, we consider a different class of theories, namely $\mathcal{N}=2$ SYM with massless matter transforming in an arbitrary representation $\cR$  of the SU$(N)$ gauge group. In this case, conformal symmetry is realized classically but it is generically broken at the quantum level, as the $\beta$-function vanishes only for specific representations associated with families of superconformal set-ups \cite{Koh:1983ir,Howe:1983wj,Fiol:2015mrp}. Consequently, both the gauge coupling and the observables acquire a dependence on an (unphysical) mass scale $\mu$ and the agreement between perturbative techniques in flat-space and on the sphere, as well as with the localization approach, is no longer obvious. 
Actually, when the $\beta$-function is non-vanishing, also the localization approach is more involved, since the one-loop determinants entering the localized partition function are divergent and require a regularization \cite{Pestun:2007rz}. As a result, observables computed by the regularized matrix model develop a dependence on an (unphysical) mass scale $M$ through the dimensionless 
combination $MR$, where $R$ is the radius of the four-dimensional sphere, and  the (expected) agreement with the field-theory perturbative expansions on $\mathbb{S}^4$ requires an appropriate choice of renormalization scheme.  

Previous investigations in this directions have been conducted in \cite{Billo:2019job}, where the authors studied the two-point functions of chiral/antichiral operators in non-conformal $\mathcal{N}=2$ SQCD with $N_f$ massless flavours. In this work, it was demonstrated that at order $g^2$, localization reproduces Feynman diagram computations of such correlators both on the sphere and on flat space, while at order $g^4$ a relatively mild  discrepancy arises. Nevertheless, field-theory observables built out of dimensionless ratios of these two-point   
functions are correctly captured by the matrix model.   

The primary goal of this work is to explore the relation between supersymmetric localization and usual perturbation theory, both in flat space and on $\mathbb{S}^4$, for a different observable, namely the expectation value of the 1/2 BPS Wilson loop operator. On the matrix model side, we  construct the regularized partition function for a representative model, i.e. $\mathcal{N}=2$ SQCD with $N_f$ massless fundamental hypermultiplets, and we generalize it to matter in any representation $\cR$. Due to the regularization procedure, the matrix model perturbative expansion is effectively organized in powers of a coupling evolving with $\log(M^2R^2)$, which we identify with the running coupling arising from the renormalization procedure in the field theory approach. Subsequently, we present the matrix model prediction for the expectation value of the 1/2 BPS Wilson loop. 

We then explicitly carry out, in dimensional regularization, the perturbative expansion of the Wilson loop v.e.v. to quartic order in the coupling, both when the theory is defined on the sphere and on flat space. After applying the renormalization procedure, with an appropriate choice of schemes, the two results agree with each other and with the matrix model.
In the process we gain a precise understanding of the relation between the couplings appearing in the matrix model and in the usual field-theoretic description. 

We also point out an effect which becomes important at higher perturbative orders, and we do so
revisiting the so-called \textit{difference theory} approach. This is an effective method to organize and compute the various Feynman diagrams \cite{Andree:2010na} of an observable which is common to $\mathcal{N}=2$ and $\mathcal{N}=4$ SYM. This approach is naturally suggested by localization and was largely employed in the past \cite{Andree:2010na,Billo:2019fbi} in superconformal set-ups. In the present case, however, we show that the renormalization procedure activates some ``evanescent'' contributions, proportional to $d-4$, resulting from the integration over the Wilson loop contour. Such contributions disappear in the difference method but are actually present in the direct evaluation of the observable starting from order $g^6$. Thus, for many reasons, it would be extremely interesting to study the comparison between localization and perturbation theory in flat space at order $g^6$. The computations are remarkably involved, mainly due to the integrations over the Wilson loop contour, but we are currently working in this direction.

The organization of the paper and its specific contents are the following. In Section \ref{sec:loc}, we review the structure of  Pestun's matrix model in general massless $\mathcal{N}=2$ set-ups and discuss the regularization of the one-loop determinants through the introduction of extra massive hypermultiplets. When these contributions are integrated out, we show that the emerging  partition function for the desired set-ups  with massless matter content is well-defined and consistently described by a coupling that runs according to the $\beta$-function of the theory. We then consider the insertion of the 1/2 BPS Wilson loops and derive the explicit prediction of localization for its perturbative expansion. We explain how the matrix model naturally suggests to compute the observable by the so-called \textit{difference theory approach}, which we revisit in detail in Section \ref{sec:SW in flat space}, where we compute the circular 1/2 BPS supersymmetric Wilson loop in flat space. Using regularization by dimensional reduction, we evaluate the \textit{bubble-like} and \textit{spider-like} diagrams and show that the divergent part of the latter contribution can be extracted with the approach suggested by localization. However, such a method does not account for additional evanescent terms which are made finite by the renormalization procedure.  The Wilson loop average on $\mathbb{S}^4$, at perturbative level, is instead considered in Section \ref{sec:Supersymmetric WL on the sphere}; we use the embedding formalism described in \cite{Belitsky:2020hzs} to regularize the Feynman diagrams, evaluating them on the $d$-dimensional sphere. Finally,  in Section \ref{sec:RNW} we discuss the renormalization of the theory both in flat space and on $\mathbb{S}^4$. The renormalized observables satisfy Callan-Symanzik equations whose solutions are expressed in terms of a running coupling which can be mapped into the coupling of the regularized matrix model approach. Upon identifying a precise renormalization scheme for the observable in flat space and on the sphere, we find, in this way, a perfect matching between the renormalized observables and the matrix model predictions at order $g^4$. In Section \ref{sec:conclusions} we draw our conclusions and highlight some future perspectives. The paper is completed with several technical Appendices containing details of the computations and some useful formulae.  

\section{Predictions from localization }
\label{sec:loc}
Supersymmetric localization \cite{Pestun:2007rz} allows to reduce the path-integral of  $\mathcal{N}=2$ super-Yang-Mills (SYM) theories compactified on $\mathbb{S}^4$ to a matrix model.  Throughout this work we are primarily interested in studying $\mathcal{N}=2$ theories with massless hypermultiplets in a representation $\mathcal{R}$ of the gauge group with non-vanishing $\beta$-function. In these set-ups, the matrix model generated by localization  has to be regularized \cite{Pestun:2007rz}. To do so, we find convenient to first construct the regularized partition function for a representative model, i.e. $\mathcal{N}=2$ SQCD with $N_f $ massless fundamental hypermultiplets, and subsequently generalize the result to arbitrary matter representations. We will conclude this section by computing the vacuum expectation value of the circular 1/2 BPS Wilson loop and showing that the predictions for this observable take the expected form of renormalized quantities. 

\subsection{The $\mathbb{S}^4$ partition function} 
Let us consider a general $\mathrm{SU}(N)$ $\mathcal{N}=2$ SYM theory compactified on $\mathbb{S}^4$.
Localization \cite{Pestun:2007rz} expresses the partition function of this theory in terms of a matrix model, i.e.   
\begin{equation}
	\label{eq:partition function}
	\mathcal{Z} = \int \mathcal{D}a \  \left|Z(\mathrm{i}a,{g}_*,{R})\right|^2,
\end{equation} 
where ${g}_*$ is the coupling constant of the theory, $R$ is the radius of the sphere and the integration is over the Coulomb moduli space. This is parametrized by the $N$ real eigenvalues $a_u$ of the $N\times N$ traceless hermitian matrix $a$ associated to the vacuum expectation value acquired by the adjoint scalar field $\phi$ of the $\mathcal{N}=2$ vector multiplet. The integration measure reads 
\begin{equation}
	\label{eq:matrix model measure}
	\mathcal{D}a=\prod_{u}^{N}\dd{a}_u\prod^N_{u<v}(a_u-a_v)^2\delta\bigg(\sum_{u=1}^{N}a_u\bigg)~,
\end{equation} 
where the delta function enforces the tracelessness condition. The Vandermonde determinant in the previous expression arises when passing from the integration over the Lie algebra $\mathfrak{g}$ to its Cartan subalgebra $\mathfrak{h}$. This means that $\mathcal{D}a$ is equivalent to the flat integration measure  $\dd{a}=\prod_{b}\dd a_b$ with $a_b$ being the components of the matrix $a=a_bT^b$ and  $T_b$ denoting the hermitian traceless generators of $\mathfrak{su}(n)$ in the fundamental representation. We normalize the latter as follows: 
\begin{equation}
	\label{eq:normalization of the generators in the fundamental representation}
	\tr T_a T_b =\dfrac{\delta_{ab}}{2}~,
\end{equation}  
namely we fix the index of the fundamental representation to be $i_F = 1/2$.
In the following, the integration measure (\ref{eq:matrix model measure}) will be understood in terms of the components $a_b$, i.e. we will work in the so-called full-Lie algebra approach, with the following normalization condition  \begin{equation}
	\int \mathcal{D} a \  \ee^{-\tr a^2 } = 1 \ .
\end{equation} 

In the localized partition function  (\ref{eq:partition function})
the integrand is the product of three different factors:
\begin{equation}
	Z = Z_{\mathrm{cl}}\, Z_{\mathrm{1-loop}}\, Z_{\mathrm{inst}}~.
\end{equation} 

The classical part arises when evaluating the original action at the saddle-points and is such that 
\begin{equation}
	\begin{split}
		\label{eq:classical}
		\left|Z_{\mathrm{cl}}(\mathrm{i}a,{g}_*,{R})\right|^2
		&\equiv \mathrm{e}^{-S_{\mathrm{cl}}(\mathrm{i} a, g_*,R)}\\ 
		&	= \mathrm{e}^{-\frac{8\pi^2 {R}^2}{{g}_*^2}\tr a^2}~.
	\end{split}
\end{equation} 
The dependence on the radius ${R}$ highlights that the classical contribution results from the coupling of the vector multiplet scalar $\phi$ with the curvature of the sphere%
\footnote{The coupling between the adjoint scalar field $\phi$ and the scalar curvature, given by ${\mathrm{Ric}}/4 \tr\phi^2$ with $\mathrm{Ric}$ being the Ricci scalar, is essential to preserve rigid supersymmetry on the sphere, see e.g. \cite{Festuccia:2011ws}. }. 

The instanton contribution $Z_{\mathrm{inst}}$ takes the form
\begin{align}
	\label{Zinstis}
	Z_{\mathrm{inst}}(\ii a, q_*,R ) = \sum_{k=0}^\infty q_*^k\, Z_k(\ii a,R)~,
\end{align} 
where $k$ is the instanton number and%
\footnote{In the following we will set the theta-angle to zero, which will be immaterial since in any case we will consider a regime in which the instanton contributions can be neglected.}
\begin{align}
	\label{qis}
	q_* = \ee^{2\pi\ii\tau_*} = \ee^{-\frac{8\pi^2}{g_*^2} + \ii \theta_*}~.
\end{align}			
The partition functions at fixed instanton number $Z_k$ 
arise from the integration over the multi-instanton (quasi)-moduli and were computed by Nekrasov \cite{Nekrasov:2002qd} via localization techniques. If the theory is superconformal, the instanton contributions are dimensionless.   

\paragraph{The one-loop determinant and the regulator theory} 

The quantity $Z_{\mathrm{1-loop}}$ is the contribution of the one-loop fluctuation determinants about the BPS locus.
As discussed in section 4 of \cite{Pestun:2007rz}, these determinants result from infinite products, which are typically divergent. In general, these products can be re-expressed in terms of well-behaved expressions built out of Barnes' $G$-function, that we will recall below. In this process some problematic $a$-dependent divergent terms\footnote{In fact, for  $\mathcal{N}=2$ SYM theories with massless hypermultiplets in the representation $\mathcal{R}$ of the gauge group, the partition function is affected by ill-defined quantities of the form $\ee^{(i_\mathcal{R}-N) (\sum_{n=1}^\infty1/n)\tr a ^2}$.} cancel only if the matter representation $\mathcal{R}$ is such that the $\beta$-function of the theory vanishes, i.e. provided that the index of the representation satisfies
\begin{equation}
	\label{condiR}
	i_{\mathcal{R}} = i_{\mathrm{adj}} = N~.
\end{equation}
This condition is also required when the matter content is massive. In this case, an $a$-independent  divergent factor proportional to the squared masses remains. This term cancels in all expectation values computed in the matrix model and the partition function itself can be redefined by removing it. However, in \cite{Pestun:2007rz},  an additional term with a finite coefficient is also  removed in the redefinition of the localized partition function, but this is a matter of choice, analogous to the choice of a particular renormalization scheme. In fact, in \cite{Russo:2013kea} such finite term is not removed and in the following,  we will follow this approach.    

To localize asymptotically free theories with $i_{\mathcal{R}}<N$, we can embed them into a larger set-up, to which we will sometimes refer as the regulator theory,  described by a matter content in a  representation $\mathcal{R}_*$ such that (\ref{condiR}) is satisfied, i.e.  $i_{\mathcal{R}_*} = N$, and with  the extra hypermultiplets being massive. As a result, in  the decoupling  limit,  in which the massive hypermultiplets are integrated out, one obtains a well-defined partition function for the desired asymptotically free set-up described by a  matter content in the representation $\mathcal{R}$ and a coupling which runs according to the $\beta$-function of the theory.

This procedure was explicitly described in \cite{Pestun:2007rz} for the pure SU$(N)$ theory, where the matter representation $\mathcal{R}$ is trivial and the regulating theory was $\mathcal{N}=2^*$, in which the extra (massive) hypermultiplet is in the adjoint, i.e. $\mathcal{R}_* = \mathrm{adj}$. 

To generalize this construction to  $\mathcal{N}=2$ set-ups with massless matter content in a general representation $\mathcal{R}$, we  find convenient to first study a well-known example, i.e. $\mathcal{N}=2$ massless SQCD, where $\mathcal{R}$ is the direct sum of $N_f$ fundamentals, with $0\leq N_f < 2 N$. If we add 
\begin{equation}
	\label{Nfprime}
	N_f^\prime = 2N-N_f
\end{equation}
hypermultiplets of mass $M$, the resulting theory, which we will denote as $\mathbb{A}^*$, has a vanishing $\beta$-function%
\footnote{The representation $\mathcal{R}_*$ is the direct sum of $2N$ fundamentals, so $i_{\mathcal{R}_0} = 2N\times 1/2 = N$ and the condition (\ref{condiR}) is satisfied.}. In fact, the one-loop coefficient $\beta_0$ of the $\beta$-function, which is the only independent one for $\mathcal{N}=2$ theories, is exactly proportional to $2N-N_f$ and vanishes in this case.  

The theory $\mathbb{A}^*$ defines a flow, triggered by a partial massive perturbation,  
from superconformal $\mathcal{N}=2$ SQCD with $2N$ massless fundamental hypermultiplets -- sometimes  denoted as theory $\mathbb{A}$ in the literature \cite{Billo:2019fbi} -- 
to  $\mathcal{N}=2$ SQCD with $N_f$ fundamental hypers. When the energy  scale 
\begin{align}
	\label{defE}
	E = 1/R
\end{align}
is much larger than $M$, the mass of the hypermultiplets can be neglected and there is no difference between the $\mathbb{A}$ and $\mathbb{A}^*$ theory. In the opposite regime $E\ll M$, the massive hypermultiplets decouple and we obtain a theory described by $N_f$ massless fundamental hypermultiplets.
Since the theory $\mathbb{A}^*$ is a relevant deformation and is endowed with $\mathcal{N}=2$ supersymmetry, we obtain a UV finite theory which can be seen as a UV regulator for our field theory set-up.
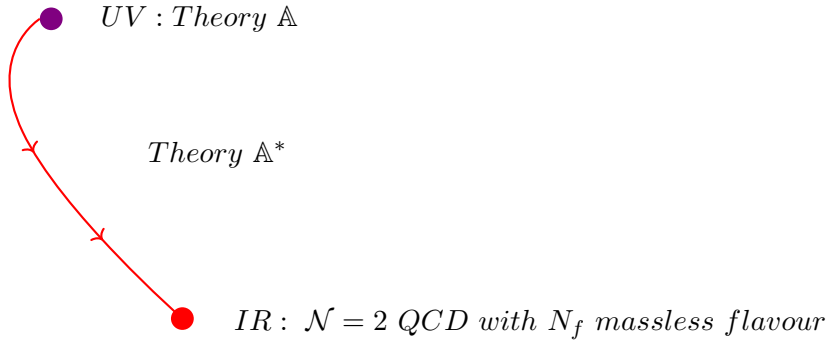
\begin{figure}
	\label{fig:flow}
	\centering
	\begin{tikzpicture}[scale=1.5]
		\begin{scope}[thick, decoration={
				markings,
				mark=at position 0.4 with {\arrow{>}},
				mark=at position 0.7 with {\arrow{>}}}
			] 
			\fill[violet] (2.1,2) circle(.1);
			\node (a) at (3.4,2) {$UV: Theory \ \mathbb{A}$};
			\node (A) at (2,2){};
			\node (B) at (-0.8,0.5){};
			\node  (C) at (1.25,-0.65) {};
			\node (b) at (3.6,0.8) {$ Theory  \ \mathbb{A}^* \ $};
			\fill[red] (3.25,-0.65) circle(.1);
			\node (c) at (6.3,-0.7) {$IR: \ \mathcal{N}=2 \ QCD \ with \ N_f \ massless \ flavour$};
			\draw[thick, red,postaction={decorate}] plot [smooth, tension=0.95] coordinates {(2,2) (1.85,1.)  (3.25,-0.65)};
		\end{scope}	  
	\end{tikzpicture}
	\caption{This picture, inspired to that in \cite{Russo:2013kea} employed to describe the flow from $\mathcal{N}=2^*$ to pure $\mathcal{N}=2$, shows the flow from superconformal $\mathcal{N}=2$ SQCD, denoted as theory $\mathbb{A}$, to the massless non-conformal $\mathcal{N}=2$ SQCD with $N_f$ massless hypermultiplets in the fundamental representation.  $\mathbb{A}^*$ is the partially massive deformation of superconformal QCD and consists of $N_f$ massless and $N_f'=2N-N_f$ massive fundamental hypermultiplets. }
\end{figure} 

The 1-loop determinant factor for the $\mathbb{A}^*$ theory takes the form 
\begin{align}
	\label{eq:one-loop determinant with massive hypermultiplets}
	\big|Z_{*,\text{1-loop}}\big|^2 
	=  \big|Z_{\text{1-loop}}\big|^2 \,  \big|Z^{(M)}_{\text{1-loop}}\big|^2~,
\end{align}
where in the previous expression the term $\big|Z_{\text{1-loop}}\big|^2$ contains the contributions from the non-conformal massless theory we're interested in, and reads \cite{Pestun:2007rz,Russo:2013kea}
\begin{equation}
	\label{Z1lis}
	\big|Z_{\text{1-loop}}\big|^2
	=  \frac{\prod_{\mathbf{\alpha}} H(\mathbf{\alpha}\cdot \mathbf{a}/E)}{
		\big[\prod_{\mathbf{w}} H(\mathbf{w}\cdot \mathbf{a}/E)\big]^{N_f}}~.
\end{equation}	
Here
$\mathbf{a}$ is an $N$-dimensional vector whose components are the eigenvalues $a_u$ of the matrix $a$, while  $\mathbf{\alpha}$ and $\mathbf{w}$ are respectively the roots of $\mathrm{SU}(N)$ and the weight vectors of the fundamental representation.
The function $H$ is defined through the Barnes' G-function as \cite{Russo:2013kea}
\begin{equation}
	\begin{split}
		H(x)&= G(1+\ii x) G(1-\ii x)\, \ee^{-(1+\gamma_E)x^2}\\
		&=  \prod_{n=1}^{\infty}\left(1+\frac{x^2}{n^2}\right)^n \mathrm{e}^{-\frac{x^2}{n}},
	\end{split}
\end{equation}
where $\gamma_E$ is the Euler's constant. This one-loop factor, as we will see, determines the interacting part of the matrix model and along with the instanton contributions is the key feature of $\mathcal{N}=2$ SYM theories\footnote{In fact, only when $\mathcal{R}=\mathrm{adj}$ the one-loop determinants become trivial. }.

The factor $|Z^{(M)}_{\text{1-loop}}|^2$ contains the contribution of the extra massive hypermultiplets and is given by		
\begin{equation}
	\label{ZMis}
	|Z^{(M)}_{\text{1-loop}}|^2 
	= \left(
	\prod_{\mathbf{w}} H\left((\mathbf{w}\cdot \mathbf{a}+M)/E\right) 
	H\left((\mathbf{w}\cdot \mathbf{a} -M)/E\right)
	\right)^{-N_f^\prime/2}~,
\end{equation}
where in the previous expression the factors containing $M$ or $-M$ result from $CPT$ conjugated states \cite{Russo:2013kea}. To determine the decoupling limit $M/E\to \infty$ of this expression it is convenient to consider the logarithm of the previous expression. One finds that 	
\begin{align}
	\label{asZmass}
	\log \big| Z_{\text{1-loop}}^{(M)}\big|^2 
	&	= -\frac{N^\prime_f}{2} \sum_{\mathbf{w}} 
	\left( \log \left(H((\mathbf{w}\cdot \mathbf{a}+M)/E) \right)
	+ \log \left(H((\mathbf{w}\cdot \mathbf{a}-M)/E) \right) \right)\\
	& = 	
	N_f^\prime {R}^2\log  MR \, \tr a^2 + \mathcal{O}(1/M^2)~,
\end{align}
where to obtain the second line we recalled that $E=1/R$ and employed the asymptotic expansion of the $H$ function for large values of its argument:   
\begin{align}
	\label{eq:asymptotic H}
	\log H(z) =
	-\frac 12 z^2 \log z^2 +
	(\frac 12 -\gamma_E) z^2 + \cO(\log z^2)~. 
\end{align} 
We also  exploited the identity 
\begin{align}
	\label{w0w2}
	\sum_{\mathbf{w}} (\mathbf{w}\cdot \mathbf{a})^2 = \tr a^2~.
\end{align}

Finally, let us now consider the instanton contributions in the $\mathbb{A}^*$ theory. From the explicit expression of the instanton partition functions for a theory with fundamental hypermultiplets, given for instance in  \cite{Russo:2013kea}, it follows that the $N_f^\prime$ massive hypermultiplets simply produce, at the leading order in their mass $M$, a factor of $M^{N^\prime_f\, k}$, namely one has
\begin{align}
	\label{Zk*}
	Z_{*,k}(\ii a, R) = M^{N^\prime_f\, k}\, Z_k(\ii a,R)~,
\end{align}
where the $Z_k$ are those of the theory encompassing the $N_f$ massless hypermultiplets only. 
Their effect is thus to promote the parameter $q_*$ in the instanton expansion (\ref{Zinstis}) to 
\begin{align}
	\label{defLambda}
	\Lambda^{N_f^\prime} \equiv M^{N_f^\prime}\, q_*~.
\end{align}
In other words at leading order for large $M$ we have
\begin{align}
	\label{Zinst*}
	Z_{*,\mathrm{inst}}(\ii a,q_*,R) = \sum_k \Lambda^{N_f^\prime k} \, Z_k(\ii a,R) = Z_{\mathrm{inst}}(\ii a, \Lambda^{N_f^\prime},R)~,
\end{align}	
where $Z_{\rm inst}$ is computed in the massless theory with $N_f$ flavours.
In the decoupling limit of the $\mathbb{A}^*$ theory we have thus altogether
\begin{align}
	\label{Z0is}
	\big|Z_*\big(\ii a, g_*, R)|^2 & = \big|Z_{*,\text{ {cl}}}(\ii a, g_*, R)\big|^2\, \big|Z_{*, \ \text{1-loop}}(\ii a, R)\big|^2 
	\, \big|Z_{*,\text{ inst}}(\ii a, q_*,R)\big|^2
	\notag \\
	& = \ee^{- \left(\frac{8\pi^2\, R^2}{g_*^2} - N_f^\prime {R}^2\log (MR) \right)\, \tr a^2}\,
	\big|Z_{\text{1-loop}}(\ii a, R)\big|^2 \, \big|Z_{\text{inst}}(\ii a, \Lambda^{N_f^\prime},R)\big|^2
	\notag +\ldots\\
\end{align}
where we took into account the expression (\ref{eq:classical}) of the classical term, which does not depend on the matter content on the theory, and suppressed corrections of order $\mathcal{O}(1/M^2)$. 

\paragraph{The perturbative matrix model}
In this work we consider the perturbative regime and we neglect all non-trivial instanton sectors, setting 
\begin{align}
	\label{Zinst1to}
	Z_{\text{inst}}\to 1~.		
\end{align}
After having discussed the perturbative behaviour, we will check that there is an energy scale range in which disregarding the instantons is indeed consistent.  

According to what suggested in \cite{Pestun:2007rz} for the pure $\mathcal{N}=2$ SYM case, to get 
the partition function for the $\mathcal{N}=2$ SQCD theory below the cut-off scale $M$ we integrate over the matrix $a$ -- as in eq. (\ref{eq:partition function}) -- the leading term in eq. (\ref{Z0is}) -- neglecting the instantons. 
This is consistent since,  employing the expansion (\ref{eq:asymptotic H}) in the expression (\ref{Z1lis}) of $\big|Z_{\text{1-loop}}\big|^2$, one can check that the proposed expression is rapidly converging for large $a$ and can thus be safely integrated.

Since our interest lies in comparing localization predictions with perturbative field theory results, we actually expand the integrand in the opposite regime, namely for $a$ small. 
Using the approach of \cite{Billo:2019fbi}, we take the logarithm of eq. (\ref{Z1lis}) to write
\begin{align}
	\label{logZ1l}
	\log \big| Z_{\text{1-loop}}\big|^2 
	& = \sum_{\alpha} \log H(\alpha\cdot \mathbf{a}/E)
	- N_f \sum_{\mathbf{w}} 
	\log H(\mathbf{w}\cdot \mathbf{a}/E)
	\notag\\
	& = - \Tr^\prime \log H( a/E)~,    
\end{align}
where we employed the short-hand notation%
\footnote{This notation was introduced in \cite{Billo:2019fbi} for a generic representation $\cR$, defining
	\begin{equation}
		\label{defTrR}
		\Tr^\prime_\cR \bullet = \Tr_\cR \bullet - \Tr_{\mathrm{adj}} \bullet~.
	\end{equation}
	In the present case, $\cR = N_f \Box$.
} 
\begin{equation}
	\label{defTr}
	\Tr^\prime \bullet = N_f \tr \bullet - \Tr_{\mathrm{adj}} \bullet~. 
\end{equation}Let us stress that this combination of traces precisely accounts for the matter content arising from the \textit{difference} between the $\mathcal{N}=2$ theories under considerations and $\mathcal{N}=4$ SYM. In other words, the matrix model suggests to compute the interacting contributions by considering the  $\mathcal{N}=2$ diagrams with matter fundamental internal lines and subtracting off the analogous ones in the adjoint representation. This method is extremely useful in the field theory approach since it drastically reduces the number of diagrams to compute and in the past, it was extensively employed in superconformal $\mathcal{N}=2$ theories \cite{Andree:2010na,Billo:2019fbi}. However, in set-ups with non-vanishing $\beta$-function the situation is more involved, as we will discuss in Section \ref{sec:SW in flat space}.  

Using the the Laurent expansion for small values of the argument of the $H$ function we have
\begin{equation}
	\label{Hsmallx}
	\log H(z) = 
	-\sum_{m=2}^{\infty}(-1)^m \dfrac{\zeta(2m-1)z^{2m}}{m}
\end{equation} 
so that, taking into account that 
\begin{align}
	\label{tradja2}
	\Tr^\prime a^2 = N_f \tr a^2 -  \Tr_{\mathrm{adj}} a^2 = (N_f  - 2N)\,\tr a^2 = - N_f^\prime \tr a^2
\end{align}
and that $E=1/R$, we can write the interaction action of the matrix model as%

\begin{align}
	\label{logZ1lex}
	S_{\text{int}}(a) \equiv - \log \big| Z_{\text{1-loop}}\big|^2
	= - \sum_{m=2}^{\infty}(-1)^m \dfrac{\zeta(2m-1)\, R^{2m}}{m}\, \Tr^\prime a^{2m} \ .
\end{align}
Note that the interaction action starts at order $a^4$.

Inserting eq. (\ref{logZ1lex}) into the leading term of eq (\ref{Z0is}) and integrating we write for the partition function of the $\mathcal{N}=2$ SQCD
\begin{align}
	\label{ZSQCD1}
	\mathcal{Z} = \int \mathcal{D}a\, 
	\ee^{-\left[\frac{8\pi^2\, R^2}{g_*^2} 
		- N_f^\prime {R}^2 \left(\log(MR) \right)\right]\, \tr a^2 
		- S_{\text{int}}} \  .
\end{align}	
The coefficient of the Gaussian term depends non trivially on the physical scale $E=1/R$ and on the UV scale $M$. Introducing 
\begin{align}
	\label{defginZ}
	\frac{1}{\tilde{g}^2(R)} = \frac{1}{g_*^2} - \frac{2N-N_f}{8\pi^2} \log MR \ ,
\end{align}
we obtain a simple expression for the $\mathbb{S}^4$-partition function of $\mathcal{N}=2$ QCD with $N_f$ massless fundamental flavours 
\begin{align}
	\label{ZSQCD}
	\mathcal{Z} = \int \mathcal{D}a\, 
	\ee^{-\frac{8\pi^2\, R^2}{\tilde{g}^2(R)}\, \tr a^2 - S_{\text{int}}(a)}~.
\end{align}	

From the field theory point of view $\tilde{g}(R)$ is the running coupling constant of SQCD with $N_f$ massless hypermultiplets evaluated at the scale $E=1/R$ and evolving with the $\beta$-function of the theory from the initial condition $g(M)= g_*$. The parameter $g_*$ is therefore interpreted as the renormalized coupling which,
by dimensional transmutation, can be expressed in terms of the RG invariant strong coupling scale 
\begin{align}
	\label{LtoMbar}
	\Lambda = M\, \ee^{-\frac{8\pi^2}{(2N-N_f)\, g^2_*}}~. 
\end{align}
It is easy to check that this coincides with the instanton weight defined in eq.(\ref{defLambda}).
At the scale $\Lambda$ perturbation theory ceases to be meaningful and the instanton contributions are no longer suppressed. Therefore,  our perturbative matrix model is only valid when the energy scale $E=1/R$ is much bigger than $\Lambda$, beside being smaller than the UV scale $M$:
\begin{align}
	\label{Rrange}
	1/\Lambda \gg R \gg 1/M~.
\end{align}

\subsection{Generalization to arbitrary representations}
In the previous section, we constructed the perturbative matrix model for the non-conformal  $\mathcal{N}=2$ SQCD with $N_f$ fundamental flavours via a massive deformation of the $\mathcal{N}=2$ superconformal SQCD. This procedure can be clearly generalized to any asymptotically free $\mathcal{N}=2$ theory with massless matter content in a representation $\mathcal{R}$ of the gauge group. In fact, starting from (\ref{ZSQCD}), it is straightforward to deduce the following expression for the perturbative  matrix model of these set-ups, i.e. \begin{align}
	\label{eq:ZN=2generic}
	\mathcal{Z} = \int \cD a\, \ee^{-\frac{8\pi^2\, R^2}{g(R)^2}\, \tr a^2 - S^\mathcal{R}_{\text{int}}(a)}~. 
\end{align}
In the previous expression  $\cD a$ is the integration measure over the full Lie algebra $\mathfrak{g}$ of $\mathrm{SU}(N)$, while $g(R)$ is the running coupling evaluated at the scale $E=1/R$ \begin{align}
	\label{defginZgenericR}
	\frac{1}{{g}^2(R)} = \frac{1}{g_*^2} +\beta_0^\mathcal{R}  \log M^2R^2~,
\end{align} and evolving from the initial condition $g_*$ in the UV via the one-loop exact $\beta$-function of theory $\beta(g)=\beta_0^\mathcal{R}g^3$. The explicit expression of the coefficient $\beta_0^\mathcal{R}$ is  \begin{equation}
	\label{eq:defbeta0}
	\beta_0^{\mathcal{R}} =\dfrac{i_{\mathcal{R}}-N}{8\pi^2}~,
\end{equation} with $i_\mathcal{R}<N$ for asymptotically free theories. In (\ref{eq:ZN=2generic}), the interacting action is defined by exponentiating the contribution of the one-loop determinants \cite{Pestun:2007rz} \begin{equation}
	\label{eq:one-loop in general massless theory}
	\left|Z^\mathcal{R}_{\text{1-loop}}\right|^2 = \dfrac{\prod_{\alpha} H( \alpha\cdot \mathbf{a}/E)}{\prod_{\mathbf{w}_{\mathcal{R}}} H(  \mathbf{w}_\mathcal{R}\cdot \mathbf{a}/E)} \ , 
\end{equation}   with $\mathbf{w}_\mathcal{R}$ being the weight vector of the representation $\mathcal{R}$. Using the Laurent expansion (\ref{Hsmallx}), it is straightforward to show that the interacting action is an obvious  generalization of  (\ref{logZ1lex}), i.e. \begin{align}
	\label{logZ1lexinrepR}
	S^\mathcal{R}_{\text{int}}(a) \equiv - \log \big| Z^\mathcal{R}_{\text{1-loop}}\big|^2
	= - \sum_{m=2}^{\infty}(-1)^m \dfrac{\zeta(2m-1)\, R^{2m}}{m}\, \Tr^\prime_\mathcal{R} a^{2m} \ ,
\end{align} where  $\Tr^\prime_\mathcal{R} = \Tr_{\mathcal{R}}-\Tr_{\text{Adj}}$. As we already noted in eq. (\ref{defTr}), this combination of traces precisely describes the matter content arising from the difference between the $\mathcal{N}=2$ theories under consideration and $\mathcal{N}=4$ SYM. From a perturbative point of view, as we already explained, this means that the interacting contributions should be described by the $\mathcal{N}=2$ diagrams with internal matter  lines in the representation $\cR$ to which we subtract off the analogous ones in which the internal lines are in the adjoint representation.  

As we discussed at the end of the previous section,  this perturbative matrix model, in which we suppressed the instanton corrections, is valid only when \begin{equation}
	1/\Lambda_\mathcal{R}\gg R\gg 1/M \ ,
\end{equation} where $\Lambda_\mathcal{R}$ is  the RG invariant strong coupling scale of our asymptotically free set-up, i.e. \begin{equation}
	\Lambda_\mathcal{R} = M \ee^{\frac{1}{2\beta_0^\mathcal{R}g_*^2}} \ .
\end{equation} In the following section, we will employ the matrix model formalism to compute the vacuum expectation value of the 1/2 BPS Wilson loop placed on the equator of the sphere. 

\subsection{Supersymmetric Wilson loop}
\label{sec:WL in matrix model}
In this section, we turn our attention to the study of the 1/2 BPS supersymmetric circular Wilson loop operator in the fundamental representation. According to \cite{Pestun:2007rz}, its expectation value can be computed via the matrix model approach as
\begin{align}
	\label{vevW}
	W 
	= \frac{1}{\cZ} \int \cD a\, \ee^{-\frac{8\pi^2\, R^2}{g^2(R)}\, \tr a^2 - S_{\text{int}}(a)}\,
	\frac 1N\, \tr\ee^{2\pi a R}~,
\end{align}
where $g(R)$ is defined in (\ref{defginZgenericR}), while the interacting action is given by (\ref{logZ1lexinrepR}). If we rescale the matrix variable by setting 
\begin{equation}
	a = \sqrt{\frac{g^2}{8\pi^2}} \frac{a^\prime}{R}~,
\end{equation} 
so that $a^\prime$ is dimensionless, 
the Jacobian factors cancels between the numerator and the denominator. We can therefore write, after renaming $a$ the new integration variable $a^\prime$, 
\begin{align}
	\label{vevWbis}
	W(g)=   \frac{1}{\cZ} \int \cD a\, \ee^{- \tr a^2 - S^\mathcal{R}_{\text{int}}(a,g)}\, \cW(a,g)
	~,		
\end{align}
where the Wilson loop operator is now
\begin{equation}
	\label{eq:Wrescaled}
	\cW(a,g) =\dfrac{1}{N} \tr \ \exp(\frac{g}{\sqrt{2}} a)
\end{equation}  
and the interaction action takes the form of an expansion in powers of $g$:
\begin{align}
	\label{Sintbis}
	S^\mathcal{R}_{\text{int}}(a,g) = - \sum_{m=2}^{\infty}\left(-\frac{g^2}{8\pi^2}\right)^m \dfrac{\zeta(2m-1)}{m} 
	\Tr_\mathcal{R}^\prime a^{2m}~. 
\end{align}
The partition function $\cZ$ is given by the same integral as in the numerator without the insertion of the Wilson loop operator.

Let us note that this is precisely the expression considered in \cite{Billo:2019fbi}, except for the fact that here the coupling $g$ is the scale-dependent running coupling and that the representation $\mathcal{R}$ corresponds to a non-zero $\beta$-function. Technically, the matrix model computation is just the same. 
Expanding eq. (\ref{vevWbis}) perturbatively in $g$, one has just to evaluate correlators of traces of powers of $a$ in a Gaussian matrix model%
\footnote{We indicate the correlator of an operator $f(a)$ in the free matrix model by $\vvev{f(a)}_0$.}. This can be done very efficiently using recursive methods, as described in \cite{Billo:2017glv}.   

It is convenient to first expand the $\exp(-S^\mathcal{R}_{\text{int}})$ insertion using eq. (\ref{Sintbis}) to obtain
\begin{align}
	\label{WexpSint}
	W(g) = W_0(g) 
	+ \left(\frac{g^2}{8\pi^2}\right)^2 \frac{\zeta(3)}{2} \vvev{W(a)\,\Tr_\mathcal{R}^\prime a^4}_{0,c}
	+ \left(\frac{g^2}{8\pi^2}\right)^3 \frac{\zeta(5)}{3} \vvev{W(a)\,\Tr_\mathcal{R}^\prime a^6}_{0,c}
	+ \ldots~,
\end{align}
where $\vvev{f(a)\,g(a)}_{0,c}$ denotes the connected correlator $\vvev{f(a)\,g(a)}_0 -\vvev{f(a)}_0 \vvev{g(a)}_0$ in the free theory. The first term, $W_0(g) \equiv \vvev{W(a)}_0$, is the expectation value in the Gaussian matrix model of the Wilson loop operator (\ref{eq:Wrescaled}). It encodes, as well-known, the expectation value of the 1/2 BPS Wilson loop in the $\cN=4$ SYM theory. Expanding eq. (\ref{vevWbis}) and taking into account that the v.e.v. of odd power traces vanishes in the gaussian model, it can be computed as
\begin{align}
	\label{expW0}
	W_0(g)
	& = 1 + \frac{g^2}{4} \frac{\vvev{\tr a^2}_0}{N} + \frac{g^4}{4! 2^2} \frac{\vvev{\tr a^4}_0}{N}
	+  \frac{g^6}{6! 2^3} \frac{\vvev{\tr a^6}_0}{N} + \ldots
	\notag\\
	& = 1 + \frac{g^2}{4} C_F +  \frac{g^4}{4! 2^2} C_F \frac{2N^2-3}{2N}
	+  \frac{g^6}{6! 2^3} C_F \frac{5(N^4-3N^2+3)}{4 N^2} + \ldots~,   
\end{align}
where we denoted with $C_F$ the quadratic Casimir of the fundamental representation, $C_F = (N^2 -1)/(2N)$. This expansion can be exactly resummed in terms of Laguerre polynomials 
\cite{Drukker:2000rr,Erickson:2000af}:  
\begin{align}
	\label{W0is}
	W_0(g) = \dfrac{1}{N}L^1_{N-1}\left(-{g}^2/4\right)\,\exp(\dfrac{{g}^2}{8}\Big(1-\dfrac{1}{N}\Big))~.
\end{align}
Here, however, $g$ is the running coupling (\ref{defginZgenericR}) of our general non-conformal $\mathcal{N}=2$ theory. 

The effects of the interaction action $S^\mathcal{R}_{\text{int}}$ in eq. (\ref{WexpSint}) start at order $g^6$. In fact, the lowest order contribution arises from 
\begin{align}
	\label{g6corr}
	\left(\frac{g^2}{8\pi^2}\right)^2 \frac{\zeta(3)}{2} \vvev{W(a)\,\Tr_\mathcal{R}^\prime a^4}_{0,c} = 
	\left(\frac{g^2}{8\pi^2}\right)^2 \frac{\zeta(3)}{2} \frac{g^2}{4N} \vvev{\tr a^2\,\Tr_\mathcal{R}^\prime a^4}_{0,c} + \cO(g^8)~,
\end{align}
since the term of order $1$ in $W(a)$ does not contribute to the connected part. To evaluate the previous expressions is necessary to reduce high order traces in the  representation $\mathcal{R}$ in terms of the fundamental ones. For a general  $\mathcal{R}$, this is a tedious calculation which can be done by means of Frobenius' theorem, as it is explained in detail in Appendix A of  \cite{Billo:2019fbi}. Here, we show a straightforward application. 

Consider again $\mathcal{N}=2$ SQCD with $N_f$ massless flavours, i.e. $\mathcal{R}=N_f\Box$. Then, eq. (\ref{g6corr}) becomes \begin{equation}
	\label{g6corrSQCD}
	\left(\frac{\tilde{g}^2}{8\pi^2}\right)^2 \frac{\zeta(3)}{2} \frac{\tilde{g}^2}{4N} \vvev{\tr a^2\,\Tr^\prime a^4}_{0,c}
\end{equation}
where we recall that  $\Tr^\prime$ is defined in (\ref{defTrR}), while $\tilde{g}$ is the running coupling (\ref{defginZ}). Using the following identity\footnote{See for instance eq. (2.29) in \cite{Billo:2019fbi} where the authors also discuss the anti-symmetric and symmetric representation of $\mathrm{SU}(N)$.}\begin{align}	
	\label{Tra2m}
	\Tr^\prime a^{2m} =  - \sum_{p=2}^{2m-2}  (-1)^p \binom{2m}{p} \tr a^p\, \tr a^{2m-p} - (2N-N_f)\tr a^{2m} ~,
\end{align} we can reduce all the relevant traces to the fundamental ones and  straightforwardly compute higher order corrections. For the simple connected part in (\ref{g6corrSQCD}) we see that the previous expression implies that 
\begin{align}
	\label{trpa4is}
	\Tr^\prime a^4 = - N_f^\prime \tr a^4 - 6 \left(\tr a^2\right)^2
\end{align}
is homogeneous of degree $4$ in $a$. As a result, we find that \cite{Billo:2017glv} 
\begin{align}
	\vvev{\tr a^2 \Tr^\prime a^4}_{0,c} = \vvev{\tr a^2 \Tr^\prime a^4}_0 - \vvev{\tr a^2}_0 		\vvev{\Tr^\prime a^4}_0 = 2 \vvev{\Tr^\prime a^4}_0~. 
\end{align}
Substituting eq. (\ref{trpa4is}) and using the recursive techniques of \cite{Billo:2017glv} one gets finally
\begin{align}
	\label{a2a4cis}
	\vvev{\tr a^2 \Tr^\prime a^4}_{0,c} = - C_F\left((2N-N_f) (2N^2-3) + 6 N (N^2+1)\right)~,
\end{align} where we recall that $C_F$ is the quadratic Casimir of the fundamental representation.
Inserting this into eq. (\ref{g6corrSQCD}) we obtain the three-loop correction to the observable induced by the interaction action   
\begin{align}
	\label{WtoW0g6}
	W(\tilde{g}) = W_0(\tilde{g}) - 
	\frac{\tilde{g}^6\, 3\zeta(3) C_F}{512 N \pi^4}(2N^3+N_f) + \frac{\tilde{g}^6\, \zeta(3) C_F N}{256 \pi^4}(N_f-2N)
	+ \cO(g^8)~.
\end{align}
The previous expression was checked against standard perturbation theory in $\mathcal{N}=2$ superconformal SQCD, i.e. $N_f=2N$, in the pioneering work \cite{Andree:2010na} by employing the difference theory approach we discussed in the previous section. In this case, where the $\beta$-function is non-vanishing, the matrix model also produces an additional correction proportional to $\zeta(3)(2N-N_f)$ whose possible origin in field theory will be discussed in the following. 

Obtaining predictions at higher orders in perturbation theory in terms of $g$ is extremely straightforward with the indicated techniques. However, our aim in this work is to compare  the matrix model prediction with the direct computation of Feynman diagrams on $\mathbb{S}^4$ and on $\mathbb{R}^4$ for a theory with general matter content in the representation $\mathcal{R}$. In carrying out this comparison, the crucial point is that the matrix model yields $W(g)$ as a function of the running coupling constant $g = g(R)$ defined in eq. (\ref{defginZgenericR}), which is equivalent to\footnote{Note that the expansion on the right-hand side of (\ref{grunis}) is only possible since we are working in the range (\ref{Rrange}) which ensures that $g_*\ll 1$.}  
\begin{align}
	\label{grunis}
	g^2 = \dfrac{g_*^2}{1 +\beta_0^\mathcal{R} g_*^2  \log M^2 R^2 }
= g_*^2 - g_*^4 \beta^\mathcal{R}_0\log M^2R^2  + \cO(g_*^6)~,
\end{align}
where we recall that $\beta_0^\mathcal{R}$ is defined in (\ref{eq:defbeta0}).

Field theory computations, which will be described in the following sections, are much harder than the matrix model ones. They are originally organized  in a perturbative series in the bare coupling $g_B$ and have to be regularized\footnote{We will employ regularization by dimensional reduction to treat UV-divergent diagrams.}. To reabsorb the divergences a renormalization scale and a renormalized coupling have to be introduced, and it is the renormalized coupling that is mapped into the parameter $g_*$ of the matrix model. From the field theory point of view, as we will show explicitly in Section \ref{sec:RNW}, the Callan-Symanzik equation implies that the Wilson loop  must actually be expressible as a function of the running coupling constant $g(R)$; this is cleverly realized in the matrix model description.   

In the present work, 
we will derive the field theory result to the order $g_*^4$ in flat space and on the sphere and we will point out some interesting higher order effects due to the non-vanishing $\beta$-function of the theory. Thus we will only check the matrix model expression up to order $g^4_*$, which is given simply by
\begin{align}
\label{Wtog4}
W(g) = W_0(g) + \cO(g^6) 
= 1 + \frac{g^2}{4} C_F+ \frac{g^4}{4! }C_F \frac{2N^2-3}{8N} + \cO(g^6)~.    		
\end{align}
Up to order $g_*^4$, inserting eq. (\ref{grunis}) we get%
\begin{align}
\label{Wtog04}
W(g) = W_0(g_*) -  g_*^4 \frac{\beta_0^\mathcal{R}\ C_F}{2} \log M R  + \cO(g_*^6)~.
\end{align} It is important to note that the logarithmically enhanced term, arising from the running coupling $g(R)$, exhibits the usual dependence on the ultraviolet cut-off $M$ of renormalized quantities. 

\section{Supersymmetric Wilson loops in flat space}
\label{sec:SW in flat space}
In this section, we study the vacuum expectation value of a circular 1/2 BPS Wilson loop in a general  $\mathrm{SU}(N)$ $\mathcal{N}=2$ SYM theory\footnote{See Appendix \ref{sec:actions in flat space} for our conventions.} with massless hypermultiplets in an arbitrary representation $\mathcal{R}$ and with a non-vanishing $\beta$-function.  In Euclidean spacetime, the operator  is defined as follows  
\begin{equation}
	\label{eq:1/2 BPS supersymmetric Wilson loop}
	\widehat\cW = \dfrac{1}{N} \text{tr} \ \mathcal{P} \exp \bigg\{ g_B \int_C \dd{s} \bigg[ \mathrm{i} A^{\mu}(x(s)) \dot{x}_{\mu}(s) + \dfrac{R}{\sqrt{2}} \Big( \Bar{\phi}(x(s)) +\phi(x(s)) \Big)\bigg]
	\bigg\} \ ,
\end{equation}  where $g_B$ is the bare coupling constant and the trace is over the fundamental representation. In the previous expression the vector multiplet scalar  $\phi$ and gauge field $A_\mu$ are integrated over the circle  $C$ of radius $R$ parametrized as  
\begin{equation}
	\label{eq:parametrization}
	x^{\mu}(s)=R(\cos{(s)}, \sin{(s)}, \mathbf{0}) \quad \text{with} \quad s\sim s+ 2\pi \ .
\end{equation} 

The vacuum expectation value of $\widehat\cW$ contains ultraviolet divergent diagrams which we regularize by employing dimensional reduction\footnote{We discuss this regularization scheme in Section \ref{sec:Feynman computations in flat space}. Here we consider a reduction from four  to $d$ dimensions, with $d<4$.} \cite{Erickson:2000af}. This regularization scheme preserves supersymmetry but it breaks (classical) conformal symmetry since $g_B$ is dimensionless only when $d=4$. We expand the dimensionally regularized observable in a power series of the bare coupling $g_B$ as follows
\begin{equation}
	\label{eq:dimensionally regularized vev in flat}
	\cW^{\rm flat} \equiv
	\vvev{\widehat{\cW}}
	= 1+ g_B^2{\cW}^{\rm flat}_2 + g_B^4{\cW}^{\rm flat}_4 + g_B^6{\cW}^{\rm flat}_6 + \ldots \ , 
\end{equation} 
where  ${\cW}^{\rm flat}_i$ are  functions of the dimension $d$ and of $R$ encoding the $i/2$-th loop correction. 

\subsection{One-loop corrections}
\label{subsec:one-loop-flat}
The lowest order term, i.e. $g_B^2{\cW}^{\rm flat}_2$, consists of two connected diagrams arising from a single exchange of a gauge field and an adjoint scalar inside the Wilson loop. In $d$ dimensions, the tree-level propagators of the scalar $\phi$ and of the gauge field $A_\mu$ in the Feynman gauge are given by 
\begin{equation}
	\label{eq:tre-level props }
	\Delta^{ab}(x_{12})
	= \delta_{ab} \dfrac{\Gamma(d/2-1)}{4\pi^{d/2}(x_{12}^2)^{d/2-1}} \ , \quad \quad 
	\Delta^{\mu\nu}_{ab}(x_{12})
	=\delta_{ab}\dfrac{\delta_{\mu \nu}\Gamma(d/2-1)}{4\pi^{d/2}(x_{12}^2)^{d/2-1}} \ .
\end{equation}
We find convenient to represent the different corrections to the expectation value of (\ref{eq:1/2 BPS supersymmetric Wilson loop}) by employing the following graphical representation  
\begin{equation}
	\begin{split}
		\mathord{ \begin{tikzpicture}[baseline=-0.65ex,scale=0.8]
				\begin{feynman}
					\vertex (A) at (1.5,0);
					\vertex (B) at (-1.5,0);
					\diagram*{
						(A) -- [photon] (B),
					};
				\end{feynman}
			\end{tikzpicture} 
		} +	\mathord{ \begin{tikzpicture}[baseline=-0.65ex,scale=0.8]
				\begin{feynman}
					\vertex (A) at (1.5,0);
					\vertex (B) at (-1.5,0);
					\diagram*{
						(B) --[ fermion] (A)
					};
				\end{feynman}
			\end{tikzpicture} 
		} = 	\mathord{ \begin{tikzpicture}[baseline=-0.65ex,scale=0.8]
				\begin{feynman}
					\vertex (A) at (1.5,0);
					\vertex (B) at (-1.5,0);
					\diagram*{
						(A) -- [photon] (B),
						(B) --[ fermion] (A)
					};
				\end{feynman}
			\end{tikzpicture} 
		} 
	\end{split}  \ ,
\end{equation} where the wavy line denotes the gauge field propagator, while the continuos one is associated with the adjoint scalar propagator. By Taylor expanding (\ref{eq:1/2 BPS supersymmetric Wilson loop}) at order $g_B^2$, we find 
\begin{equation}
	\begin{split}
		\label{eq:ladder g2}
		g_B^2 \cW^{\rm flat}_2 =  
		\mathord{ \begin{tikzpicture}[baseline=-0.65ex,scale=0.8]
				\draw [black] (0,0) circle [radius=1.5cm];
				\begin{feynman}
					\vertex (A) at (0,1.5);
					\vertex (B) at (0,-1.5);
					\diagram*{
						(A) -- [photon] (B),
						(A) --[ fermion] (B)
					};
				\end{feynman}
			\end{tikzpicture} 
		} &=\bigg( \dfrac{g_B^2\Gamma(d/2-1) C_F}{2}\bigg) \times \bigg(\oint_C \dfrac{\dd^2{s}}{4 \pi^{d/2}} \dfrac{R^2 - \dot{x}_1 \cdot \dot{x}_2}{[x_{12}^2]^{d/2-1}}\bigg)\\
		&=g_B^2\bigg( \dfrac{\Gamma(d/2-1)\Gamma(5/2-d/2)}{\Gamma(3-d/2)}\bigg) \dfrac{C_F(4\pi R^2)^{2-d/2}}{4\pi^{1/2}}~,
	\end{split}
\end{equation}
having integrated over the contour by employing the parametrization of the coordinates (\ref{eq:parametrization}) and  the master integral (\ref{eq:master integral}) in Appendix \ref{sec:Useful formulae}. This expression, which is regular for $d\to 4$, is identical to the one-loop correction in the $\cN=4$ theory since it arises from the same diagrams.  

\subsection{Two-loop corrections}
\label{sec:Insertion}
The two-loop correction $g_B^4\cW_4^{\rm flat}$ to the Wilson loop vacuum expectation value  contains three classes of diagrams:
\begin{equation}
	\label{w4flatis}
	g_B^4 \cW_4^{\rm flat} = \Sigma_\cR + \Sigma_2 + \Sigma_{\rm ladder}~,
\end{equation}
where
\begin{equation}
	\label{eq:sigma1 e 2}
	\Sigma_\mathcal{R} = \mathord{
		\begin{tikzpicture}[radius=2.cm, baseline=-0.65ex,scale=0.6]
			\draw [black] (0,0) circle [];
			\filldraw[color=gray!80, fill=gray!15](0,0) circle (1);	
			\draw [black]   (0,0) circle [radius=1.];
			\begin{feynman}
				\vertex (A) at (0,2.);
				\vertex (C) at (0, 1.);
				\vertex (D) at (0, -1.);
				\vertex (B) at (0,-2.);
				\vertex (d) at (0.1,0.3) {\text{\footnotesize 1-loop }\normalsize} ;
				\vertex (d) at (0.1,-0.3) {\text{\footnotesize $\cR$ }\normalsize} ;
				\diagram*{
					(A) -- [photon] (C),
					(A) --[ fermion] (C),
					(D) -- [photon] (B),
					(D) --[ fermion] (B)
				};
			\end{feynman},
	\end{tikzpicture} } \ , \quad 
	\Sigma_2=	\mathord{
		\begin{tikzpicture}[radius=2.cm, scale=0.6, baseline=-0.65ex]
			\draw [black] (0,0) circle [];
			\begin{feynman}
				\vertex (A) at (0,2.);
				\vertex (C) at (0,0);
				\vertex (D) at (-1.5, -1.3);
				\vertex (B) at (1.5, -1.3);
				\diagram*{
					(A) -- [photon] (C),
					(A) --[ fermion] (C),
					(C) -- [photon] (B),
					(C) --[ fermion] (D),
					(C) --[ photon] (D)
				};
			\end{feynman} 
	\end{tikzpicture} } \ , \quad 
	\Sigma_{\rm ladder} = 
	\mathord{
		\begin{tikzpicture}[baseline=-0.65ex,scale=0.6]
			\draw [black] (0,0) circle [radius=2cm];
			\begin{feynman}
				\vertex (A) at (-0.75,1.85);
				\vertex (B) at (-0.75,-1.85);
				\vertex (C) at (0.75,1.85);
				\vertex (D) at (0.75,-1.85);
				\diagram*{
					(A) -- [photon] (B),
					(A) --[ fermion] (B),
					(C) -- [photon] (D),
					(C) --[ fermion] (D)
				};
			\end{feynman}
		\end{tikzpicture} 
	}~.
\end{equation} 
The diagrams $\Sigma_\cR$ and $\Sigma_2$ have ultraviolet singularities when $d\to 4 $, while the ladder-like contribution is perfectly finite in this limit. In  (\ref{eq:sigma1 e 2}) the diagrams $\Sigma_2$ and $\Sigma_{\rm ladder}$ entirely result from the $\mathcal{N}=2$ vector multiplet and consequently, are in common with the $\cN=4$ theory. On the other hand, the {bubble-like} class $\Sigma_\cR$  contains the one-loop corrections to the adjoint scalar and gauge field propagator. In these diagram the matter hypermultiplet fields in the representation $\cR$ of the $\mathcal{N}=2$ theories appear in the virtual loops; in the $\cN=4$ theory these hypers\footnote{Here we refer to the fact that in the $\mathcal{N}=2$ language, $\mathcal{N}=4$ SYM can be regarded as $\mathcal{N}=2$ vector multiplet coupled to a single adjoint hypermultiplet.} would transform in the adjoint representation and we would have $\Sigma_{\rm adj}$. In this section, we consider in detail all these corrections separately. 

$\Sigma_2$ was originally computed in \cite{Erickson:2000af} in the context of $\mathcal{N}=4$ SYM. In that case, the authors showed that the divergent part of $\Sigma_2$  cancel out $\Sigma_{\rm adj}$  exactly, and when $d=4$ the two-loop correction is given by $\Sigma_{\rm ladder}$. 
As we will shortly see,  in generic $\mathcal{N}=2$ set-ups the situation is more involved at the field theory level.

\paragraph{Expectations based on localization} The localization description of the  $\mathcal{N}=2$ set-ups under consideration contains a non-trivial one-loop part, which suggests (see eq.(\ref{logZ1lexinrepR})) that the  interacting correction $\Sigma_\cR + \Sigma_2$ should be equivalently described by the diagrammatic difference
$g_B^4 \Delta\cW_4^{\rm flat}$ between the $\mathcal{N}=2$ and the $\mathcal{N}=4$ theory, namely by 
\begin{equation}
	\label{eq:bubble-exchange differen flat}
	\begin{split}
		g_B^4\Delta \cW^{\rm flat }_4 & 
		= \mathord{
			\begin{tikzpicture}[radius=2.cm, baseline=-0.65ex,scale=0.7]
				\draw [black] (0,0) circle [];
				\filldraw[color=gray!80, fill=gray!15](0,0) circle (1);	
				\draw [black]   (0,0) circle [radius=1.];
				\begin{feynman}
					\vertex (A) at (0,2.);
					\vertex (C) at (0, 1.);
					\vertex (D) at (0, -1.);
					\vertex (d) at (0.1,0.3) {\text{\footnotesize 1-loop }\normalsize} ;
					\vertex (d) at (0.1,-0.3) {\text{\footnotesize $\cR$ }\normalsize} ;
					\vertex (B) at (0,-2.);
					\diagram*{
						(A) -- [photon] (C),
						(A) --[ fermion] (C),
						(D) -- [photon] (B),
						(D) --[ fermion] (B)
					};
				\end{feynman}
		\end{tikzpicture} } -
		\mathord{
			\begin{tikzpicture}[radius=2.cm, baseline=-0.65ex,scale=0.7]
				\draw [black] (0,0) circle [];
				\filldraw[color=gray!80, fill=gray!15](0,0) circle (1);	
				\draw [black]   (0,0) circle [radius=1.];
				\begin{feynman}
					\vertex (A) at (0,2.);
					\vertex (C) at (0, 1.);
					\vertex (D) at (0, -1.);
					\vertex (B) at (0,-2.);\vertex (d) at (0.1,0.3) {\text{\footnotesize 1-loop }\normalsize} ;
					\vertex (d) at (0.1,-0.3) {\text{\footnotesize $\rm adj$ }\normalsize} ;
					\diagram*{
						(A) -- [photon] (C),
						(A) --[ fermion] (C),
						(D) -- [photon] (B),
						(D) --[ fermion] (B)
					};
				\end{feynman}
		\end{tikzpicture} }~.
	\end{split}
\end{equation} 
If the $\cN=2$ set-ups under consideration were superconformal, namely if the matter representation $\cR$ were such that $i_\cR = N$ and the $\beta$-function vanished, all the divergences would cancel out and  $\cW_4^{\rm flat}$ would be finite for $d\to 4$. Therefore, it would be perfectly legitimate to set $\Sigma_2 = - \Sigma_{\rm adj}$ \cite{Erickson:2000af} and consequently, it would be true that
\begin{equation}
	\label{w4isdiff}
	g_B^4\cW_4^{\rm flat} = g_B^4 \Delta \cW_4^{\rm flat} + \Sigma_{\rm ladder}~.
\end{equation} However, when the $\beta$-function is non-vanishing, eq. (\ref{eq:bubble-exchange differen flat}) only captures the divergent part of $\Sigma_2+\Sigma_\cR$ but it does not account for evanescent contributions which are hidden in $\Sigma_2$.  To clarify these aspects, we first compute (\ref{eq:bubble-exchange differen flat}) in detail and subsequently, we will compare the result with a direct evaluation of $\Sigma_2+\Sigma_\cR$. 

Let us begin with computing the one loop correction to the adjoint scalar propagator in the difference theory approach. The relevant diagrams are  \begin{equation}
	\label{eq:one-loop corrections adjoint}
	\begin{split}
		\mathord{
			\begin{tikzpicture}[scale=0.7, baseline=-0.65ex]
				\filldraw[color=gray!80, fill=gray!15](0,0) circle (1);	
				\draw [black] (0,0) circle [radius=1cm];
				\begin{feynman}
					\vertex (A) at (-2,0);
					\vertex (C) at (-1,0);
					\vertex (d) at (0.1,0.3) {\text{\footnotesize 1-loop }\normalsize} ;
					\vertex (d) at (0.1,-0.3) {\text{\footnotesize $\cR$ }\normalsize} ;
					\vertex (B) at (1, 0);
					\vertex (D) at (2, 0);
					\diagram*{
						(A) -- [fermion] (C),
						(B) --[fermion] (D),
					};
				\end{feynman}
			\end{tikzpicture} 
		} &= \mathord{\begin{tikzpicture}[baseline=-0.65ex,scale=0.7]
				\begin{feynman}
					\vertex (a) at (-1,0)  ;
					\vertex (b) at (4,0)  ;
					\vertex (c) at (0.5,0) ;
					\vertex (d) at (2.5,0) ;
					\vertex (e) at (1.5, 1.3) {$A A$} ;  
					\diagram*{
						(a) -- [fermion] (c),
						(c) -- [photon, half left] (d),
						(c) -- [fermion] (d),
						(d) -- [fermion] (b),
					};
				\end{feynman}
			\end{tikzpicture}
		} + \mathord{\begin{tikzpicture}[baseline=-0.65ex,scale=0.7]
				\begin{feynman}
					\vertex (a) at (-1,0)  ;
					\vertex (b) at (4,0)  ;
					\vertex (c) at (0.5,0) ;
					\vertex (d) at (2.5,0) ;
					\vertex (e) at (1.5, 1.3) {$\psi\Bar{\psi}$} ;
					\vertex (e) at (1.5, -1.3) {$\lambda\Bar{\lambda}$};
					\diagram*{
						(a) -- [fermion] (c),
						(c) -- [fermion, half left, thick] (d),
						(c) -- [fermion, half right, thick] (d),
						(d) -- [fermion] (b),
					};
				\end{feynman}
			\end{tikzpicture}
		}
		+ 	\mathord{\begin{tikzpicture}[baseline=-0.65ex,scale=0.7]
				\begin{feynman}
					\vertex (a) at (-1,0) ;
					\vertex (b) at (4,0) ;
					\vertex (c) at (0.5,0) ;
					\vertex (d) at (2.5,0) ;
					\vertex (e) at (1.5, 1.3) {$\eta\Bar{\eta}$};
					\vertex (e) at (1.5, -1.3) {$\Tilde{\eta}\Bar{\Tilde{\eta}}$};
					\diagram*{
						(a) -- [fermion] (c),
						(c) -- [anti charged scalar, half left, thick] (d),
						(c) -- [anti charged scalar, half right, thick] (d),
						(d) -- [fermion] (b),
					};
				\end{feynman}
			\end{tikzpicture}
		} \ ,
	\end{split} 
\end{equation} where in the previous expression $\psi$ and $\lambda$ are the two adjoint \textit{gauginos} of the $\mathcal{N}=2$ vector multiplet, while $\eta$ and $\tilde{\eta}$ are the two-component Weyl fermions of the $\mathcal{N}=2$ hypermultiplets in the representation $\cR$ of the gauge group. In the difference theory approach, the first two diagrams are in common with the $\mathcal{N}=4$ theory and consequently, they do not contribute. Using the results outlined in Appendix \ref{sec:Feynman computations in flat space}, we find that in momentum space the correction reads \begin{equation}
	\label{eq:one-loop correction adjoint scalar difference}
	\begin{split}
		\mathord{
			\begin{tikzpicture}[scale=0.7, baseline=-0.65ex]
				\filldraw[color=gray!80, fill=gray!15](0,0) circle (1);	
				\draw [black] (0,0) circle [radius=1cm];
				\begin{feynman}
					\vertex (A) at (-2,0);
					\vertex (C) at (-1,0);
					\vertex (d) at (0.1,0.3) {\text{\footnotesize 1-loop }\normalsize} ;
					\vertex (d) at (0.1,-0.3) {\text{\footnotesize $\cR$ }\normalsize} ;
					\vertex (B) at (1, 0);
					\vertex (D) at (2, 0);
					\diagram*{
						(A) -- [fermion] (C),
						(B) --[fermion] (D),
					};
				\end{feynman}
			\end{tikzpicture} 
		} - 	\mathord{
			\begin{tikzpicture}[scale=0.7, baseline=-0.65ex]
				\filldraw[color=gray!80, fill=gray!15](0,0) circle (1);	
				\draw [black] (0,0) circle [radius=1cm];
				\begin{feynman}
					\vertex (A) at (-2,0);
					\vertex (C) at (-1,0);
					\vertex (d) at (0.1,0.3) {\text{\footnotesize 1-loop }\normalsize} ;
					\vertex (d) at (0.1,-0.3) {\text{\footnotesize $\rm adj$ }\normalsize} ;
					\vertex (B) at (1, 0);
					\vertex (D) at (2, 0);
					\diagram*{
						(A) -- [fermion] (C),
						(B) --[fermion] (D),
					};
				\end{feynman}
			\end{tikzpicture} 
		} &= \mathord{\begin{tikzpicture}[baseline=-0.65ex,scale=0.7]
				\begin{feynman}
					\vertex (a) at (-1,0) ;
					\vertex (b) at (4,0) ;
					\vertex (c) at (0.5,0) ;
					\vertex (d) at (2.5,0) ;
					\vertex (e) at (1.5, 1.3) {$\eta\Bar{\eta}$};
					\vertex (e) at (1.5, -1.3) {$\Tilde{\eta}\Bar{\Tilde{\eta}}$};
					\diagram*{
						(a) -- [fermion] (c),
						(c) -- [anti charged scalar, half left, thick] (d),
						(c) -- [anti charged scalar, half right, thick] (d),
						(d) -- [fermion] (b),
					};
				\end{feynman}
			\end{tikzpicture}
		} - \mathord{\begin{tikzpicture}[baseline=-0.65ex,scale=0.7]
				\begin{feynman}
					\vertex (a) at (-1,0) ;
					\vertex (b) at (4,0) ;
					\vertex (c) at (0.5,0) ;
					\vertex (d) at (2.5,0) ;
					\vertex (e) at (1.5, 1.5) {$\psi_3\Bar{\psi}_3$};
					\vertex (e) at (1.5, -1.5) {$\psi_2\Bar{\psi}_2$};
					\diagram*{
						(a) -- [fermion] (c),
						(c) -- [anti fermion, half left, thick] (d),
						(c) -- [anti fermion, half right, thick] (d),
						(d) -- [fermion] (b),
					};
				\end{feynman}
			\end{tikzpicture}
		} \\ 
		&=-\dfrac{4g^2 }{p^{6-d}}\frac{\Gamma(d/2)\Gamma(2-d/2)\Gamma(d/2-1)}{\Gamma(d-1)(4\pi)^{d/2}} \Tr^\prime_{\mathcal{R}} {T}^a{T}^b \ , 
	\end{split}
\end{equation}  where we denoted the adjoint Weyl fermions of the $\mathcal{N}=4$ theory as $\psi_{j}$. Moreover, the colour factor is encoded in the following trace \begin{equation}
	\label{eq:trace}
	\Tr^\prime_{\mathcal{R}} {T}^a{T}^b\equiv\left(\Tr_\cR T^a T^b-\Tr_{\text{adj}}T^a T^b\right)=(i_\mathcal{R}-N)\delta^{ab} \ , 
\end{equation}
which  precisely reproduces the prediction of the interaction action in the localization approach (\ref{logZ1lexinrepR}). We now consider the one-loop correction to the gauge field propagator in the difference theory approach. The relevant diagrams in the $\mathcal{N}=2 $ theories are \begin{equation}
	\label{eq:one-loop diagrams gauge field}
	\begin{split}
		\mathord{
			\begin{tikzpicture}[scale=0.6, baseline=-0.65ex]
				\filldraw[color=gray!80, fill=gray!15](0,0) circle (1);	
				\draw [black] (0,0) circle [radius=1cm];
				\begin{feynman}
					\vertex (A) at (-2,0);
					\vertex (C) at (-1,0);
					\vertex (d) at (0.1,0.3) {\text{\footnotesize 1-loop }\normalsize} ;
					\vertex (d) at (0.1,-0.3) {\text{\footnotesize $\cR$ }\normalsize} ;
					\vertex (B) at (1, 0);
					\vertex (D) at (2, 0);
					\diagram*{
						(A) -- [photon] (C),
						(B) --[photon] (D),
					};
				\end{feynman}
			\end{tikzpicture} 
		}& = \mathord{\begin{tikzpicture}[baseline=-0.65ex,scale=0.6]
				\begin{feynman}
					\vertex (a) at (-0.5,0)  ;
					\vertex (b) at (3.5,0)  ;
					\vertex (c) at (0.5,0) ;
					\vertex (d) at (2.5,0) ;
					\vertex (e) at (1.5, 1.5) {$A A$};
					\diagram*{
						(a) -- [photon] (c),
						(c) -- [photon, half left] (d),
						(c) -- [photon, half right] (d),
						(d) -- [photon] (b),
					};
				\end{feynman}
			\end{tikzpicture}
		} + \mathord{\begin{tikzpicture}[baseline=-0.65ex,scale=0.6]
				\begin{feynman}
					\vertex (a) at (3.5,0)  ;
					\vertex (b) at (7.5,0)  ;
					\vertex (c) at (4.5,0) ;
					\vertex (d) at (6.5,0) ;
					\vertex (e) at (5.5, 1.5) {$c \bar{c}$};
					\diagram*{
						(a) -- [photon] (c),
						(c) -- [ghost, half left,thick] (d),
						(c) -- [ghost, half right,thick] (d),
						(d) -- [photon] (b),
					};
				\end{feynman}
			\end{tikzpicture}
		} + \mathord{\begin{tikzpicture}[baseline=-0.65ex,scale=0.6]
				\begin{feynman}
					\vertex (a) at (-0.5,0)  ;
					\vertex (b) at (3.5,0)  ;
					\vertex (c) at (0.5,0) ;
					\vertex (c1) at (0.6,0) ;
					\vertex (d) at (2.5,0) ;
					\vertex (d1) at (2.4,0) ;
					\vertex (e) at (1.5, 1.5) {$\bar{\phi}\phi$};
					\diagram*{
						(a) -- [photon] (c),
						(c) -- [fermion, half left] (d),
						(c) -- [anti fermion, half right] (d),
						(d) -- [photon] (b),
					};
				\end{feynman}
			\end{tikzpicture}
		} \\
		&+\mathord{\begin{tikzpicture}[baseline=-0.65ex,scale=0.6]
				\begin{feynman}
					\vertex (a) at (-0.5,0)  ;
					\vertex (b) at (3.4,0)  ;
					\vertex (c) at (0.5,0) ;
					\vertex (c1) at (0.6,0) ;
					\vertex (d) at (2.5,0) ;
					\vertex (d1) at (2.4,0) ;
					\vertex (e) at (1.5, 1.5) {$A_iA_i$};
					\diagram*{
						(a) -- [photon] (c),
						(c1) -- [scalar, half left] (d1),
						(c1) -- [scalar, half right] (d1),
						(d1) -- [photon] (b),
					};
				\end{feynman}
			\end{tikzpicture}
		} +  \mathord{\begin{tikzpicture}[baseline=-0.65ex,scale=0.6]
				\begin{feynman}
					\vertex (a) at (-0.5,0)  ;
					\vertex (b) at (3.5,0)  ;
					\vertex (c) at (0.5,0) ;
					\vertex (d) at (2.5,0) ;
					\vertex (e) at (1.5, 1.5) {$\psi\bar{\psi}/\lambda\bar{\lambda}$};
					\diagram*{
						(a) -- [photon] (c),
						(c) -- [fermion, half left,thick] (d),
						(c) -- [anti fermion, half right,thick] (d),
						(d) -- [photon] (b),
					};
				\end{feynman}
		\end{tikzpicture}}+  
		\mathord{\begin{tikzpicture}[baseline=-0.65ex,scale=0.6]
				\begin{feynman}
					\vertex (a) at (-0.5,0) ;
					\vertex (b) at (3.5,0)  ;
					\vertex (c) at (0.5,0) ;
					\vertex (d) at (2.5,0) ;
					\vertex (e) at (1.5, 1.5) {$\eta\bar{\eta}/\tilde{\eta}\bar{\tilde{\eta}}$};
					\diagram*{
						(a) -- [photon] (c),
						(c) -- [charged scalar, half left,thick] (d),
						(c) -- [anti charged scalar, half right,thick] (d),
						(d) -- [photon] (b),
					};
				\end{feynman}
			\end{tikzpicture}
		} + 	 	\mathord{\begin{tikzpicture}[baseline=-0.65ex,scale=0.6]
				\begin{feynman}
					\vertex (a) at (-0.5,0) ;
					\vertex (b) at (3.5,0)  ;
					\vertex (c) at (0.5,0) ;
					\vertex (d) at (2.5,0) ;
					\vertex (e) at (1.5, 1.5) {$\bar{q}q/\bar{\tilde{q}}\tilde{q}$};
					\diagram*{
						(a) -- [photon] (c),
						(c) -- [charged scalar, half left] (d),
						(c) -- [anti charged scalar, half right] (d),
						(d) -- [photon] (b),
					};
				\end{feynman}
			\end{tikzpicture}
		} \  ,
	\end{split} 
\end{equation} where in the previous expression $A_i$, with $i=1,\ldots 4-d$ are the adjoint real scalar fields resulting from dimensional reduction, while $q$ and $\tilde{q}$ are the complex scalars of the $\mathcal{N}=2 $ hypers in the representation $\mathcal{R}$ of the gauge group. In the difference theory approach, only the contribution resulting from the matter content diagrams survive.  
Using again the results in Appendix \ref{sec:Feynman computations in flat space}, we find that 
\begin{equation}
	\label{eq:one-loop correction gauge field difference}
	\begin{split}
		\mathord{
			\begin{tikzpicture}[scale=0.7, baseline=-0.65ex]
				\filldraw[color=gray!80, fill=gray!15](0,0) circle (1);	
				\draw [black] (0,0) circle [radius=1cm];
				\begin{feynman}
					\vertex (A) at (-2,0);
					\vertex (C) at (-1,0);
					\vertex (d) at (0.1,0.3) {\text{\footnotesize 1-loop }\normalsize} ;
					\vertex (d) at (0.1,-0.3) {\text{\footnotesize $\cR$ }\normalsize} ;
					\vertex (B) at (1, 0);
					\vertex (D) at (2, 0);
					\diagram*{
						(A) -- [photon] (C),
						(B) --[photon] (D),
					};
				\end{feynman}
			\end{tikzpicture} 
		} &- 	\mathord{
			\begin{tikzpicture}[scale=0.7, baseline=-0.65ex]
				\filldraw[color=gray!80, fill=gray!15](0,0) circle (1);	
				\draw [black] (0,0) circle [radius=1cm];
				\begin{feynman}
					\vertex (A) at (-2,0);
					\vertex (C) at (-1,0);
					\vertex (d) at (0.1,0.3) {\text{\footnotesize 1-loop }\normalsize} ;
					\vertex (d) at (0.1,-0.3) {\text{\footnotesize $\rm adj$ }\normalsize} ;
					\vertex (B) at (1, 0);
					\vertex (D) at (2, 0);
					\diagram*{
						(A) -- [photon] (C),
						(B) --[photon] (D),
					};
				\end{feynman}
			\end{tikzpicture} 
		} \\ 
		&=-\dfrac{4g^2 }{(4\pi)^{d/2}}\frac{\Gamma(d/2)\Gamma(2-d/2)\Gamma(d/2-1)}{\Gamma(d-1)}\dfrac{P_{\mu \nu}}{p^{6-d}} \Tr_{\mathcal{R}}^\prime {T}^a{T}^b \ .
	\end{split}
\end{equation} The previous expression, up to spacetime indices encoded in the transverse projector $P_{\mu \nu}=\delta_{\mu \nu}-p_\mu p_\nu/p^2$, coincides with the correction of the adjoint scalar (\ref{eq:one-loop correction adjoint scalar difference}), as expected from supersymmetry. Finally, exploiting translation invariance and the Fourier transform (\ref{eq:Fourier transform for massless propagators}), we obtain the configuration space expression eq.s (\ref{eq:one-loop correction adjoint scalar difference}) and (\ref{eq:one-loop correction gauge field difference}). For instance, using (\ref{eq:trace}) we find that the one-loop correction to the adjoint scalar propagator in the difference theory becomes \begin{equation}
	-\dfrac{4g^2 }{p^{6-d}}\frac{\Gamma(d/2)\Gamma(2-d/2)\Gamma(d/2-1)}{\Gamma(d-1)(4\pi)^{d/2}} \Tr_{\mathcal{R}}^\prime {T}^a{T}^b  \quad \to \quad - \dfrac{(i_{\mathcal{R}}-N) \delta^{ab} g_B^2\Gamma^2(d/2-1)}{2^5\pi^d(2-d/2)(d-3)[x_{12}^2]^{d-3}} \ .
\end{equation} Analogously, the net result for the one-loop correction to the gauge field propagator (\ref{eq:one-loop correction gauge field difference})  is given by the previous expression multiplied by $\delta_{\mu \nu}$. Indeed, upon Fourier transform,  the contribution in the transverse projector involving $p_\mu p_\nu/p^2$ gives rise to total derivatives which vanish when integrated along the Wilson loop. 
As a result, we find that 
\begin{equation}
	\label{eq:unrenormalized combination of spider+bubble }
	\begin{split} 
		g_B^4	\Delta \cW^{\rm flat}_4	&= - \Bigg( \dfrac{ C_F g^4_B
			(i_\mathcal{R}-N) \Gamma^2(d/2-1) }{16\pi^2(d-3)(2-d/2)} \Bigg) \times  \oint_C \dfrac{\dd{s_1}\dd{s_2}}{4 \pi^{d-2}} \dfrac{R^2 - \dot{x}_1 \cdot \dot{x}_2}{[x_{12}^2]^{d-3}}       \\
		&= -\beta^{\mathcal{R}}_0\times\Bigg( \dfrac{g_B^4C_F\Gamma^2(d/2-1)\Gamma(9/2-d)R^{8-2d}}{2^{2d-6} \pi^{d-2-3/2}(2-d/2)(d-3)\Gamma(5-d)}\Bigg) \ ,
	\end{split}
\end{equation} where to obtain the second line we employed the parametrization (\ref{eq:parametrization}) and the master integral (\ref{eq:master integral}). Moreover, the previous expression is proportional to the coefficient $\beta_0^\cR$, defined in (\ref{eq:defbeta0}), which fully characterizes the one-loop exact $\beta$-function of these theories.

\paragraph{Direct evaluation of $\Sigma_2+\Sigma_\cR$}
We now want to evaluate $\Sigma_\cR+\Sigma_2$ in a direct way, without considering the difference theory approach. Firstly, we derive  $\Sigma_\mathcal{R}$ in the $\mathcal{N}=2$ theories we are considering. Working again in momentum space and using the results in Appendix \ref{sec:Feynman computations in flat space}, we find that (\ref{eq:one-loop corrections adjoint})
becomes  \begin{equation}
	\label{eq:one-loop momentum adjoint}
	\begin{split}
		\mathord{
			\begin{tikzpicture}[scale=0.7, baseline=-0.65ex]
				\filldraw[color=gray!80, fill=gray!15](0,0) circle (1);	
				\draw [black] (0,0) circle [radius=1cm];
				\begin{feynman}
					\vertex (A) at (-2,0);
					\vertex (C) at (-1,0);
					\vertex (d) at (0.1,0.3) {\text{\footnotesize 1-loop }\normalsize} ;
					\vertex (d) at (0.1,-0.3) {\text{\footnotesize $\cR$ }\normalsize} ;
					\vertex (B) at (1, 0);
					\vertex (D) at (2, 0);
					\diagram*{
						(A) -- [fermion] (C),
						(B) --[fermion] (D),
					};
				\end{feynman}
			\end{tikzpicture} 
		} &= -\dfrac{g^2 \delta^{ab}4i_\mathcal{R}}{(4\pi)^{d/2}p^{6-d}}\frac{\Gamma(d/2)\Gamma(2-d/2)\Gamma(d/2-1)}{\Gamma(d-1)} \ .
	\end{split} 
\end{equation} The previous expression coincides with (\ref{eq:one-loop correction adjoint scalar difference}), but it is characterized by the colour coefficient $i_\mathcal{R}$ rather than  ($i_\mathcal{R}-N)$. This is due to the fact that first two diagrams in eq. (\ref{eq:one-loop corrections adjoint}), which are not present in the difference theory but in principle contribute to $\Sigma_\cR$, actually cancel each other.  

Analysing the corrections to the gauge field propagator, we find that (\ref{eq:one-loop diagrams gauge field}) is given by eq. (\ref{eq:one-loop momentum adjoint}) multiplied by the projector $P_{\mu\nu}$, as expected from supersymmetry, i.e. 

\begin{equation}
	\label{eq:one-loop momentum gauge}
	\begin{split}
		\mathord{
			\begin{tikzpicture}[scale=0.7, baseline=-0.65ex]
				\filldraw[color=gray!80, fill=gray!15](0,0) circle (1);	
				\draw [black] (0,0) circle [radius=1cm];
				\begin{feynman}
					\vertex (A) at (-2,0);
					\vertex (C) at (-1,0);
					\vertex (d) at (0.1,0.3) {\text{\footnotesize 1-loop }\normalsize} ;
					\vertex (d) at (0.1,-0.3) {\text{\footnotesize $\cR$ }\normalsize} ;
					\vertex (B) at (1, 0);
					\vertex (D) at (2, 0);
					\diagram*{
						(A) -- [photon] (C),
						(B) --[photon] (D),
					};
				\end{feynman}
			\end{tikzpicture} 
		} &= -\dfrac{g^2 \delta^{ab}4i_\mathcal{R}}{(4\pi)^{d/2}p^{6-d}}\frac{\Gamma(d/2)\Gamma(2-d/2)\Gamma(d/2-1)}{\Gamma(d-1)} P_{\mu \nu} \ .
	\end{split} 
\end{equation}  To compute $\Sigma_\mathcal{R}$ we Fourier transform the one-loop corrections (\ref{eq:one-loop momentum adjoint}) and (\ref{eq:one-loop momentum gauge}) to configuration space and we insert them in the Wilson loop. Obviously, the result coincides with the difference theory expression (\ref{eq:unrenormalized combination of spider+bubble }) with just $i_\cR$ instead of $i_\cR - N$:
\begin{equation}
	\begin{split}
		\label{eq:bubble-exchange diagram}
		\Sigma_\cR &=  
		\Bigg( \dfrac{ C_F (-i_{\mathcal{R}})g_B^4 \Gamma^2(d/2-1) }{16\pi^2(d-3)(2-d/2)} \Bigg) \times \Bigg( \oint_C \dfrac{\dd{s_1}\dd{s_2}}{4 \pi^{d-2}} \dfrac{R^2 - \dot{x}_1 \cdot \dot{x}_2}{[x_{12}^2]^{d-3}}\Bigg) \ .
	\end{split}
\end{equation} 

We now consider $\Sigma_2$ in (\ref{eq:sigma1 e 2}). This term results from two diagrams with internal vertex, associated wit the pure gauge and the gauge-scalar interaction which can be extracted from the action (\ref{eq:pure super-Yang-Mills N=2}). The expression for $\Sigma_2$ is \cite{Erickson:2000af}  
\begin{equation}
	\label{eq:spider con integrazione su s}
	\begin{split}
		\Sigma_2 &= - \dfrac{g_B^4(N^2-1)}{4} \oint \dd^3{s} \  \epsilon(s_1,s_2,s_3)  \ (R^2-\dot{x}_1\cdot \dot{x}_3) \times \Big( \dot{x}_2\cdot \partial_{x_1} \int \dd^d{y}\prod_{i=1}^3\Delta(x_i-y)\Big) \\
		&= \dfrac{g_B^4 C_FN \Gamma(d-2) }{R^{2d-8}2^{d+4}\pi^d} \int_0^1 \dd F \ \oint\dd^3{s} \ \epsilon(s_1,s_2,s_3)  \ \dfrac{(\alpha(1-\alpha)\sin{(s_{12})}-\alpha\gamma\sin{(s_{32})})} {(1 - \cos{(s_{13})})^{-1}[Q]^{d-2}} \ , 
	\end{split}
\end{equation} where in the previous expression $\epsilon(s_1,s_2,s_3)$ is the path ordering symbol equal to $1$ when  $s_1>s_2>s_3$, while 
\begin{equation}
	\label{defDelta}
	\Delta(x_i-y)
	= \dfrac{\Gamma(d/2-1)}{4\pi^{d/2}[(x-y)^2]^{d/2-1}}~.
\end{equation} 
Note that, despite the notation, $\Delta(x)$ is actually a function of the norm $x^2$ only.  
To obtain the second line in eq. (\ref{eq:spider con integrazione su s}) we employed the Feynman parameters, evaluated the integral over the internal vertex and defined 
\begin{equation}
	\begin{split}
		\dd F &= \dd\alpha\dd\beta\dd\gamma\delta(1-\alpha-\beta-\gamma)(\alpha\beta\gamma)^{d/2-2}\, ,\\
		Q &=  \alpha\beta(1-\cos s_{12} )+\alpha\gamma(1 - \cos{s_{13}}) + \beta\gamma(1 - \cos{s_{23}}) \ .
	\end{split}
\end{equation}
The key observation here is that the three-point functions inside the Wilson loop contour in $\Sigma_2$, see eq. (\ref{eq:sigma1 e 2}), do not contain singularities as long as we keep the external points separated. However, as it was showed in  \cite{Erickson:2000af},  the integration over the coordinates $s_i$ leads to a short-distance divergence which is precisely (minus) the bubble-like diagram of the $\mathcal{N}=4$ theory with an additional evanescent contribution\footnote{See Section 4.2 for more details about this point.}, i.e.  
\cite{Erickson:2000af}
\begin{equation}	
	\label{eq:net result spider}
	\begin{split}
		\Sigma_2 &= -\Sigma_{\mathrm{adj}} +  \dfrac{g_B^4 C_FN \Gamma(d-2) (d-4) }{R^{2d-8}2^{d+4}\pi^d (d-3)} \int_0^1 \dd F \ \oint\dd^3{s} \ \epsilon(s_1,s_2,s_3)  \ \dfrac{\sin s_{31}} {Q^{d-3}}\\
		&\equiv-\Sigma_{\mathrm{adj}}+g_B^4\Sigma_{\mathrm{Ev}} \ . 
	\end{split}
\end{equation} 
In the previous expression $\Sigma_{\mathrm{adj}}$ is the {bubble-like} diagram of the $\mathcal{N}=4$ theory, which can be obtained from (\ref{eq:bubble-exchange diagram}) setting $\cR=\mathrm{adj}$ so that $i_\mathrm{adj}=N$. This function has a (UV) pole in $d=4$, while  $\Sigma_{\mathrm{Ev}}$ is completely finite in every dimension $d$ and goes to zero when $d\to 4$. This second contribution is an interesting example of evanescent function generated by the integration over the contour. 

Combining the previous expression with the divergent contributions $\Sigma_\cR$ (\ref{eq:bubble-exchange diagram}), we find that  
we find altogether
\begin{equation}
	\begin{split}
		\Sigma_\mathcal{R} + \Sigma_2 &= \Sigma_\mathcal{R}-\Sigma_{\rm adj} +g_B^4\Sigma_{\mathrm{Ev}}\\
		&= g_B^4\Delta \cW^{\rm flat}_4 + g_B^4\Sigma_{\mathrm{Ev}} \ ,
	\end{split}
\end{equation}  
where $	g_B^4\Delta \cW^{\rm flat}_4$ describes the diagrammatic difference (\ref{eq:bubble-exchange differen flat}) at order $g_B^4$ between the $\mathcal{N}=2 $ and the $\mathcal{N}=4$ theory and is explicitly given by (\ref{eq:unrenormalized combination of spider+bubble }). 

\paragraph{The ladder diagram}
Let us consider the explicit expression of $\Sigma_{\rm ladder}$. Using the tree-level propagators (\ref{eq:tre-level props }) and the parametrization (\ref{eq:parametrization}), we find  
\begin{equation}
	\begin{split}
		\label{eq:ladder diagram at order g_0^4}
		\Sigma_{\rm ladder} =	\mathord{
			\begin{tikzpicture}[baseline=-0.65ex,scale=0.8]
				\draw [black] (0,0) circle [radius=1.5cm];
				\begin{feynman}
					\vertex (A) at (-0.75,1.3);
					\vertex (B) at (-0.75,-1.3);
					\vertex (C) at (0.75,1.3);
					\vertex (D) at (0.75,-1.3);
					\diagram*{
						(A) -- [photon] (B),
						(A) --[ fermion] (B),
						(C) -- [photon] (D),
						(C) --[ fermion] (D)
					};
				\end{feynman}
			\end{tikzpicture} 
		} = \Bigg( \dfrac{g_B^4\Gamma^2(d/2-1)}{2^{6}\pi^dR^{2d-8}N}\Bigg) \times \Big( \cI_1(d)+\cI_2(d)+\cI_3(d)\Big)  \ . 
	\end{split}
\end{equation}  The sum in the previous expression involves three different functions of the dimension $d$ resulting from the following  nested integrals  
\begin{equation}
	\label{eq:nested}
	\begin{split}
		\mathcal{I}_1(d)&= \dfrac{(N^2-1)^2}{N} \oint_{\mathcal{D}} \dd^4{s} \dfrac{(1-\cos(s_{12}))(1-\cos(s_{34}))}{(4\sin^2{\dfrac{s_{12}}{2}}4\sin^2{\dfrac{s_{34}}{2}})^{d/2-1}}\ , \\
		\mathcal{I}_2(d)&= -\dfrac{N^2-1}{N} \oint_{\mathcal{D}} \dd^4{s} \dfrac{(1-\cos(s_{13}))(1-\cos(s_{24}))}{(4\sin^2{\dfrac{s_{13}}{2}}4\sin^2{\dfrac{s_{24}}{2}})^{d/2-1}} \ , \\
		\mathcal{I}_3(d)&= \dfrac{(N^2-1)^2}{N} \oint_{\mathcal{D}} \dd^4{s} \dfrac{(1-\cos(s_{14}))(1-\cos(s_{23}))}{(4\sin^2{\dfrac{s_{14}}{2}}4\sin^2{\dfrac{s_{23}}{2}})^{d/2-1}} \ ,
	\end{split}
\end{equation} where $\mathcal{D}$ is the ordered region  $s_1>s_2>s_3>s_4$. The previous expression is finite%
\footnote{It is straightforward to note that when $d\to 4$ the integrands in (\ref{eq:nested}) are constant.}
when $d\to 4 $; for arbitrary $d$ the nested integration is a potential source of an evanescent function $\Sigma^\prime_{\rm Ev}$ proportional to $d-4$, whose detailed form we do not need here%
\footnote{The explicit expression of $\Sigma_{\rm Ev}$ in eq. (\ref{eq:net result spider}) was given just as an example.
}. Therefore, we can write 
\begin{equation}
	\label{ladderis} 	
	\Sigma_{\mathrm{ladder}}= g_B^4\widetilde{\Sigma}_{\rm ladder } + g_B^4\Sigma^\prime_{\rm Ev}\ ,
\end{equation} where we introduced \begin{equation}
	\label{eq:ladder finite}
	\widetilde{\Sigma}_{\rm ladder }= \dfrac{C_F}{192N}(2N^2-3) \ .
\end{equation}

Including the previous contribution, the final result for the dimensionally regularized vacuum expectation value of (\ref{eq:1/2 BPS supersymmetric Wilson loop}) up to order $g_B^4$  reads \begin{equation}
	\begin{split}
		\label{eq:finally expression dimensionally regularized vev in flat space}
		\mathcal{W}^{\mathrm{flat}}= 1+ g_B^2{\cW}_2 
		+ g_B^4 \left(\Delta \cW^{\rm flat}_4  + \widetilde{\Sigma}_{\rm ladder} + \Sigma_{\mathrm{Ev}} + \Sigma^\prime_{\mathrm{Ev}}\right)+ \ldots \ .
	\end{split}
\end{equation}  

\section{Supersymmetric Wilson loop on the sphere}
\label{sec:Supersymmetric WL on the sphere}
In this section, we study the expectation value of the 1/2 BPS Wilson loop on the four-sphere $\mathbb{S}^4$ in a general $\mathcal{N}=2$ SYM theory with massless hypermultiplets in the representation $\mathcal{R}$ of the gauge group. For a generic choice of local coordinates $x^{\mu}$ on the manifold, the 1/2 BPS Wilson loop operator takes the following form \cite{Belitsky:2020hzs}
\begin{equation}
	\label{eq:1/2 BPS supersymmetric Wilson loop on S}
	\widehat{\mathcal{W}}_S = \dfrac{1}{N} \text{tr} \ \mathcal{P} \exp \bigg\{ {g}_B \int_{{C}} \dd{s} \bigg[ \mathrm{i} A^{\mu}(x(s)) \dot{x}_{\mu}(s) + \dfrac{{R}}{\sqrt{2}}(\phi+\bar{\phi})(x(s)) \bigg]
	\bigg\}\ .
\end{equation} In the previous expression, the gauge field $A_\mu(x(s))$ and the vector multiplet adjoint scalar $\phi(x(s))$  are integrated over a great circle ${C}$ of radius $R$ and  we contracted the spacetime indices via the metric tensor $g_{\mu \nu}(x)$, with $\mu=1,\ldots,4$. 

To regularize the divergent contributions in the vacuum expectation value of $\widehat{\cW}_S$, we find convenient to compactify the theory on $\mathbb{S}^d$ via the embedding formalism following \cite{Belitsky:2020hzs,Drummond:1975yc}. In this approach, the $d$-dimensional sphere $\mathbb{S}^d$ is regarded as a  submanifold of $\mathbb{R}^{d+1}$ defined by the equation
\begin{equation}
	\label{eq:Sd}
	X^M X_M={R}^2 ~, 
\end{equation} 
with $X^M$ being flat coordinates on $\mathbb{R}^{d+1}$. 
Employing the stereographic projection
\begin{equation}
	\label{eq:stereographic projection}
	X_{\mu} = {R}\dfrac{2x_{\mu}{R}}{x^2+{R}^2}\ ,  \quad \quad 
	X_{d+1} ={R} \dfrac{{R}^2-x^2}{{R}^2+x^2}
\end{equation} 
we can parametrize this sphere with the coordinates $x^\mu$, where $\mu=1,\ldots, d$, associated with the hyperplane $X_{d+1}=0$ that passes through the equator. We choose the embedding in such a way that the circle $C$ lays in this hyperplane. In the previous expression, by $x^2$ we mean  $x^\mu x_\mu$.   

In terms of the embedding coordinates the 1/2 BPS Wilson loop (\ref{eq:1/2 BPS supersymmetric Wilson loop on S}) takes a form analogous to that in flat space \cite{Belitsky:2020hzs}:
\begin{equation}
	\label{eq:1/2 BPS supersymmetric Wilson loop on S embedding formalism}
	\widehat{\cW}_S= \dfrac{1}{N} \text{tr} \ \mathcal{P} \exp \bigg\{ {g}_B \int_{{C}} \dd{s} \bigg[ \mathrm{i} A^{M}(X(s)) \dot{X}_{M}(s) + \dfrac{{R}}{\sqrt{2}}(\phi+\bar{\phi})(X(s))\bigg]
	\bigg\} \ ,
\end{equation} 
where $A_M(X(s))$ is related to the gauge field  $A_\mu(x)$ on the sphere by
\begin{equation}
	\label{eq:pullback}
	A_{\mu}(x)= \dfrac{\partial X^M}{\partial{x}^{\mu}}A_{M}(X)
\end{equation}
and the loop $C$ is parameterized by 
\begin{equation}
	\label{eq:parametrization on the sphere}
	X^M(s) ={R}(\cos(s),\sin(s),\mathbf{0}) \ . 
\end{equation}  

Expanding in power series of the bare coupling we write
\begin{equation}
	\begin{split}
		\label{eq:dimensionally regularized expectation value on S}
		\vvev{\widehat{\cW}_S} = \mathcal{W}=  1+ {g}_B^2\cW_2 + {g}_B^4 \cW_4 + \ldots \ , 
	\end{split}
\end{equation} where  $\cW_i$ are again functions of the dimension $d$ and of the radius $R$ encoding the $i/2$-th loop correction.  Note that 
in arbitrary dimension $d$ the functions 
$\cW_i$ are not, in general, expected to coincide with their flat space counterparts appearing in (\ref{eq:dimensionally regularized vev in flat}).

\subsection{One-loop correction}
Let us begin with computing the first correction, i.e. $g_B^2\cW_2$. The tree-level propagators on the sphere for the massless scalar field $\phi(X)$ and of the gauge field $A_M(X)$ in the Feynman gauge are given by
\cite{Belitsky:2020hzs} (see also eq. \ref{eq:scalar prop on S} in Appendix \ref{sec:actionsonS})
\begin{align}
	\label{propsphere}
	D^{ab}(X_{12}) & =  \dfrac{\delta^{ab}\Gamma(d/2-1)}{4\pi^{d/2}(X_{12}^2)^{d/2-1}} = \delta^{ab} \Delta(X_{12})
	\ , 
	\notag\\
	D^{ab}_{MN}(X_{12}) & =  \dfrac{\delta^{ab}\delta_{MN}\Gamma(d/2-1)}{4\pi^{d/2}(X_{12}^2)^{d/2-1}} 
	= \delta^{ab}\delta_{MN} \Delta(X_{12})	\ ,
\end{align} where we recall that the function $\Delta(X)$ is defined in (\ref{defDelta}).
Taking the relevant Wick contractions leads then to 
\begin{equation}
	\begin{split}
		\label{eq:ladder g2 on S}
		g_B^2 \cW_2 = 
		\mathord{ \begin{tikzpicture}[baseline=-0.65ex,scale=0.8]
				\draw [black] (0,0) circle [radius=1.5cm];
				\begin{feynman}
					\vertex (A) at (0,1.5);
					\vertex (B) at (0,-1.5);
					\diagram*{
						(A) -- [photon] (B),
						(A) --[ charged scalar] (B)
					};
				\end{feynman}
			\end{tikzpicture} 
		} &=\bigg( \dfrac{{g}_B^2\Gamma(d/2-1) C_F}{2}\bigg) \times \bigg(\oint_C \dfrac{\dd^2{s}}{4 \pi^{d/2}} \dfrac{{R}^2 - \dot{X}_1 \cdot \dot{X}_2}{[X_{12}^2]^{d/2-1}}\bigg)  \ .
	\end{split}
\end{equation}  
Using the parametrization (\ref{eq:parametrization on the sphere}), we find that actually  
$\cW_2$ coincides with the flat space expression $\cW_2^{\rm flat}$ given in eq. (\ref{eq:ladder g2}). 
\subsection{Two-loop corrections}
\label{sec:radiative correction on S}
In this section, we study the two-loop correction to the expectation value of (\ref{eq:1/2 BPS supersymmetric Wilson loop on S embedding formalism}), i.e. $g_B^4\cW_4$. The diagrams we have to consider are analogous to those we studied in flat space in (\ref{eq:sigma1 e 2}). Thus, we set
$g_B^4\cW_4=\Sigma_\cR^S+\Sigma_2^S+\Sigma_{\mathrm{ladder}}^S$, where 
\begin{equation}
	\label{eq:sigma on the sphere}
	\Sigma^S_\mathcal{R} = \mathord{
		\begin{tikzpicture}[radius=2.cm, baseline=-0.65ex,scale=0.7]
			\draw [black] (0,0) circle [];
			\filldraw[color=gray!80, fill=gray!15](0,0) circle (1);	
			\draw [black]   (0,0) circle [radius=1.];
			\begin{feynman}
				\vertex (A) at (0,2.);
				\vertex (C) at (0, 1.);
				\vertex (D) at (0, -1.);
				\vertex (B) at (0,-2.);
				\vertex (d) at (0.1,0.3) {\text{\footnotesize 1-loop }\normalsize} ;
				\vertex (d) at (0.1,-0.3) {\text{\footnotesize $\cR$ }\normalsize} ;
				\diagram*{
					(A) -- [photon] (C),
					(A) --[ fermion] (C),
					(D) -- [photon] (B),
					(D) --[ fermion] (B)
				};
			\end{feynman},
	\end{tikzpicture} } \ , \quad 
	\Sigma^S_2=	\mathord{
		\begin{tikzpicture}[radius=2.cm, scale=0.7, baseline=-0.65ex]
			\draw [black] (0,0) circle [];
			\begin{feynman}
				\vertex (A) at (0,2.);
				\vertex (C) at (0,0);
				\vertex (D) at (-1.5, -1.3);
				\vertex (B) at (1.5, -1.3);
				\diagram*{
					(A) -- [photon] (C),
					(A) --[ fermion] (C),
					(C) -- [photon] (B),
					(C) --[ fermion] (D),
					(C) --[ photon] (D)
				};
			\end{feynman} 
	\end{tikzpicture} } \ , \quad 
	\Sigma^S_{\rm ladder} = 
	\mathord{
		\begin{tikzpicture}[baseline=-0.65ex,scale=0.7]
			\draw [black] (0,0) circle [radius=2cm];
			\begin{feynman}
				\vertex (A) at (-0.75,1.85);
				\vertex (B) at (-0.75,-1.85);
				\vertex (C) at (0.75,1.85);
				\vertex (D) at (0.75,-1.85);
				\diagram*{
					(A) -- [photon] (B),
					(A) --[ fermion] (B),
					(C) -- [photon] (D),
					(C) --[ fermion] (D)
				};
			\end{feynman}
		\end{tikzpicture} 
	}~.
\end{equation}   

The direct evaluation of $\Sigma_2^{S}$ on $\mathbb{S}^d$ is prohibitively complicated. However, as we explained in the previous section, we can extract their divergent part by taking the diagrammatic difference between our $\mathcal{N}=2$ set-up and $\mathcal{N}=4$ SYM. 
Let us stress again that this method does not keep track of possible evanescent functions contained in $\Sigma_2^S$, such as that appearing in (\ref{eq:net result spider}). In other words, we will have
\begin{equation}
	\label{Ssev}
	\Sigma_\cR^S + \Sigma_2^S = g_B^4\left(\Delta\cW_4 + \Sigma_{\mathrm{Ev}}^S\right)\ ,
\end{equation}
where $\Delta\cW_4$ encodes the divergent part of $\Sigma_\cR^S + \Sigma_2^S$ and arises from the difference method, while $\Sigma_{\mathrm{Ev}}^S$ is a regular function which vanishes when $d\to 4$.  As we will explain in detail in Section \ref{sec:RNW}, $\Sigma_{\mathrm{Ev}}^S$ only contributes at higher orders in perturbation theory and consequently, we will not need here its explicit expression.

On the $d$-dimensional sphere, the diagrammatic difference between the $\mathcal{N}=2$ theories under consideration and $\mathcal{N}=4$ SYM can be cast in the following way  
\begin{equation}
	\label{eq:radiative corrections on the sphere}
	\begin{split}
		g_B^4\Delta \cW_4 & =\mathord{
			\begin{tikzpicture}[radius=2.cm, baseline=-0.65ex,scale=0.7]
				\draw [black] (0,0) circle [];
				\filldraw[color=gray!80, fill=gray!15](0,0) circle (1);	
				\draw [black]   (0,0) circle [radius=1.];
				\begin{feynman}
					\vertex (A) at (0,2.);
					\vertex (C) at (0, 1.);
					\vertex (D) at (0, -1.);
					\vertex (d) at (0.1,0.3) {\text{\footnotesize 1-loop }\normalsize} ;
					\vertex (d) at (0.1,-0.3) {\text{\footnotesize $\cR$ }\normalsize} ;
					\vertex (B) at (0,-2.);
					\diagram*{
						(A) -- [photon] (C),
						(A) --[ charged scalar] (C),
						(D) -- [photon] (B),
						(D) --[ charged scalar] (B)
					};
				\end{feynman}
		\end{tikzpicture} } -
		\mathord{
			\begin{tikzpicture}[radius=2.cm, baseline=-0.65ex,scale=0.7]
				\draw [black] (0,0) circle [];
				\filldraw[color=gray!80, fill=gray!15](0,0) circle (1);	
				\draw [black]   (0,0) circle [radius=1.];
				\begin{feynman}
					\vertex (A) at (0,2.);
					\vertex (C) at (0, 1.);
					\vertex (D) at (0, -1.);
					\vertex (B) at (0,-2.);\vertex (d) at (0.1,0.3) {\text{\footnotesize 1-loop }\normalsize} ;
					\vertex (d) at (0.1,-0.3) {\text{\footnotesize $\rm adj$ }\normalsize} ;
					\diagram*{
						(A) -- [photon] (C),
						(A) --[ charged scalar] (C),
						(D) -- [photon] (B),
						(D) --[ charged scalar] (B)
					};
				\end{feynman}
		\end{tikzpicture} } \\
		& = \dfrac{{g}^2_B}{2N} \tr T^a T^b \oint \dd^2{s}\big[ R^2 {D}^{(1)}_{ab}(X_{12})- \dot{X}^1_M\dot{X}^2_{N}{D}_{ab}^{(1),MN} (X_{12})\big]\ ,
	\end{split}
\end{equation}
where we denoted with $D^{(1)}_{ab}(X_{12})$ and  ${D}_{ab}^{(1),MN} (X_{12})$ the one-loop scalar and gauge propagators in the difference theory.

The diagrams we have to compute to determine the one-loop corrections in the difference theory approach are a generalization of those we considered in flat space (see eq.s (\ref{eq:one-loop corrections adjoint}) and (\ref{eq:one-loop diagrams gauge field})). In Euclidean space, the analysis of these diagrams is simplified by employing translation invariance and by going to the momentum space. On the sphere, the Fourier analysis uses the spherical harmonics. However, we find more convenient to evaluate the diagrams in configuration space, where we can exploit the integration over the Wilson loop contour to neglect total derivatives. 

We begin with considering ${D}^{(1)}_{ab}$. As we already explained in Section \ref{sec:Insertion}, in the difference theory approach only the diagrams resulting from the matter content are relevant. Therefore, ${D}^{(1)}_{ab}$ receives corrections from the following fermionic loops:
\begin{equation}
	\label{eq:expression for D on S main corpus}
	\begin{split}
		{D}_{ab}^{(1)}(X_{12}) &=	\mathord{\begin{tikzpicture}[baseline=-0.65ex,scale=0.7]
				\begin{feynman}
					\vertex (a) at (-1,0) ;
					\vertex (b) at (4,0) ;
					\vertex (c) at (0.5,0) ;
					\vertex (d) at (2.5,0) ;
					\vertex (e) at (1.5, 1.4) {$\eta\Bar{\eta}$};
					\vertex (e) at (1.5, -1.4) {$\Tilde{\eta}\Bar{\Tilde{\eta}}$};
					\diagram*{
						(a) -- [fermion] (c),
						(c) -- [anti charged scalar, half left, thick] (d),
						(c) -- [anti charged scalar, half right, thick] (d),
						(d) -- [fermion] (b),
					};
				\end{feynman}
			\end{tikzpicture}
		} - \mathord{\begin{tikzpicture}[baseline=-0.65ex,scale=0.7]
				\begin{feynman}
					\vertex (a) at (-1,0) ;
					\vertex (b) at (4,0) ;
					\vertex (c) at (0.5,0) ;
					\vertex (d) at (2.5,0) ;
					\vertex (e) at (1.5, 1.4) {$\psi_3\Bar{\psi}_3$};
					\vertex (e) at (1.5, -1.4) {$\psi_2\Bar{\psi}_2$};
					\diagram*{
						(a) -- [fermion] (c),
						(c) -- [anti fermion, half left, thick] (d),
						(c) -- [anti fermion, half right, thick] (d),
						(d) -- [fermion] (b),
					};
				\end{feynman}
			\end{tikzpicture}
		} \\
		&=4\left(\Tr^\prime_{\mathcal{R}} {T}_a{T}_b\right){g}_B^2\int \dd Z_1\dd Z_2\, \Delta(X_1-Z_1) \,\Delta(Z_2-X_2)\, {D}^2(Z_{12})\, Z^2_{12}\\
		&\equiv \left(-\Tr^\prime_{\mathcal{R}} {T}_a{T}_b\right){g}_B^2\,I(X_{12}) \ , 
	\end{split}
\end{equation} 
where the function $D(Z_{12})$ is given in (\ref{eq:propagator Majorana in S}) and we recall that the primed trace $\Tr^\prime_\cR$ is defined in (\ref{eq:trace}). The  integration measure on the $d$-dimensional sphere is
given by $\dd Z = \dd^{d}z\sqrt{g(z)}$ if $z$ are local coordinates for the point $Z$  and $g(z)$ is the determinant of the metric tensor $g_{\mu \nu}(z)=\partial_\mu Z^M\partial_\nu Z^N\delta_{MN}$.

The massless integral $I(X_{12})$ we defined in the previous expression was originally computed in \cite{Billo:2019job} and we refer to Appendix $C$ of this work for more details. In terms of the embedding coordinates $X_i$ the result reads
\begin{equation}
	\label{eq:definition of I main body}
	\begin{split}
		I(X_{12}) 
		&=\dfrac{\Gamma^2(d/2-1)}{2^5\pi^d(X_{12}^2)^{d-3}(2-d/2)(d-3)}
		-\dfrac{\Gamma^2(d/2-1)(2-d/2)}{2^5\pi^d(X_{12}^2)^{d-3}(d-3)}F(X_{12}) +\ldots
	\end{split}
\end{equation} where $F(X_{12})$ is the following function of the embedding coordinates
\begin{equation}
	\label{eq:definition of rho}
	F(X_{12}) = \text{Li}_2(1-U(X_{12}))+ \dfrac{U(X_{12})\log U(X_{12})}{U(X_{12})-1}+\dfrac{1}{2}\big(\log(U(X_{12}))-1\big)^2+\Big(\dfrac{\pi^2}{6}-\dfrac{1}{2}\Big) \ ,
\end{equation} with $U(X_{12})=4R^2/X_{12}^2$.  

The one loop correction to gauge field propagator in the difference theory approach receives corrections from the following  scalar and fermionic loops:
\begin{equation}
	\label{eq:fermionic loops on the sphere}
	\begin{split}
		& \mathord{\begin{tikzpicture}[baseline=-0.65ex,scale=0.65]
				\begin{feynman}
					\vertex (a) at (-1,0)  ;
					\vertex (b) at (4,0)  ;
					\vertex (c) at (0.5,0) ;
					\vertex (d) at (2.5,0) ;
					\vertex (e) at (1.5, 1.45) {$(\eta,\tilde{\eta})$};
					\diagram*{
						(a) -- [photon] (c),
						(c) -- [charged scalar, half left,thick] (d),
						(c) -- [anti charged scalar, half right,thick] (d),
						(d) -- [photon] (b),
					};
				\end{feynman}
			\end{tikzpicture}
		}
		+	\mathord{\begin{tikzpicture}[baseline=-0.65ex,scale=0.65]
				\begin{feynman}
					\vertex (a) at (-1,0)  ;
					\vertex (b) at (4,0)  ;
					\vertex (c) at (0.5,0) ;
					\vertex (d) at (2.5,0) ;
					\vertex (e) at (1.5, 1.38) {$(q,\tilde{q})$};
					\diagram*{
						(a) -- [photon] (c),
						(c) -- [charged scalar, half left] (d),
						(c) -- [anti charged scalar, half right] (d),
						(d) -- [photon] (b),
					};
				\end{feynman}
			\end{tikzpicture}
		}	- \mathord{\begin{tikzpicture}[baseline=-0.65ex,scale=0.65]
				\begin{feynman}
					\vertex (a) at (3,0);
					\vertex (b) at (8,0)  ;
					\vertex (c) at (4.5,0) ;
					\vertex (d) at (6.5,0) ;
					\vertex (e) at (5.5, 1.45) {$(\psi_2,\psi_3)$} ;
					\diagram*{
						(a) -- [photon] (c),
						(c) -- [fermion, half left,thick] (d),
						(c) -- [anti fermion, half right,thick] (d),
						(d) -- [photon] (b),
					};
				\end{feynman}
			\end{tikzpicture}
		}
		- \mathord{\begin{tikzpicture}[baseline=-0.65ex,scale=0.65]
				\begin{feynman}
					\vertex (a) at (3,0);
					\vertex (b) at (8,0)  ;
					\vertex (c) at (4.5,0) ;
					\vertex (d) at (6.5,0) ;
					\vertex (e) at (5.5, 1.38) {$(\phi_2,\phi_3)$};
					\diagram*{
						(a) -- [photon] (c),
						(c) -- [fermion, half left] (d),
						(c) -- [anti fermion, half right] (d),
						(d) -- [photon] (b),
					};
				\end{feynman}
			\end{tikzpicture}
		}~.
	\end{split}
\end{equation} 
The detailed computations of these diagrams are again summarised in Section \ref{sec:Feynman diagrams on the sphere}
and the final result is simply%
\footnote{This form is expected from gauge invariance and supersymmetry. In fact, if one considers separately the fermionic and bosonic loops, they contain (with opposite signs) additional term which are not proportional to the scalar propagator, see Appendix \ref{sec:Feynman diagrams on the sphere}.}
\begin{equation}
	\label{eq:expectation based on susy and gauge invar on the sphere}
	{D}^{(1),MN}_{ab}(X_{12}) = \delta^{MN}{D}_{ab}^{(1)}(X_{12}) \ .
\end{equation}

It is now extremely straightforward to compute the dimensionally regularized correction to the vacuum expectation value of the Wilson loop given by (\ref{eq:radiative corrections on the sphere}). Employing (\ref{eq:expectation based on susy and gauge invar on the sphere}) and (\ref{eq:expression for D on S main corpus}) and taking the colour taccording (\ref{eq:trace}), we find that 
\begin{equation}
	\label{eq:divergent contribution on the sphere-0}
	{g}_B^4 \Delta\cW_4		=-4\pi^2\beta_0^{\mathcal{R}}{{g}_B^4C_F}\oint\dd^2s\big[{R}^2-\dot{X}_1\cdot\dot{X}_2\big]I(X_{12})  \ . \\
\end{equation}
This expression vanishes in superconformal set-ups as it is proportional to the one-loop coefficient  $\beta^{\mathcal{R}}_0$ of the $\beta$-function defined in (\ref{eq:defbeta0}). Using the explicit expression of the massless integral $I(X_{12})$, given by  (\ref{eq:definition of I main body}), and the parametrization (\ref{eq:parametrization on the sphere}) we get
\begin{equation}
	\label{eq:divergent contribution on the sphere}
	\Delta\cW_4	= \Delta\cW^{\rm flat}_4 + \delta\cW_4\ ,
\end{equation}
where the additional term w.r.t. flat space
\begin{equation}
	\label{deltaW4is}
	\delta\cW_4 = (4-d)\dfrac{C_F\beta^{\mathcal{R}}_0\Gamma^2(d/2-1)}{16\pi^{d-2}(d-3)}\oint\dd^2s\dfrac{{R}^2-\dot{X}_1\cdot\dot{X}_2}{|X_{12}|^{2d-6}}F(X_{12})  + \ldots  \ 
\end{equation}	
vanishes for $d\to 4$ and will only contribute at higher orders in perturbation theory.

Finally, the ladder-like contribution at order $g_B^4$, namely $\Sigma_{\mathrm{ladder}}^S$ in (\ref{eq:sigma on the sphere}), takes the same form of the flat space one:
\begin{equation}
	\label{eq:laddersphereis}
	\Sigma_{\mathrm{ladder}}=\Sigma_{\mathrm{ladder}}^S \ ,
\end{equation}
as one can verify with a straightforward calculation employing the propagators of the gauge field and of the adjoint scalar, i.e. (\ref{eq:propagator gauge field on S}) and (\ref{eq:scalar prop on S}). 

Altogether, the corrections of order $g_B^4$ we determined on the sphere differ from those in flat space only by evanescent terms:
\begin{equation}
	\label{ev4}
	\cW_4 = \cW_4^{\rm flat} + \cO(d-4)~.
\end{equation}
Since at the one-loop level we found that $\cW_2 = \cW_2^{\rm flat}$, it follows that the dimensionally regularized observable on the sphere (\ref{eq:dimensionally regularized expectation value on S}) and that computed in flat space (\ref{eq:dimensionally regularized vev in flat}) satisfy
\begin{equation}
	\label{WisWflat}
	\cW = \cW^{\rm flat} + \ldots\ ,
\end{equation}
where the dots stand for terms of order $g_B^4$ that vanish for $d\to 4$ and terms of order $g_B^6$ and higher.

\section{Renormalization}
\label{sec:RNW}
The two-loop corrections $\cW_4^{\rm flat}$ and $\cW_4$ to the vacuum expectation value of the supersymmetric Wilson loop operator in flat space  and on the sphere  are divergent for $d\to 4$ since the $\Delta\cW_4^{\rm flat}$ and $\Delta\cW_4$ contributions,
given by (\ref{eq:unrenormalized combination of spider+bubble }) and  (\ref{eq:divergent contribution on the sphere}) respectively, contain simple poles $1/(4-d)$. 

These ultraviolet singularities are the only divergences affecting the regularized observables in the limit $d\to 4 $. To obtain a finite result we have to apply the renormalization procedure. For smooth curves, such as the contours on which we placed the 1/2 BPS Wilson loop in flat space  (\ref{eq:parametrization}) and on the sphere (\ref{eq:parametrization on the sphere}), the UV singularities are reabsorbed just by the charge renormalization $Z_g$, i.e. by setting  
\cite{Korchemsky:1987wg}   
\begin{equation}
	\label{eq:renormalized g}
	\begin{split}
		g_B&= Z_g(d) g_*(\mu) \mu^{2-d/2} \\
	\end{split} 
\end{equation} where $\mu$ is the renormalization scale and $g_*(\mu)$ is the renormalized coupling. The ultraviolet divergent contributions in $Z_g$ are obtained by usual subtraction operations and are the same in flat space and on the sphere \cite{Drummond:1975yc}, since in the short-distance limit the two spaces are undistinguishable. Expressing the dimensionally regularized observables in terms of the renormalized coupling via  eq. (\ref{eq:renormalized g}), all the divergences are removed and we can take the limit $d\to 4$, defining  
\begin{equation}
	\label{eq:renormalized observables}
	\begin{split}
		W^{\rm flat} &=  \lim_{d\to4} \cW^{\rm flat}~.
	\end{split}
\end{equation}
Analogously, we define $W$ as the limit $d\to 4$ of $\cW$ on the sphere. 
Let us note that $W^{\rm flat}$, as well as $W$, can depend on $\mu$ via $g_*(\mu)$ and on the dimensionless quantity $\mu R$. The overall dependence on $\mu$ is unphysical and consequently, the following Callan-Symanzik equation  
\begin{align}
	\label{eq:CS equation}
	\Big(\beta(g_*)\dfrac{\partial}{\partial g_*}+ R \dfrac{\partial}{\partial R}\Big){W}^{\rm flat}=0 
\end{align}
must hold. 
Here $\beta(g_*) = \mu\partial_\mu g_*$ is the $\beta$-function of the theory. This CS equation implies that $W^{\rm flat}$ is expressible as a function of the running coupling $g(R)$, which satisfies the evolution equation
\begin{equation}
	\label{eq:running ghat}
	\dfrac{\dd{g}}{\dd \log \mu R}  =\beta(g)
\end{equation} 
and determines its value at the energy scale $1/R$ in terms of the initial condition $g(\mu) = g_*(\mu)$. Thus, we must have
\begin{equation}
	\label{eq:solution C.S. on the sphere}
	{W}^{\rm flat}= {W}^{\rm flat}(g(R))~,
\end{equation}
and similarly for the sphere quantity $W(g(R))$. Recalling that the (one-loop exact) $\beta$-function of a $\mathcal{N}=2$ SYM theory with hypermultiplets in the representation $\mathcal{R}$ is $\beta(g)= g^3\beta_0^{\mathcal{R}}$, with the coefficient $\beta_0^{\mathcal{R}}$  defined in (\ref{eq:defbeta0}), we can easily integrate (\ref{eq:running ghat}) to obtain  
\begin{equation}
	\label{eq:ghat}
	g^2(R)=\dfrac{g_*^2}{1+2\beta_0^{\mathcal{R}}g_*^2\ln (\mu R)} \ .
\end{equation} 
Let us observe that, if we identity the renormalization scale $\mu$ with the mass scale $M$ we introduced in the localization approach, this equation coincides with the coupling  (\ref{grunis}) appearing in the regularized  matrix model associated with these theories. 

\subsection{Results to order $g_*^4$}
\label{subsec:resg4}
We now derive the perturbative expansion, up to order $g_*^4$, of the renormalized observables ${W}^{\rm flat}$ and ${W}$, which we will compare with the localization prediction (\ref{Wtog04}). We will begin with deriving the results in $\mathbb{R}^4$.

Standard computations in flat space show that, to order $g_*^2$, the charge renormalization is given by 
\begin{equation}
	\label{eq:subtraction terms on the sphere}
	{Z}_g(d) = 1+\beta_0^\mathcal{R}g_*^2\left( \dfrac{1}{{4-d}} + K_1\right) + \mathcal{O}(g_*^4) \ ,
\end{equation} where $K_1$ characterizes the renormalization scheme.
Using this expression in eq. (\ref{eq:renormalized g}) we can express the results of Section \ref{sec:SW in flat space} in terms of  $g_*$. 
For the lowest-order correction,  
given by (\ref{eq:ladder g2}), we have
\begin{equation}
	\label{eq:renormalized ladder in flat}
	\begin{split}
		g_B^2\cW_2^{\rm flat}\ \to \  \dfrac{C_F}{4} g_*^2+ \dfrac{\beta^\mathcal{R}_0g_*^4C_F}{2(4-d)} + g_*^4\beta^\cR_0\dfrac{C_F}{4}\Big(  \gamma_E + 2K_1 + \log{\mu^2R^2\pi} \Big) + \mathcal{O}(g_*^6)  \ .
	\end{split}
\end{equation} Analogously, the divergent contribution at two-loop accuracy,  
defined in (\ref{eq:unrenormalized combination of spider+bubble }), yields  
\begin{equation}
	\label{eq:renomalized bubble + spider}
	{g}_B^4 \Delta \cW^{\rm flat}_4 \ \to  \ -\dfrac{C_F g_*^4\beta^\mathcal{R}_0}{2(4-d)}- \dfrac{C_F g_*^4\beta^\mathcal{R}_0}{2}\Big( 1 + \gamma_E +\log{\pi R^2\mu^2} \Big) +\cO(g_*^6) \ .
\end{equation} 
Combining together eq. (\ref{eq:renormalized ladder in flat}) with eq. (\ref{eq:renomalized bubble + spider}), we find that the UV divergences cancel each other out, leaving a well-defined quantity at order $g_*^4$ as expected.
Thus, we can take the limit $d\to 4$ and  find that the final expression for the renormalized observable at order ${g}_*^4$ is  
\begin{equation}
	\label{eq:renormalized W in flat, perturbative expression}
	W^{\rm flat}= \dfrac{C_F g_*^2}{4} +\dfrac{C_F g_*^4}{192N}(2N^2-3) -\dfrac{C_F g_*^4}{4}\beta^\mathcal{R}_0
	(2+\gamma_E-2K_1 +\log \mu^2{R}^2 \pi ) +\mathcal{O}(g_*^6) \ ,
\end{equation}
where  we also included the ladder contribution (\ref{eq:ladder finite}), proportional to $g_*^4$. 

The analysis on the sphere is analogous. Let us begin with observing that, up to order $g_B^4$, the dimensionally regularized observable $\cW$ defined on the sphere coincides with $\cW^{\rm flat}$, except for terms that vanish for $d\to 4$ (see eq. (\ref{WisWflat})). As a result, the renormalization procedure gives us again (\ref{eq:renormalized W in flat, perturbative expression}). Indeed, when rewriting $g_B$ in terms of the renormalized coupling $g_*$ via eq. (\ref{eq:renormalized g}), at order $g_*^4$ we only have to replace $Z_g\sim 1$. Taking the limit $d\to 4$, the additional contributions in (\ref{WisWflat}) proportional to $d-4$ vanish and we have
\begin{equation}
	\label{eq:comparison sphere and flat}
		W=W^{\rm flat}+\mathcal{O}(g_*^6)
	\ .
\end{equation}

Our final task is to compare the renormalized observables with the localization prediction (\ref{Wtog04}). In particular, using (\ref{expW0}) and (\ref{grunis}), we can cast the matrix model prediction as follows  
\begin{equation}
	\label{eq:UV term of the matrix model}
	W(g_*) =	\dfrac{g_*^2C_F}{4} +\dfrac{g_*^4C_F}{192N}(2N^2-3) -  g_*^4 \frac{\beta_0^\mathcal{R}\ C_F}{2} \log M R   +\cO(g_*^6)~.
\end{equation}

Setting $\mu=M$ we note that \textit{both} the renormalized observable on the sphere and in flat space, i.e. (\ref{eq:comparison sphere and flat}) and  (\ref{eq:renormalized W in flat, perturbative expression}), perfectly matches the previous expression if we fix the renormalization scheme as follows 
\begin{equation}
	K_1=\dfrac{2+\gamma_E + \log \pi }{2} \ .
\end{equation} It is interesting to note that the choice of the renormalization scheme is identical in flat space and on the sphere. This stems from the fact that, at this perturbative order, the finite parts resulting from the renormalization procedure are identical
even if the diagrams are defined on spaces connected by conformal symmetry -- which is broken at the quantum level.     
In the following section, we will discuss further  effects resulting from the renormalization procedure which, starting from order $g_*^6$, make the analysis in these models even more subtle.

\subsection{Observations regarding higher perturbative orders}
\label{subsec:higher}
The replacement of the bare coupling constant $g_B$ via (\ref{eq:renormalized g}) has to be applied to all the contributions appearing in the dimensionally regularized observables. As a result, evanescent corrections, i.e. terms proportional to $(d-4)$, are activated by the poles of the charge renormalization $Z_g$ and contribute to the renormalized observables at higher orders in perturbation theory. Through out this work we encountered different evanescent contributions on the four-sphere and in flat space. Since they have a different origin, we discuss them separately.  
\paragraph{An interesting effect in flat space}
In Section \ref{sec:Insertion}, we showed that the regularized v.e.v. $\cW$ is given, to order $g_B^4$, by eq. (\ref{eq:finally expression dimensionally regularized vev in flat space}) and contains the evanescent contributions $\Sigma_{\mathrm{Ev}}$ and $\Sigma_{\mathrm{Ev}}^\prime$
originating from the integration over the Wilson loop contour in the $\Sigma_2$ and $\Sigma_{\rm ladder}$ diagrams.

Once the bare coupling $g_B$ is expressed in terms of the renormalized one via (\ref{eq:renormalized g}), the $(d-4)$ factor in
the evanescent contributions is compensated by the simple pole in $Z_g$ and we obtain finite corrections of order $g_*^6$. For instance, in Section \ref{sec:Insertion} we gave the explicit integral expression for the contribution $\Sigma_{\mathrm{Ev}}$. In this case, the  renormalization procedure yields the following $g_*^6$ term:
\begin{equation}
	\label{eq:evanescent finite spider flat}
	\begin{split}
		\dfrac{g_*^6 \beta_0^\mathcal{R} C_FN }{2^{6}\pi^4} \int_0^1 \dd\alpha\dd\beta\dd\gamma\delta(1-\alpha-\beta-\gamma) \ \oint\dd^3{s} \ \epsilon(s_1,s_2,s_3)  \ \dfrac{\sin s_{13}}{Q} + \ldots \ .  
	\end{split}
\end{equation} 
The integral appearing in the previous expression is finite \cite{Erickson:2000af} and the entire correction vanishes in superconformal models. 
Moreover, this term, as well as the analogous one coming from $\Sigma_{\mathrm{Ev}}^\prime$, is not captured by the difference method we described in Section \ref{sec:SW in flat space}, since it does not arise from $\Delta\cW_4^{\rm flat}$. In this work, where we restrict ourselves to study the observable at two-loop accuracy, i.e. at order $g_*^4$, we can ignore these effects. However, at the next perturbative order, such corrections might play a non-trivial role precisely because they do not stem from the difference theory diagrams, which, na\"ively, are the only ones suggested by the matrix model description. Yet, the matrix model prediction at order $g^6$,  see%
\footnote{Actually, eq. (\ref{WtoW0g6}), describes the three-loop prediction of the matrix model in $\mathcal{N}=2$ SQCD with $N_f$ massless fundamental flavour. In this theory $\cR=N_f \Box$ and  $\beta_0^\cR=(N_f-2N)/16\pi^2$. With these identifications it is straightforward to see that the colour structure of (\ref{eq:evanescent finite spider flat}) coincides with that appearing in the third term on the right-hand side of (\ref{WtoW0g6}) characterized by the $\zeta(3)$-Riemann function.} eq. (\ref{WtoW0g6}), contains a term proportional to $\zeta(3)$ characterized by the same colour structure of (\ref{eq:evanescent finite spider flat}). 
At this stage of the analysis, is not possible to conclude whether corrections such as that appearing in eq. (\ref{eq:evanescent finite spider flat}) are ultimately captured by the matrix model without computing all contributions at order $g^6$. The above observations show in any case that
the relation between perturbation theory in flat space and the matrix model in non-conformal models is not obvious and extremely subtle. 

\paragraph{The four-sphere} The two-loop correction $g_B^4\cW_4$ to the dimensionally regularized observable on the sphere was computed in Section \ref{sec:radiative correction on S} by the difference theory approach,  which enabled us to extract the divergent part of the spider-like diagram $\Sigma^S_2$ in eq. (\ref{eq:sigma on the sphere}). Nevertheless, employing this method we did not obtain the explicit expression of evanescent term $\Sigma_{\rm Ev}^S$ of eq. (\ref{Ssev}) which, along with the evanescent correction resulting from the ladder-like diagram $\Sigma_{\mathrm{ladder}}^S$ in eq. (\ref{eq:laddersphereis}), are expected to produce at order $g_*^6$ similar effects to those we described in flat space.

Let us now consider the further evanescent term  $\delta\cW_4$ defined in eq. (\ref{deltaW4is}). This correction results from the structure of the virtual loop on the sphere and is not present in flat space. After replacing the bare coupling constant $g_B$ in terms of the renormalized one via (\ref{eq:renormalized g}), we generate the following contribution
\begin{equation}
	\label{eq:renormalization diverg on sphere}
	\begin{split}
		g_*^6\dfrac{C_F(\beta^{\mathcal{R}}_0)^2}{2}(2\zeta(2) + \log^22-\log 2 +1) \ .
	\end{split}
\end{equation} 
To integrate over the Wilson loop contour the function $F(X_{12})$, defined in (\ref{eq:definition of rho}), we employed the trigonometric integrals outlined in Appendix \ref{sec:trigonometric integrals}. 
The role played by the $g_*^6$ correction (\ref{eq:renormalization diverg on sphere}) is distinctly different from that originating from the evanescent terms such as that  in eq. (\ref{eq:evanescent finite spider flat}). In fact, the contribution appearing in  (\ref{eq:renormalization diverg on sphere}) is quadratic in $\beta_0^\cR$ rather than linear. 
It will thus combine with the other finite parts\footnote{Let us note that these additional finite parts resulting from the renormalization procedure will contribute with $\log^2\mu R$ to the renormalized observable.} arising from the renormalization of the divergences proportional to $(\beta_0^\cR)^2$. Consequently, 
these finite parts on the sphere will differ from those in flat space. As a result, the renormalization scheme we will have to employ to match the result on the sphere to the matrix model will be different from the one needed in flat space.

\section{Conclusions and future perspectives}
\label{sec:conclusions}
In this work, we considered the perturbative computation of the circular 1/2 BPS supersymmetric Wilson loop in 
SU($N$) ${\mathcal N}= 2$ theories with massless matter in a generic representation $\mathcal {R}$ of the gauge group and thus with a non-vanishing $\beta$-function. 
Our aim was to compare the Feynman diagram results, derived both in the Euclidean space $\mathbb{R}^4$ and on the sphere $\mathbb{S}^4$, with the expressions obtained by  Pestun's matrix model generated via supersymmetric localization. 

With non-zero $\beta$ function the localization approach requires a regularization and the net effect is that the resulting matrix model is expressd in terms of a running coupling, which can be identified with the running coupling constant usually introduced in the field-theoretic description.   

Our analysis shows that at quartic order in the renormalized coupling $g_*$, by selecting a suitable renormalization scheme, the matrix model prediction matches precisely 
the perturbative results both on the four sphere and on flat space, which we computed in dimensional regularization. 
Thus at order $g_*^4$, despite conformal symmetry being broken at the quantum level, placing the $\cN=2$ theory  on the four sphere or on flat space yields an equivalent outcome.  
 
We also point out 
that in theories with non-vanishing $\beta$-function the approach based on the difference theory, strongly suggested by the matrix model technique, effectively simplifies the analysis of diagrams with internal vertices but does not account for certain ``evanescent'' contributions which vanishes when $d \to 4 $. In superconformal set-ups, these terms can be safely ignored since the bare coupling does not 
experience divergent subtractions. In general, however, they become relevant at
higher orders in perturbation theory since the renormalization procedure makes them finite. In other words, our analysis seems to suggest that
the difference theory method can be safely employed to simplify a specific perturbative order, provided that the evanescent
contributions generated at lower orders are correctly considered.

In a follow-up paper, our goal will be to present an explicit comparison between the two approaches at order $g_*^6$, further clarifying what information we can extract from the matrix model approach about perturbation theory in flat space and exploring in more details the role of the evanescent terms. It could be also interesting to study the role of the evanescent corrections for the correlators of chiral/antichiral  local operators, following the approach pioneered in \cite{Billo:2019job}. More ambitiously, one could try to go beyond the perturbative sector and  discuss the relation between the instanton contributions derived from localization and the theory in flat-space. This involves a decompactification limit that has been proved extremely subtle in ${\cal N}=2^*$ case \cite{Russo:2019lgq}.

\vskip 1cm
\noindent {\large {\bf Acknowledgments}}
\vskip 0.2cm
We would like to thank M. Frau, F. Galvagno, R. R. John, A. Lerda, A. Miscioscia, I. Pesando and D. Seminara for many useful discussions. This research is partially
supported by the MIUR PRIN contract 2020KR4KN2 “String Theory as a bridge between
Gauge Theories and Quantum Gravity” and by the INFN projects ST\&FI “String Theory
\& Fundamental Interactions” and GAST "Gauge Theory And Strings".
\vskip 1cm

\appendix

\section{Useful formulae}
\label{sec:Useful formulae}
In this section we collect some formulae which are useful  to reproduce the computations of this work. We extensively employ the following Fourier transform in the Euclidean spacetime \begin{equation}
	\label{eq:Fourier transform for massless propagators}
	\int \dfrac{ \dd^d{p}}{(2\pi)^d} \dfrac{e^{\mathrm{i}p\cdot x}}{(p^2)^s} = \dfrac{\Gamma(d/2-s)} {4^s \pi^{d/2} \Gamma(s)} \dfrac{1}{(x^2)^{d/2-s}} \ .
\end{equation}

One-loop integrals are easily evaluated using the Feynman parameters \begin{equation}
	\label{eq:Feynman parameters}
	\prod_i A_i^{-n_i}=\frac{\Gamma\left(\sum n_i\right)}{\prod_i \Gamma\left(n_i\right)} \int_0^1 d x_1 \cdots d x_k x_1^{n_1-1} \cdots x_k^{n_k-1} \frac{\delta\left(1-\sum x_i\right)}{\left[\sum_i A_i x_i\right]^{\sum n_i}} \ ,
\end{equation} which obviously implies that \begin{equation}
	\frac{\prod_i \Gamma\left(n_i\right)}{\Gamma\left(\sum n_i\right)} = \int_0^1 d x_1 \cdots d x_k x_1^{n_1-1} \cdots x_k^{n_k-1} \delta\left(1-\sum x_i\right)  \ .
\end{equation}  

Finally, to evaluate trigonometric integrals we also exploit the following master integral
\begin{equation}
	\label{eq:master integral}
	\begin{split}
		\mathcal{M}(a,b,c) &= \int_{0}^{2\pi} \dd^3\mathbb{\tau} \Big( \sin^2{\dfrac{\tau_{12}}{2}}\Big)^a \Big( \sin^2{\dfrac{\tau_{13}}{2}}\Big)^b\Big( \sin^2{\dfrac{\tau_{23}}{2}}\Big)^c\\
		&=8 \pi^{3/2} \dfrac{\Gamma(a+1/2)\Gamma(b+1/2)\Gamma(c+1/2)\Gamma(1+a+b+c)}{\Gamma(1+a+c)\Gamma(1+b+c)\Gamma(1+a+b)} \ ,
	\end{split}
\end{equation} which was solved explicitly in Appendix G of \cite{Bianchi:2016vvm}.

\section{Field theory set-ups and conventions}
\label{sec:Conventions}
Our conventions are an “Euclideanized” version of those of Wess-Bagger \cite{Wess:1992cp} and follow those of \cite{Billo:2017glv,Billo:2018oog,Billo:2019fbi}. In Euclidean space the spin group is $\mathrm{Spin}(4)\simeq \mathrm{SU}(2)_{\alpha}\otimes \mathrm{SU}(2)_{\dot{\alpha}}$.  Chiral spinors carry  undotted indices  $\alpha, \beta,\ldots$, while  anti-chiral spinors carry dotted indices $\dot{\alpha},\dot{\beta},\dots\ $, which are contracted as follows 
\begin{equation}
	\psi\chi\equiv\psi^{\alpha}\chi_{\alpha} \ , \quad \quad \bar{\psi}\bar{\chi}\equiv\bar{\psi}_{\dot{\alpha}}\bar{\chi}^{\dot{\alpha}} \ .
\end{equation} In the following, we raise and lower indices as follows 
\begin{equation}
	\label{eq:Fierz identities}
	\psi^{\alpha}=\epsilon^{\alpha \beta}\psi_{\beta}, \quad \quad \bar{\psi}^{\dot{\alpha}}=\epsilon^{\dot{\alpha}\dot{\beta}}\bar{\psi}_{\dot{\beta}} \ ,
\end{equation}
where $\epsilon^{12}=\epsilon_{21}=\epsilon^{\dot{1}\dot{2}}=\epsilon_{\dot{2}\dot{1}}=1$. Let us note in Euclidean spacetime spinors satisfy \textit{pseudoreality} conditions, i.e. \begin{equation}
	(\psi_\alpha)^\dagger = \psi^\alpha \ .
\end{equation} 

The matrices $(\bar{\sigma}^{\mu})^{\dot{\alpha}\alpha}$ and $(\sigma^{\mu})_{\alpha \dot{\beta}}$ are defined as follows
\begin{equation}
	\label{eq:sigma matrices}
	\sigma^{\mu}=(\vec{\tau},-\mathrm{i}1) \ , \quad \quad \bar{\sigma}^{\mu}= (-\vec{\tau}, -\mathrm{i}1) \ ,
\end{equation}
where $\vec{\tau}$ are the ordinary Pauli matrices. Furthermore, these matrices are such that
\begin{equation}(\bar{\sigma}^{\mu})^{\dot{\alpha}\alpha}=\epsilon^{\dot{\alpha}\dot{\beta}}\epsilon^{\alpha \beta}(\sigma^{\mu})_{\beta \dot{\beta}}
\end{equation}
and satisfy the Clifford algebra 
\begin{equation}
	\sigma^{\mu}\bar{\sigma}^{\nu}+ \sigma^{\nu}\bar{\sigma}^{\mu}=-2\delta^{\mu \nu}\mathbf{1} \ .
\end{equation} The previous expression obviously implies that \begin{equation}
\Tr \sigma^\mu\bar{\sigma}^\nu=-2\delta^{\mu \nu} \ .
\end{equation}

\subsection{Euclidean actions in flat space}
\label{sec:actions in flat space}
We consider $\mathcal{N}=2$ super-Yang-Mills with gauge group $\mathrm{SU}(N)$ and massless matter content in the representation $\mathcal{R}$. The associated Lie algebra is $\mathfrak{su}(n)$, spanned by hermitian generators $T^a$ with $a=1,\ldots,N^2-1$ and in the fundamental representation. They satisfy  \begin{equation}
	[T^a, T^b]=\mathrm{i}f^{abc}T^c \ , \quad \tr T_a T_b= \dfrac{\delta_{ab}}{2} \ ,
\end{equation} 
that is the Dynkin index of the fundamental representation is $i_F=1/2$.

In the $\mathcal{N}=2$ language the vector multiplet consists of the following field content \begin{equation}
V_{\mathcal{N}=2}= (A_\mu,\psi_\alpha,\lambda_\alpha,\phi) \ ,
\end{equation} where $\lambda_{\alpha}$ and $\psi_{\alpha}$ are two-component Weyl spinors known as \textit{gauginos}. The dynamics in Euclidean space of this supermultiplet is described by the following  gauged-fixed action 
\begin{equation}
	\label{eq:pure super-Yang-Mills N=2}
	\begin{split}
		S^{\mathrm{gauge}}_0= &\int \dd^4{x} \ \text{Tr} \bigg[ -\dfrac{1}{2} F_{\mu \nu}F^{\mu \nu} -2\mathrm{i} \lambda \sigma^{\mu} D_{\mu}\Bar{\lambda} -2\mathrm{i}  \psi \sigma^{\mu} D_{\mu}\Bar{\psi}
		-2 D_{\mu}\Bar{\phi}D^{\mu}\phi -2\partial_{\mu}\Bar{c}D^{\mu}c \bigg] \ , \\
		S_{\mathrm{int}}=	&\int \dd^4{x} \ \text{Tr} \bigg[2\mathrm{i}g_B \sqrt{2}\Big( \Bar{\phi}\big\{\lambda^\alpha,\psi_\alpha\big\}-\phi\big\{\Bar{\psi}_{\dot{\alpha}},\Bar{\lambda}^{\dot{\alpha}}\big\} \Big) -\xi (\partial_{\mu}A^{\mu})^2 -g_B^2\big[\phi,\bar{\phi} \big]^2\bigg] \ ,
	\end{split}
\end{equation} where in the previous expression we denoted with $c$ the ghost field. Let us note that with these conventions the actions are negative defined and consequently, they appear as $\mathrm{e}^S$ in the path integral. The field-strength and the adjoint covariant derivatives are defined as follows \begin{equation}
	\begin{split}
		F_{\mu \nu}=\partial_{\mu} A_{\nu} -\partial_{\nu}A_{\mu} -\mathrm{i} g_B [A_{\mu},A_{\nu}]\ , \quad 
		D_{\mu}=A_{\mu} -\mathrm{i}g_B[A_{\mu}, \bullet ] \ .
	\end{split}
\end{equation}

In the $\mathcal{N}=2$ language matter sits in the hypermultiplets. These transform in a generic representation of the gauge group $\mathcal{R}$ and their the field content is \begin{equation}
H^\cR_{\mathcal{N}=2} =(q,\tilde{q},\eta_\alpha,\Tilde{\eta}_\alpha) \ ,
\end{equation} where $q$ and $\tilde{q}$ are complex scalars, while $\eta$ and $\tilde{\eta}$ are two-component Weyl fermions. The action  reads
\begin{equation}
	\label{eq:actions matter in flat space}
	\begin{split}
		S_0^Q = \int \dd^4{x} \bigg[  &-D_{\mu}\Bar{q} D^{\mu}q -\mathrm{i} \Bar{\eta}\Bar{\sigma}^{\mu}D_{\mu}\eta   - D_{\mu}{\Tilde{q}} D^{\mu}\bar{\Tilde{q}} -\mathrm{i} {\tilde{\eta}}{\sigma}^{\mu}D_{\mu}\bar{\tilde{\eta}} \bigg] \\
		S_{\mathrm{int}}^Q = \int \dd^4{x} \ \bigg[ & \mathrm{i}\sqrt{2}g_B \Big( \Tilde{q}\Bar{\lambda}\Bar{\tilde{\eta}} -\tilde{\eta}\lambda \Bar{\Tilde{q}}\Big) + \mathrm{i}\sqrt{2}g_B \Big( \Bar{\eta}\Bar{\phi}\Bar{\tilde{\eta}}-\tilde{\eta}\phi\eta \Big) + \mathrm{i}\sqrt{2}g_B \Big( \bar{\eta}\bar{\psi}\bar{\tilde{q}} -\tilde{q}\psi\eta \Big)\\
		& + \mathrm{i}\sqrt{2}g_B \Big( \bar{q}\bar{\psi}\bar{\tilde{\eta}}-\tilde{\eta}\psi q \Big) +\mathrm{i}\sqrt{2}g_B \Bigl(\Bar{q}\lambda \eta -\bar{\eta} \bar{\lambda} q\Bigr)
		+V(\phi,\tilde{q},q) \bigg] \ ,
	\end{split}
\end{equation}  where we denoted with $V(\phi,\tilde{q},q)$ the scalar potential describing quartic interactions \begin{equation}
	\label{eq:scalar potential}
	V(\phi,\tilde{q},q)=-2g_B^2\Big(\tilde{q}\phi\bar{\phi}\bar{\tilde{q}}+\bar{q}\bar{\phi}\phi q +\left(\bar{q}T^a_\mathcal{R}\bar{\tilde{q}}\right)\left(\tilde{q}T^a_\cR q\right)\Big)\ ,
\end{equation}
where in the previous expression $T^a_\cR$ denotes the generators of the Lie algebra $\mathfrak{su}(n)$ in the representation $\cR$ of the gauge group. The covariant derivatives for a field transforming in this representation is defined as  \begin{equation}
	D_{\mu}=\partial_{\mu} -\mathrm{i}g_B A^a_{\mu} T^a_{\mathcal{R}}\ .
\end{equation}


In the Feynman gauge, i.e. $\xi=1$, the propagator for the gauge field in momentum space reads\begin{equation}
	\begin{split}
		\mathord{ \begin{tikzpicture}[baseline=-0.65ex,scale=0.8]
				\begin{feynman}
					\vertex (A) at (2,0) ;
					\vertex (A1) at (2,-0.5) {$A_\mu^a$};
					\vertex (B) at (-1.5,0);
					\vertex (B1) at (-1.5,-0.5) {$A^b_\nu$};
					\diagram*{
						(B) --[ photon] (A)
					};
				\end{feynman}
			\end{tikzpicture} 
		} =\dfrac{\delta^{ab}}{p^2} \delta_{\mu \nu} \ ,
	\end{split}
\end{equation} while that of the adjoint scalar $\phi$ is  \begin{equation}
	\begin{split}
	\mathord{ \begin{tikzpicture}[baseline=-0.65ex,scale=0.8]
				\begin{feynman}
					\vertex (A) at (2,0) ;
					\vertex (A1) at (2,-0.5) {$\phi^a$};
					\vertex (B) at (-1.5,0);
					\vertex (B1) at (-1.5,-0.5) {$\bar{\phi}^b$};
					\diagram*{
						(B) --[ fermion] (A)
					};
				\end{feynman}
			\end{tikzpicture} 
		} =\dfrac{\delta^{ab}}{p^2} \ .
	\end{split}
\end{equation} Finally, we consider the propagators for the Weyl spinors. As a concrete example we consider the adjoint \textit{gaugino} $\lambda$. The two relevant Wick contractions are defined as follows \begin{equation}
	\label{eq:Wick's contraction for fermions}
	\big<\lambda^a_{\alpha}(x)\bar{\lambda}^b_{\dot{\alpha}}(y)\big>_0 \ ,\quad \quad \big<\bar{\lambda}_b^{\dot{\alpha}}(y)\lambda^{\alpha}_a(x)\big>_0 \ .
\end{equation} We use the convention where the particle flow arrows always goes from the dotted index to the undotted one. As a result, in momentum space we write the first contraction of the previous expression as follows  \begin{equation}
\begin{split}
	\mathord{ \begin{tikzpicture}[baseline=-0.65ex,scale=0.8]
			\begin{feynman}
				\vertex (A) at (2,0) ;
				\vertex (A1) at (2,-0.5) {$\alpha, \ a $} ;
				\vertex (B) at (-1.5,0);
				\vertex (B1) at (-1.5,-0.5){$\dot{\alpha}, \ b$} ;
				\diagram*{
					(B) --[ fermion, momentum={[arrow style=black]\( p \)}] (A)
				};
			\end{feynman}
		\end{tikzpicture} 
	} =\dfrac{\delta^{ab} {\sigma}_{\alpha \dot{\alpha}}\cdot p}{p^2} \ ,
\end{split}
\end{equation} while the second one is obtained from the previous one by raising the indices as  explained in (\ref{eq:sigma matrices}). 
  
\subsection{Actions on $\mathbb{S}^4$}
\label{sec:actionsonS}
In this section we consider in detail the expression of the $\mathcal{N}=2$ SYM theories on $S^4$. Choosing some local coordinates $x^{\mu}$ on the sphere and denoting with $g_{\mu \nu}(x)$ the metric tensor,  the general form of the action is \begin{equation}
	S^{\mathcal{N}=2}_{S^4} =\int \dd^4{x}\sqrt{g(x)} \big(\mathcal{L}_0+\mathcal{L}_{\mathrm{int}}\big)\ ,
\end{equation} where $\mathcal{L}_0$ contains minimally-coupled kinetics terms of the fields, while $\mathcal{L}_{\mathrm{int}}$ encodes the interacting terms. We begin with considering in detail the Lagrangian for pure $\mathcal{N}=2$ SYM. We have,  \begin{equation}
	\label{eq:action on the sphere }
	\begin{split}
		\mathcal{L}^{\mathrm{gauge}}_0 &= \text{Tr} \Biggl[-\dfrac{1}{2} F_{\mu \nu}F^{\mu \nu} -2\mathrm{i} \lambda^{\alpha}
		\slashed{\mathcal{D}}_{\alpha\dot{\alpha}}\Bar{\lambda}^{\dot{\alpha}} -2\mathrm{i} \psi^{\beta} \slashed{\mathcal{D}}_{\beta \dot{\beta}}\Bar{\psi}^{\dot{\beta}}
		-2 \mathcal{D}_{\mu}\Bar{\phi}\mathcal{D}^{\mu}\phi -2\dfrac{\phi\bar{\phi}}{{R}^2}\Biggl] \ , \\
		\mathcal{L}^{gauge}_{\mathrm{int}} &=\text{Tr} \Biggl[2\mathrm{i}\sqrt{2}{g}_B\Big( \Bar{\phi}\big\{\lambda^{\alpha},\psi_{\alpha}\big\} -\phi\big\{\Bar{\psi}_{\dot{\alpha}},\Bar{\lambda}^{\dot{\alpha}}\big\} \Big)  -{g}_B \big[\phi,\bar{\phi}\big]^2 \Biggl] \ .
	\end{split}
\end{equation}  The last term in $\mathcal{L}_0$ denotes the conformally invariant coupling of the complex scalar field with the curvature which is essential to preserves rigid supersymmetry on the sphere.  The two gauginos, i.e. $\lambda_{\alpha}$ and $\psi_{\alpha}$, along with their chiral partners $\bar{\lambda}_{\dot{\alpha}}$ and $\bar{\psi}_{\dot{\alpha}}$, transform in the adjoint representation of the gauge group and their covariant derivatives are defined as follows \begin{equation}
	\slashed{\mathcal{D}}_{\alpha\dot{\alpha}}=\slashed{\nabla}_{\alpha\dot{\alpha}}-\mathrm{i} {g}_B[\slashed{A}_{\alpha\dot{\alpha}},\cdot].
\end{equation} where $\slashed{\nabla}$ is the appropriate spinor derivative. Since its expression is not necessary in this work we omit it and we refer to Appendix C of \cite{Belitsky:2020hzs} for more details.

We now consider the Lagrangians for the massless hypermultiplets. The minimally coupled kinetic terms and the interacting vertices are given by  \begin{equation}
	\label{eq:action on the sphere matter}
	\begin{split}
		\mathcal{L}^Q_0 &=  -\mathcal{D}_\mu\Bar{q} \mathcal{D}^{\mu}q -2\dfrac{\bar{q}q}{R^2} -\mathrm{i} \Bar{\eta}\bar{\slashed{\mathcal{D}}}\eta  - \mathcal{D}_{\mu}{\Tilde{q}} \mathcal{D}^{\mu}\bar{\Tilde{q}}-2\dfrac{\tilde{q}\bar{\tilde{q}}}{R^2}-\mathrm{i} {\tilde{\eta}}\slashed{\mathcal{D}}\bar{\tilde{\eta}}  \ ,  \\
		\mathcal{L}_{\mathrm{int}}^Q&=	 \mathrm{i}\sqrt{2}g_B \Big( \Tilde{q}\Bar{\lambda}\Bar{\tilde{\eta}} -\tilde{\eta}\lambda \Bar{\Tilde{q}}\Big) + \mathrm{i}\sqrt{2}g_B \Big( \Bar{\eta}\Bar{\phi}\Bar{\tilde{\eta}}-\tilde{\eta}\phi\eta \Big) + \mathrm{i}\sqrt{2}g_B \Big( \bar{\eta}\bar{\psi}\bar{\tilde{q}} -\tilde{q}\psi\eta \Big) + \mathrm{i}\sqrt{2}g_B \Big( \bar{q}\bar{\psi}\bar{\tilde{\eta}}-\tilde{\eta}\psi q \Big)  \\
		& +\mathrm{i}\sqrt{2}g_B\Bigl(\Bar{q}\lambda \eta -\bar{\eta} \bar{\lambda} q\Bigr) +V(\phi,\tilde{q},q) \ , 
	\end{split}
\end{equation} where in the previous expression $V(\phi,\tilde{q},q)$ denotes the quartic scalar potential given by (\ref{eq:scalar potential}). Covariant derivatives act on scalar fields in the usual way, i.e.  \begin{equation}
	\mathcal{D}_{\mu}=\nabla_\mu-\mathrm{i}{g}_BT\cdot A_\mu\ ,
\end{equation} where $\nabla_\mu$ is the standard covariant derivative.

We conclude this section by summarizing the different expressions  for the tree-level propagators on the sphere. Consider a generic massless scalar field $\phi$ in some representation $\mathcal{R}$ of the gauge group with dimension $\dim\cR$. Then tree-level propagator on the sphere expressed in terms of the embedding coordinates $X_M$ introduced in Section \ref{sec:Supersymmetric WL on the sphere}  is \cite{Drummond:1975yc,Belitsky:2020hzs,Adler:1972qq}
\begin{equation}
	\label{eq:scalar prop on S}
	\begin{split}
		D_{\phi}(X_{12}) 
		&= \mathbf{1}\Bigg(\dfrac{\Gamma(d/2-1)}{4\pi^{d/2}(X_{12}^2)^{d/2-1}}\Bigg) \ ,
	\end{split}
\end{equation} where $\mathbf{1}$ is the $\dim\cR\times\dim\cR$ unit matrix. Considering the gauge field $A^a_M(X)$, which is related to that on the sphere $A^a_\mu$ by the pull-back (\ref{eq:pullback}), it turns out that in the Feynman gauge the propagator reads   \cite{Belitsky:2020hzs,Drummond:1975yc,Adler:1972qq} \begin{equation}
	\label{eq:propagator gauge field on S}
	D^{ab}_{MN}(X_{12})=\delta_{MN} \delta^{ab}	D_{\phi}(X^2_{12}) \ .
\end{equation}
Finally, we consider the expression for the spinor field propagators. Their derivation is  involved and we again refer to Appendix C of \cite{Belitsky:2020hzs} for more details. Following the formalism of the authors, we find convenient to arrange each Weyl fermion, along with its chiral counterpart, into a four dimensions spinor. As a concrete example, we take one of the two gauginos of the $\mathcal{N}=2$ vector-supermultiplet, i.e.  $\lambda$ and $\bar{\lambda}$. Introducing   $\Lambda=(\lambda_{\alpha},\bar{\lambda}^{\dot{\alpha}})$, where we recall that $\mathrm{SU}(2)$ chiral indices are raised and lowered as explained in Appendix (\ref{sec:Conventions}), we write the spinor propagator as follows  \cite{Belitsky:2020hzs}
\begin{equation}
	\begin{split}
		\label{eq:propagator Majorana in S}
		\big<\Lambda(X_1)\Lambda^{\dagger}(X_2)\big>&=D_{\Lambda}\left(X_{12}\right)\\
		&= \frac{\Gamma\left(\frac{d}{2}\right)}{2 \pi^{d/ 2}} \frac{U_1^{-1} \slashed{X}_{12} U_2}{\left|X_{12}\right|^d},\\
		&=D\left(X_{12}\right)U_1^{-1}\slashed{X}_{12}U_2 \ .
	\end{split}
\end{equation} In the previous expression we employed the standard slashed notation to denote $\slashed{X}=\Gamma^M X_M$, with $\Gamma^M$ being the gamma matrices in $\mathbb{R}^{d+1}$ satisfying \begin{equation}
	\Tr(\Gamma^M\Gamma^N)=2\delta^{MN} \ .
\end{equation} The rotation matrices appearing in the propagator (\ref{eq:propagator Majorana in S}) satisfy the following properties\begin{equation}
	\label{eq:rotation matrices}
	U \Gamma^{i} U^{-1}=e_\mu^{i}(x) \frac{\partial}{\partial x_\mu} \slashed{X}\  ,  \quad U \Gamma^{d+1} U^{-1}=\slashed{X} \ , 
\end{equation} where $e_\mu^{i}(x)$ are the vielbeins on $S^d$, such that 
\begin{equation}
	\label{viel}
		e_\mu^{i}(x)e_{\nu,i}(x)=g_{\mu \nu}(x)	
\end{equation}			
with $i=1,...,d$ being tangent-space indices. The explicit expression for the matrices $U$ is not important for our perturbative calculations but their explicit expressions can be found in Appendix C of \cite{Belitsky:2020hzs}.

\section{Computing  Feynman diagrams}
\label{sec:perturbative computations}
\subsection{Perturbative computation in flat space}
\label{sec:Feynman computations in flat space}
In this section we consider in detail the one-loop flat space diagrams appearing in Section \ref{sec:Insertion}. The ultraviolet divergent contributions are regularized by dimensional reduction \cite{Erickson:2000af}, where we reduce the theory from four to $d$ dimensions, with $d<4$. In this scheme the gauge field becomes a $d$-dimensional vector and we denote the $4-d$ adjoint real scalars introduced by the reduction as $A_i$, with $i = 1,\ldots,4-d$.

All the diagrams we will consider in the following are elementary and can be evaluated straightforwardly by means of the expressions in Appendix \ref{sec:Useful formulae}.

  We begin with considering the one loop corrections to adjoint scalar field $\phi$.  Using the actions in Appendix \ref{sec:actions in flat space} we find the three  contributions we reported in the main body  in eq. (\ref{eq:one-loop corrections adjoint}). In the Feynman gauge, i.e. $\xi=1$, the self-energy diagrams resulting from the $\mathcal{N}=2$ vector multiplet  are given by the gauge-scalar interaction 
  	\begin{equation}
  		\begin{split}
  			\mathord{\begin{tikzpicture}[baseline=-0.65ex,scale=0.65]
  					\begin{feynman}
  						\vertex (a) at (-1,0) ;
  						\vertex (b) at (4,0)  ;
  						\vertex (c) at (0.5,0) ;
  						\vertex (d) at (2.5,0) ;
  						\vertex (e) at (1.5, 1.3) {$A A $} ;  
  						\diagram*{
  							(a) -- [fermion] (c),
  							(c) -- [photon, half left] (d),
  							(c) -- [fermion] (d),
  							(d) -- [fermion] (b),
  						};
  					\end{feynman}
  				\end{tikzpicture}
  			} &=\dfrac{g_B^2 \delta^{ab}N}{(4\pi)^{d/2}}\dfrac{ \Gamma(d/2)\Gamma(2-d/2)\Gamma(d/2-1)}{\Gamma(d)p^{6-d}}4(d-1) \ ,
  		\end{split}
  	\end{equation}   and from the fermion loop in which $\phi$ interacts with the two {gauginos} $\psi$ and $\lambda$ 	\begin{equation}
  	\begin{split}
  		\mathord{\begin{tikzpicture}[baseline=-0.65ex,scale=0.65]
  				\begin{feynman}
  					\vertex (a) at (-1,0) ;
  					\vertex (b) at (4,0)  ;
  					\vertex (c) at (0.5,0) ;
  					\vertex (d) at (2.5,0) ;
  					\vertex (e) at (1.5, 1.3) {$\psi\Bar{\psi}$};
  					\vertex (e) at (1.5, -1.3) {$\lambda\Bar{\lambda}$};
  					\diagram*{
  						(a) -- [fermion] (c),
  						(c) -- [fermion, half left, thick] (d),
  						(c) -- [fermion, half right, thick] (d),
  						(d) -- [fermion] (b),
  					};
  				\end{feynman}
  			\end{tikzpicture}
  		}
  		=-\dfrac{g_B^2 \delta^{ab}N}{(4\pi)^{d/2}}\dfrac{\Gamma(d/2)\Gamma(2-d/2)\Gamma(d/2-1)}{\Gamma(d)p^{6-d}}4(d-1) \ .
  	\end{split}
  \end{equation}  

The previous diagrams cancel each other out identically, meaning that the one-loop correction to adjoint scalar propagator only depends on the $\mathcal{N}=2$ hypermultiplet fields in the representation $\mathcal{R}$. Indeed, the Yukawa-like interaction between the adjoint scalar $\phi$ and the two two-component Weyl fermions $\eta$ and $\tilde{\eta}$ gives the following diagram \begin{equation}
	\label{eq:fermion loop in rep R}
	\begin{split}
		\mathord{\begin{tikzpicture}[baseline=-0.65ex,scale=.65]
				\begin{feynman}
					\vertex (a) at (-1,0)  ;
					\vertex (b) at (4,0)  ;
					\vertex (c) at (0.5,0) ;
					\vertex (d) at (2.5,0) ;
					\vertex (e) at (1.5, 1.4) {$\eta\Bar{\eta}$};
					\vertex (e) at (1.5, -1.4) {$\Tilde{\eta}\Bar{\Tilde{\eta}}$};
					\diagram*{
						(a) -- [fermion] (c),
						(c) -- [anti charged scalar, half left, thick] (d),
						(c) -- [anti charged scalar, half right, thick] (d),
						(d) -- [fermion] (b),
					};
				\end{feynman}
			\end{tikzpicture}
		}
		=-i_{\mathcal{R}}\dfrac{g_B^2 \delta^{ab}}{(4\pi)^{d/2}}\dfrac{\Gamma(d/2)\Gamma(2-d/2)\Gamma(d/2-1)}{\Gamma(d)p^{6-d}}4(d-1) \ .
	\end{split}
\end{equation} The previous expression is  the result we presented in the main body (\ref{eq:one-loop momentum adjoint}). 

Let us now consider the gauge field.  The self-energy diagrams resulting from the $\mathcal{N}=2$ vector multiplet 
receive contributions from the gauge and ghost fields: 
\begin{equation}
	\begin{split}
		\mathord{\begin{tikzpicture}[baseline=-0.65ex,scale=0.57]
				\begin{feynman}
					\vertex (a) at (-1,0) ;
					\vertex (b) at (4,0)  ;
					\vertex (c) at (0.5,0) ;
					\vertex (d) at (2.5,0) ;
					\vertex (e) at (1.5, 1.3) {$A A$};
					\diagram*{
						(a) -- [photon] (c),
						(c) -- [photon, half left] (d),
						(c) -- [photon, half right] (d),
						(d) -- [photon] (b),
					};
				\end{feynman}
			\end{tikzpicture}
		}
		&+ \mathord{\begin{tikzpicture}[baseline=-0.65ex,scale=0.57]
				\begin{feynman}
					\vertex (a) at (3,0)  ;
					\vertex (b) at (8,0)  ;
					\vertex (c) at (4.5,0) ;
					\vertex (d) at (6.5,0) ;
					\vertex (e) at (5.5, 1.3) {$c\bar{c}$};
					\diagram*{
						(a) -- [photon] (c),
						(c) -- [ghost, half left,thick] (d),
						(c) -- [ghost, half right,thick] (d),
						(d) -- [photon] (b),
					};
				\end{feynman}
			\end{tikzpicture}
		}=\dfrac{g^2 \delta^{ab}N}{(4\pi)^{d/2}}\dfrac{ \Gamma(d/2)\Gamma(2-d/2)\Gamma(d/2-1)}{\Gamma(d)}(3d-2)\dfrac{P_{\mu \nu}}{p^{6-d}}\ ,
	\end{split}
\end{equation}
from one complex and $4-d$ real adjoint scalar fields:
\begin{equation}
\begin{split}
	\mathord{\begin{tikzpicture}[baseline=-0.65ex,scale=0.57]
			\begin{feynman}
				\vertex (a) at (-1,0)  ;
				\vertex (b) at (4,0)  ;
				\vertex (c) at (0.5,0) ;
				\vertex (d) at (2.5,0) ;
				\vertex (e) at (1.5, 1.5) {$\phi \Bar{\phi}$};
				\diagram*{
					(a) -- [photon] (c),
					(c) -- [fermion, half left] (d),
					(c) -- [anti fermion, half right] (d),
					(d) -- [photon] (b),
				};
			\end{feynman}
		\end{tikzpicture}
	}
	&+ \mathord{\begin{tikzpicture}[baseline=-0.65ex,scale=0.57]
			\begin{feynman}
				\vertex (a) at (3,0)  ;
				\vertex (b) at (8,0)  ;
				\vertex (c) at (4.5,0) ;
				\vertex (d) at (6.5,0) ;
				\vertex (e) at (5.5, 1.3) {$A_iA_i$};
				\diagram*{
					(a) -- [photon] (c),
					(c) -- [scalar, half left] (d),
					(c) -- [scalar, half right] (d),
					(d) -- [photon] (b),
				};
			\end{feynman}
		\end{tikzpicture}
	}=\dfrac{g^2 \delta^{ab}N}{(4\pi)^{d/2}}\dfrac{ \Gamma(d/2)\Gamma(2-d/2)\Gamma(d/2-1)}{\Gamma(d)}(d-6)\dfrac{P_{\mu \nu}}{p^{6-d}} \ , 
\end{split}
\end{equation} 
and from two two-component Weyl spinors in the adjoint representations, $\lambda$ and $\psi$:
\begin{equation}
\begin{split}
	\mathord{\begin{tikzpicture}[baseline=-0.65ex,scale=0.57]
			\begin{feynman}
				\vertex (a) at (-1,0)  ;
				\vertex (b) at (4,0)  ;
				\vertex (c) at (0.5,0) ;
				\vertex (d) at (2.5,0) ;
				\vertex (e) at (1.5, 1.5) {$\lambda\Bar{\lambda}$};
				\diagram*{
					(a) -- [photon] (c),
					(c) -- [fermion, half left,thick] (d),
					(c) -- [anti fermion, half right,thick] (d),
					(d) -- [photon] (b),
				};
			\end{feynman}
		\end{tikzpicture}
	}
	&+ \mathord{\begin{tikzpicture}[baseline=-0.65ex,scale=0.57]
			\begin{feynman}
				\vertex (a) at (3,0) ;
				\vertex (b) at (8,0)  ;
				\vertex (c) at (4.5,0) ;
				\vertex (d) at (6.5,0) ;
				\vertex (e) at (5.5, 1.5) {$\psi\Bar{\psi}$};
				\diagram*{
					(a) -- [photon] (c),
					(c) -- [fermion, half left,thick] (d),
					(c) -- [anti fermion, half right,thick] (d),
					(d) -- [photon] (b),
				};
			\end{feynman}
		\end{tikzpicture}
	}=\dfrac{g^2 \delta^{ab}N}{(4\pi)^{d/2}}\dfrac{ \Gamma(d/2)\Gamma(2-d/2)\Gamma(d/2-1)}{\Gamma(d)}(8-4d)\dfrac{P_{\mu \nu}}{p^{6-d}} \ .
\end{split}
\end{equation} Each of the previous expressions is proportional to the transverse projector $P_{\mu \nu } = \delta_{\mu \nu }-p_\mu p_\nu/p^2$, which ensures the gauge invariance of each diagram. As it occurred for the adjoint scalar field the corrections resulting from the $\mathcal{N}=2 $ vector multiplet cancel out.

 We now consider the diagrams resulting from the hypermultiplets in the representation $\mathcal{R}$ of the gauge group. In particular, we find that the gauge field propagator is corrected by the fermion loops involving the two-component Weyl spinors $\eta$ and $\tilde{\eta}$ 
	\begin{equation}
		\begin{split}
			\mathord{\begin{tikzpicture}[baseline=-0.65ex,scale=0.57]
					\begin{feynman}
						\vertex (a) at (-1,0)  ;
						\vertex (b) at (4,0)  ;
						\vertex (c) at (0.5,0) ;
						\vertex (d) at (2.5,0) ;
						\vertex (e) at (1.5, 1.5) {$\eta\Bar{\eta}$};
						\diagram*{
							(a) -- [photon] (c),
							(c) -- [charged scalar, half left,thick] (d),
							(c) -- [anti charged scalar, half right,thick] (d),
							(d) -- [photon] (b),
						};
					\end{feynman}
				\end{tikzpicture}
			}
			&+ \mathord{\begin{tikzpicture}[baseline=-0.65ex,scale=0.57]
					\begin{feynman}
						\vertex (a) at (3,0) ;
						\vertex (b) at (8,0)  ;
						\vertex (c) at (4.5,0) ;
						\vertex (d) at (6.5,0) ;
						\vertex (e) at (5.5, 1.5) {$\tilde{\eta}\bar{\tilde{\eta}}$};
						\diagram*{
							(a) -- [photon] (c),
							(c) -- [charged scalar, half left,thick] (d),
							(c) -- [anti charged scalar, half right,thick] (d),
							(d) -- [photon] (b),
						};
					\end{feynman}
				\end{tikzpicture}
			}=\dfrac{g^2 \delta^{ab}4i_\mathcal{R}}{(4\pi)^{d/2}}\dfrac{ \Gamma(d/2)\Gamma(2-d/2)\Gamma(d/2-1)}{\Gamma(d)}(2-d)\dfrac{P_{\mu \nu}}{p^{6-d}} \ , 
		\end{split}
	\end{equation}
and from the loops of the complex scalar fields of the $\mathcal{N}=2$ hypermultiplets:
 \begin{equation}
		\begin{split}
			\mathord{\begin{tikzpicture}[baseline=-0.65ex,scale=0.57]
					\begin{feynman}
						\vertex (a) at (-1,0) ;
						\vertex (b) at (4,0)  ;
						\vertex (c) at (0.5,0) ;
						\vertex (d) at (2.5,0) ;
						\vertex (e) at (1.5, 1.5) {$q \Bar{q}$};
						\diagram*{
							(a) -- [photon] (c),
							(c) -- [charged scalar, half left] (d),
							(c) -- [anti charged scalar, half right] (d),
							(d) -- [photon] (b),
						};
					\end{feynman}
				\end{tikzpicture}
			}
			&+ \mathord{\begin{tikzpicture}[baseline=-0.65ex,scale=0.57]
					\begin{feynman}
						\vertex (a) at (3,0) ;
						\vertex (b) at (8,0)  ;
						\vertex (c) at (4.5,0) ;
						\vertex (d) at (6.5,0) ;
						\vertex (e) at (5.5, 1.5) {$\Tilde{q}\Bar{\Tilde{q}}$};
						\diagram*{
							(a) -- [photon] (c),
							(c) -- [anti charged scalar, half left] (d),
							(c) -- [charged scalar, half right] (d),
							(d) -- [photon] (b),
						};
					\end{feynman}
				\end{tikzpicture}
			}=- 4 \dfrac{g^2 \delta^{ab} i_{\mathcal{R}} }{(4\pi)^{d/2}}\dfrac{ \Gamma(d/2)\Gamma(2-d/2)\Gamma(d/2-1)}{\Gamma(d)}\dfrac{P_{\mu \nu}}{p^{6-d}} \ .
		\end{split}
	\end{equation}

Combining together the previous diagrams we note that,  up to the transverse projector $P_{\mu\nu}$,  the final result matches (\ref{eq:fermion loop in rep R}), confirming the expectation based on supersymmetry and  reproduced eq. (\ref{eq:one-loop momentum gauge}) in the main body.

\subsection{Perturbative computations on $\mathbb{S}^d$}
\label{sec:Feynman diagrams on the sphere}
In this Appendix we discuss in more detail ${D}_{ab}^{(1)}(X_{12})$ and ${D}^{(1),MN}_{ab}(X_{12})$, i.e. the one loop corrections to adjoint scalar and gauge field propagator in the difference theory approach on the sphere. These quantities enter the dimensionally regularized at order $g_B^4$ and  have been defined  (\ref{eq:radiative corrections on the sphere}). In the following we will evaluate all the relevant diagrams by employing the embedding formalism introduced in Section (\ref{sec:Supersymmetric WL on the sphere}).

For simplicity, we begin with discussing the correction to the adjoint scalar propagator. In the difference theory we have to determine
 \begin{equation}
	\label{eq:fermion loop in rep R on S}
	\begin{split}
		{D}_{ab}^{(1)}(X_{12})=
	\mathord{\begin{tikzpicture}[baseline=-0.65ex,scale=0.7]
			\begin{feynman}
				\vertex (a) at (-1,0) ;
				\vertex (b) at (4,0) ;
				\vertex (c) at (0.5,0) ;
				\vertex (d) at (2.5,0) ;
				\vertex (e) at (1.5, 1.4) {$\eta\Bar{\eta}$};
				\vertex (e) at (1.5, -1.4) {$\Tilde{\eta}\Bar{\Tilde{\eta}}$};
				\diagram*{
					(a) -- [fermion] (c),
					(c) -- [anti charged scalar, half left, thick] (d),
					(c) -- [anti charged scalar, half right, thick] (d),
					(d) -- [fermion] (b),
				};
			\end{feynman}
		\end{tikzpicture}
	} - \mathord{\begin{tikzpicture}[baseline=-0.65ex,scale=0.7]
			\begin{feynman}
				\vertex (a) at (-1,0) ;
				\vertex (b) at (4,0) ;
				\vertex (c) at (0.5,0) ;
				\vertex (d) at (2.5,0) ;
				\vertex (e) at (1.5, 1.4) {$\psi_3\Bar{\psi}_3$};
				\vertex (e) at (1.5, -1.4) {$\psi_2\Bar{\psi}_2$};
				\diagram*{
					(a) -- [fermion] (c),
					(c) -- [anti fermion, half left, thick] (d),
					(c) -- [anti fermion, half right, thick] (d),
					(d) -- [fermion] (b),
				};
			\end{feynman}
		\end{tikzpicture}
	} \ ,
	\end{split}
\end{equation} where we recall that $\eta$ and $\tilde{\eta}$ are two-component Weyl fermions in the representation $\mathcal{R}$, while the second diagram arises from the adjoint fermions of the $\mathcal{N}=4$ theory. This second contribution can be simply deduced by setting $\mathcal{R}=\rm adj$, which implies $i_\cR=N$, in the first diagram\footnote{Let us recall that $\mathcal{N}=4$ SYM can bee seen as an $\mathcal{N}=2$ theory with a single adjoint hypermultiplet.}.
Thus, it is sufficient to determine the fermionic loop of the $\mathcal{N}=2$ theory. Using the matter actions in (\ref{eq:action on the sphere matter}) it is straightforward to deduce that in configuration space we have 
\begin{equation}
	\begin{split}
			\mathord{\begin{tikzpicture}[baseline=-0.65ex,scale=0.7]
				\begin{feynman}
					\vertex (a) at (-1,0) ;
					\vertex (b) at (4,0) ;
					\vertex (c) at (0.5,0) ;
					\vertex (d) at (2.5,0) ;
					\vertex (e) at (1.5, 1.4) {$\eta\Bar{\eta}$};
					\vertex (e) at (1.5, -1.4) {$\Tilde{\eta}\Bar{\Tilde{\eta}}$};
					\diagram*{
						(a) -- [fermion] (c),
						(c) -- [charged scalar, half left, thick] (d),
						(c) -- [charged scalar, half right, thick] (d),
						(d) -- [fermion] (b),
					};
				\end{feynman}
			\end{tikzpicture}
		} 	= -g_B^2\left(\Tr_{\mathcal{R}} \mathrm{T}^a\mathrm{T}^b\right)\int \dd Z_1\dd Z_2 \Delta(X_1-Z_1)\Delta(Z_2-X_2)\Pi_\phi(Z_{12})\ , 
		\end{split}
\end{equation} where the function $\Delta(X)$ is defined in (\ref{defDelta}) and the minus sign results from the fermionic statistics. The  polarization operator is given by 
 \begin{equation}
 	\begin{split}
	\label{eq:self-energy on S}
	\Pi_{\phi}(Z_{12})
	&=	 \Tr\Big(D_{\Psi_{\eta}}(Z_1, Z_2)D_{\Psi_{\tilde{\eta}}}\left(Z_2, Z_1\right)\Big) \ ,
	\end{split}
\end{equation} 
where  the  $\Psi_{\eta}=(\eta_{\alpha},\bar{\eta}^{\dot{\alpha}})$ and $\Psi_{\tilde{\eta}}=(\tilde{\eta}_{\alpha},\bar{\tilde{\eta}}^{\dot{\alpha}})$ are four-dimensional spinors whose propagators $D_{\Psi_{\eta}}(Z_1, Z_2)$ and $D_{\Psi_{\tilde{\eta}}}\left(Z_2, Z_1\right)$ are defined in (\ref{eq:propagator Majorana in S}). Subtracting off the analogous contribution of the $\mathcal{N}=4$ theory we find that
\begin{equation}
	\begin{split}
	D^{ab}_{(1)}(X_{12})&=-\left(\Tr_{\mathcal{R}}^\prime \mathrm{T}^a\mathrm{T}^b\right)g_B^2\int \dd Z_1\dd Z_2 \Delta (X_1-Z_1)\Delta(Z_2-X_2) \Pi_\phi(Z_{12})\\
	&=4{g}_B^2 \left(\Tr_{\mathcal{R}}^\prime \mathrm{T}^a\mathrm{T}^b\right)\int \dd Z_1\dd Z_2 \Delta(X_1-Z_1)\Delta(Z_2-X_2){D}^2(Z_{12})Z^2_{12}\\
	&\equiv -{g}_B^2\left(\Tr_{\mathcal{R}}^\prime \mathrm{T}^a\mathrm{T}^b\right)I(X_{12}) \ ,
	\end{split}
\end{equation} where to obtain the second line we employed the fermion propagators (\ref{eq:propagator Majorana in S}) and recalled that $\Tr_{\mathcal{R}}^\prime \mathrm{T}^a\mathrm{T}^b$ is given by (\ref{eq:trace}). The previous expression reproduces eq. (\ref{eq:expression for D on S main corpus}).

The one-loop corrections to the gauge propagator in the difference theory approach, i.e. ${D}^{(1),M N}_{ab}(X_{12})$, receives contribution from both fermions and complex scalar fields  and consequently, we find convenient to analyse each contribution separately. The fermionic self-energies are schematically given  by \begin{equation}
	\begin{split}
\mathord{\begin{tikzpicture}[baseline=-0.65ex,scale=0.7]
		\begin{feynman}
			\vertex (a) at (-0.5,0) ;
			\vertex (b) at (3.5,0)  ;
			\vertex (c) at (0.5,0) ;
			\vertex (d) at (2.5,0) ;
			\vertex (e) at (1.5, 1.5) {$(\eta,\tilde{\eta})$};
			\diagram*{
				(a) -- [photon] (c),
				(c) -- [charged scalar, half left,thick] (d),
				(c) -- [anti charged scalar, half right,thick] (d),
				(d) -- [photon] (b),
			};
		\end{feynman}
	\end{tikzpicture}
} 
		- \mathord{\begin{tikzpicture}[baseline=-0.65ex,scale=0.7]
				\begin{feynman}
					\vertex (a) at (-0.5,0) ;
					\vertex (b) at (3.5,0)  ;
					\vertex (c) at (0.5,0) ;
					\vertex (d) at (2.5,0) ;
					\vertex (e) at (1.5, 1.5) {$(\psi_2,\psi_3)$};
					\diagram*{
						(a) -- [photon] (c),
						(c) -- [fermion, half left,thick] (d),
						(c) -- [fermion, half right,thick] (d),
						(d) -- [photon] (b),
					};
				\end{feynman}
			\end{tikzpicture}
		}  \ ,
	\end{split} 
\end{equation} where the dashed loop denotes the two self-energies in which the gauge field interacts with the Weyl fermions $\eta$ and $\tilde{\eta}$ in the representation $\mathcal{R}$ of the gauge group, while the second loop in which the fermions $\psi_I$ run, with $I=2,3$, is associated to the $\mathcal{N}=4$ theory. Let consider in detail one of the fermionic loops resulting from the $\mathcal{N}=2$ theory. We have   \begin{equation}
	\begin{split}
		\mathord{\begin{tikzpicture}[baseline=-0.65ex,scale=0.6]
				\begin{feynman}
					\vertex (a) at (-1,0)  ;
					\vertex (b) at (4,0)  ;
					\vertex (c) at (0.5,0) ;
					\vertex (d) at (2.5,0) ;
					\vertex (e) at (1.5, 1.38) {$\eta\Bar{\eta}$};
					\diagram*{
						(a) -- [photon] (c),
						(c) -- [charged scalar, half left,thick] (d),
						(c) -- [anti charged scalar, half right,thick] (d),
						(d) -- [photon] (b),
					};
				\end{feynman}
			\end{tikzpicture}
		} &=\dfrac{{g}_B^2}{2}\left(\Tr_{\mathcal{R}} \mathrm{T}^a\mathrm{T}^b\right)\int \dd Z_1 \dd Z_2 \Delta(X_1-Z_1)\Delta(Z_2-X_1) \Pi_{\rm fermion}^{MN}(Z_1,Z_2)\ .
	\end{split}
\end{equation} The self-energy operator of the previous diagram is more involved than that we analysed in the case of the adjoint scalar field. In particular, it reads   \begin{equation}
	\begin{split}
		\Pi_{\rm fermion}^{MN}(Z_1,Z_2)&=\big( e^{\mu}_{i}(z_1)\partial_{\mu}Z^M_1\big)\big( e^{\nu}_{j}(z_2)\partial_{\mu}Z^N_2\big) \Tr\Gamma^{i}D_{\Psi_{\eta}}(Z_1,Z_2) \Gamma^{j}D_{\Psi_{\eta}}(Z_2,Z_1)\\
		&=Q^{MR}(Z_1)Q^{NS}(Z_2)D^2(Z_{12}^2)\Tr\Gamma_R\slashed{Z}_{12} \Gamma_S\slashed{Z}_{21} \ ,
	\end{split}
\end{equation} where $i,j=1,...,d$ are tangent-space indices.
Let us also observe that the previous expression is extremely similar to that first computed in Section 4.1 of \cite{Belitsky:2020hzs} in the case of $\mathcal{N}=2^*$ SYM. Here, however, our fields are massless. To obtain the second line we exploited the properties of the matrices $U_i$ in their covariant form (\ref{eq:rotation matrices}) and introduced the symmetric tensor\footnote{To obtain the explicit expression of $Q$ one can decompose it as $Q=\delta^{MN}A(Z^2)+Z^MZ^NB(Z^2)$. Contracting this expression with $\delta^{MN}$ leads to the desired result.}\begin{equation}
	\begin{split}
		Q^{MN}(Z) &= \partial^z_{\mu}Z^{M}\partial^z_{\nu}Z^Ng^{\mu \nu}(z)\\
		&= \delta^{MN}-\dfrac{Z^MZ^N}{R^2} \ ,
	\end{split}
\end{equation} which is a projector onto the orthogonal plane to $Z^M$, meaning that $Z_MQ^{MN}(Z)=0$. After expanding out the trace and relabelling the integration variables we have \begin{equation}
	\begin{split}
		\label{eq:expanded trace}
		Q^{MR}(Z_1)Q^{NS}{(Z_2)}\Tr\Gamma_R\slashed{Z}_{12} \Gamma_S\slashed{Z}_{21}&=4\Big(Q^{MS}{(Z_1)}Q_{S}^{N}{(Z_2)}Z_{12}^2+2\ Z_2^RQ^{M}_{R}{(Z_1)} \ Z_1^SQ^{N}_{S}{(Z_2)}\Big)\\
		&=4\Big[Z_{12}^2\delta^{MN} + 2Z_2^MZ_1^N-4Z_1^MZ_1^N+Z_1^MZ_2^NZ_1\cdot Z_2\Big]  \ .
	\end{split}
\end{equation} To further simplify the previous expression we recall that these corrections enter the vacuum expectation value of the 1/2 BPS Wilson loop contracted with the velocities $\dot{X}^M_1$ and $\dot{X}^N_2$. This means that contributions of the form $ \dot{X}_1\cdot Z_1$ provide us with total derivatives which vanish upon the integration over the contour. As a result, only the first two terms survive in (\ref{eq:expanded trace}). Taking into account the analogous contribution resulting from the interaction with $\tilde{\eta}$ and subtracting off the diagrams in the $\mathcal{N}=4$ theory, we find that \begin{equation}
\label{eq:fermioninc loop sphere appendix}
	\begin{split}
\mathord{\begin{tikzpicture}[baseline=-0.65ex,scale=0.7]
		\begin{feynman}
			\vertex (a) at (-0.5,0) ;
			\vertex (b) at (3.5,0)  ;
			\vertex (c) at (0.5,0) ;
			\vertex (d) at (2.5,0) ;
			\vertex (e) at (1.5, 1.5) {$(\eta,\tilde{\eta})$};
			\diagram*{
				(a) -- [photon] (c),
				(c) -- [charged scalar, half left,thick] (d),
				(c) -- [anti charged scalar, half right,thick] (d),
				(d) -- [photon] (b),
			};
		\end{feynman}
	\end{tikzpicture}
} 
&- \mathord{\begin{tikzpicture}[baseline=-0.65ex,scale=0.7]
		\begin{feynman}
			\vertex (a) at (-0.5,0) ;
			\vertex (b) at (3.5,0)  ;
			\vertex (c) at (0.5,0) ;
			\vertex (d) at (2.5,0) ;
			\vertex (e) at (1.5, 1.5) {$(\psi_2,\psi_3)$};
			\diagram*{
				(a) -- [photon] (c),
				(c) -- [fermion, half left,thick] (d),
				(c) -- [fermion, half right,thick] (d),
				(d) -- [photon] (b),
			};
		\end{feynman}
	\end{tikzpicture}
}=\delta^{MN}{D}^{(1)}_{ab}(X_{12}) \\ 
&+ 8{g}_B^2\left(\Tr_{\cR}^\prime\mathrm{T}^a\mathrm{T}^b\right)\int \dd Z_1 \dd Z_2 \Delta(X_1-Z_1)\Delta(Z_2-X_2) \ D^2(Z_{12})\big(Z_2^MZ_1^N\big) \ .
\end{split} 
\end{equation}  Let us note that the contribution involving $Z_2^MZ_1^N$  has to be retained since it does not vanish when contracted with the velocities.  We now show that self-energies resulting from the scalar fields cancel this term out.  Diagrammatically, we depict the situation as follows  \begin{equation}
	\begin{split}	\mathord{\begin{tikzpicture}[baseline=-0.65ex,scale=0.7]
				\begin{feynman}
					\vertex (a) at (-0.5,0) ;
					\vertex (b) at (3.5,0)  ;
					\vertex (c) at (0.5,0) ;
					\vertex (d) at (2.5,0) ;
					\vertex (e) at (1.5, 1.5) {$(q,\tilde{q})$};
					\diagram*{
						(a) -- [photon] (c),
						(c) -- [charged scalar, half left] (d),
						(c) -- [anti charged scalar, half right] (d),
						(d) -- [photon] (b),
					};
				\end{feynman}
			\end{tikzpicture}
		} 
		- \mathord{\begin{tikzpicture}[baseline=-0.65ex,scale=0.7]
				\begin{feynman}
					\vertex (a) at (-0.5,0) ;
					\vertex (b) at (3.5,0)  ;
					\vertex (c) at (0.5,0) ;
					\vertex (d) at (2.5,0) ;
					\vertex (e) at (1.5, 1.5) {$(\phi_2,\phi_3)$};
					\diagram*{
						(a) -- [photon] (c),
						(c) -- [fermion, half left] (d),
						(c) -- [fermion, half right] (d),
						(d) -- [photon] (b),
					};
				\end{feynman}
			\end{tikzpicture}
		}\ ,
	\end{split}
\end{equation} in analogy to the case of fermions. Again, the dashed loop accounts for the self-energies resulting from the $\mathcal{N}=2$ theory, where the gauge field interacts with the complex scalars $q$ and $\tilde{q}$, while the continuos loop represent the two loops of the $\mathcal{N}=4$ theory in which we denoted the complex scalars $\phi_2$ and $\phi_3$. Let us first determine one of the scalar loop of the $\cN=2$ theory. We find, 
 \begin{equation}
		\begin{split}
			\mathord{\begin{tikzpicture}[baseline=-0.65ex,scale=0.6]
					\begin{feynman}
						\vertex (a) at (-1,0)  ;
						\vertex (b) at (4,0)  ;
						\vertex (c) at (0.5,0) ;
						\vertex (d) at (2.5,0) ;
						\vertex (e) at (1.5, 1.38) {$q\Bar{q}$};
						\diagram*{
							(a) -- [photon] (c),
							(c) -- [charged scalar, half left] (d),
							(c) -- [anti charged scalar, half right] (d),
							(d) -- [photon] (b),
						};
					\end{feynman}
				\end{tikzpicture}
			} =\left(\Tr_\cR\mathrm{T}^a\mathrm{T}^b \right) g_B^2\int \dd Z_1  \dd Z_2 \Delta(X_1-Z_1)\Delta(Z_2-X_1) \Pi^{MN}_{\mathrm{scalar}}(Z_1,Z_2) \ .
		\end{split}
	\end{equation} In particular, the self-energy operator of the diagrams involving the scalar fields reads \begin{equation}
		\begin{split}
			\Pi^{MN}_{\mathrm{scalar}}(Z_1,Z_2)&=2Q^{MT}_1 Q^{NS}_2\Bigg[\partial_T^{Z_1}\partial_S^{Z_2}\Delta(Z_{12})\Delta(Z_{12})-\partial_T^{Z_{1}}\Delta(Z_{12})\partial_{S}^{Z_2}\Delta(Z_{21})
			\Bigg]\\
			&=2Q^{MT}_1 Q^{NS}_2\left[\partial_T^{Z_1}\left(\partial_S^{Z_2}\Delta(Z_{12})\Delta(Z_{12})\right)-2\partial_T^{Z_{1}}\Delta(Z_{12})\partial_{S}^{Z_2}\Delta(Z_{21}) \right]\\
			&=-4D^2(Z_{12})Z_2^MZ_1^N \ ,
		\end{split}
	\end{equation} where  $Q_1^{MT}\equiv Q^{MT}(Z_1)$ and we recall that $D(Z_{12})$ is defined in (\ref{eq:propagator Majorana in S}). In the third line we neglected vanishing contributions and the factor $2$ in front of the first line results from the bosonic statistics of the gauge field. If we combine this result with the analogous one resulting from $\tilde{q}$ and we take difference with the $\mathcal{N}=4$ theory, we find that 
	 \begin{equation}
		\begin{split}
	\mathord{\begin{tikzpicture}[baseline=-0.65ex,scale=0.7]
			\begin{feynman}
				\vertex (a) at (-0.5,0) ;
				\vertex (b) at (3.5,0)  ;
				\vertex (c) at (0.5,0) ;
				\vertex (d) at (2.5,0) ;
				\vertex (e) at (1.5, 1.5) {$(q,\tilde{q})$};
				\diagram*{
					(a) -- [photon] (c),
					(c) -- [charged scalar, half left] (d),
					(c) -- [anti charged scalar, half right] (d),
					(d) -- [photon] (b),
				};
			\end{feynman}
		\end{tikzpicture}
	} 
	&- \mathord{\begin{tikzpicture}[baseline=-0.65ex,scale=0.7]
			\begin{feynman}
				\vertex (a) at (-0.5,0) ;
				\vertex (b) at (3.5,0)  ;
				\vertex (c) at (0.5,0) ;
				\vertex (d) at (2.5,0) ;
				\vertex (e) at (1.5, 1.5) {$(\phi_2,\phi_3)$};
				\diagram*{
					(a) -- [photon] (c),
					(c) -- [fermion, half left] (d),
					(c) -- [fermion, half right] (d),
					(d) -- [photon] (b),
				};
			\end{feynman}
		\end{tikzpicture}
	}\\
	 & = -  8{g}_B^2\left(\Tr_{\cR^\prime}\mathrm{T}^a\mathrm{T}^b\right) \int \dd Z_1 \dd Z_2  \Delta(X_1-Z_1)\Delta(Z_2-X_2) D^2(Z_{12})Z_2^MZ_1^N  \ .
		\end{split}
	\end{equation} Combining this result with (\ref{eq:fermioninc loop sphere appendix}) we find that, \begin{equation}
	{D}^{(1),MN}_{ab}(X_{12})=\delta^{M N}{D}_{ab}^{(1)}(X_{12}) \ ,
\end{equation}  reproducing eq. (\ref{eq:expectation based on susy and gauge invar on the sphere}) in the main body.

\section{Trigonometric Integrals}
\label{sec:trigonometric integrals}
In this section we evaluate in detail the trigonometric integrals appearing in $\delta\cW_4$, defined in (\ref{eq:renormalization diverg on sphere}). Setting $\alpha=d-4$, with $d$ being the space-time dimension, we define   \begin{equation}
	\label{eq:integral}
	\begin{split}
\mathcal{I}(\alpha)& = 	\oint \dd^2{s} \ \dfrac{{R}^2-\dot{X}_1\cdot\dot{X}_2}{(X^2_{12})^{\alpha+1}}  \ F(X_{12})\\
&=\oint \dd^2{s} \ \dfrac{{R}^2-\dot{X}_1\cdot\dot{X}_2}{(X^2_{12})^{\alpha+1}} \Bigg[ \dfrac{U\log U}{U-1}+\dfrac{1}{2}\big(\log (U )-1\big)^2+ \text{Li}_2(1-U)+\Big(\dfrac{\pi^2}{6}-\dfrac{1}{2}\Big)\Bigg] \ ,
\end{split}
\end{equation} where $F(X_{12})$ is given in eq. (\ref{eq:definition of rho}) and we recall that $U=4R^2/X_{12}^2$ and that the embedding coordinates $X_i\equiv X(s_i)$ are defined in  (\ref{eq:parametrization on the sphere}). We will first evaluate the previous expression as a function of $\alpha$ and we will eventually take the limit $d \to 4 $, i.e. $\alpha\to 0$.

To compute the different contribution we will rely on the Fourier expansion of the function $[\sin^2 \frac{x}{2}]^b$ \cite{Beccaria:2017rbe},  \begin{equation}
	\label{eq:Fourier series sin}
	\dfrac{1}{\left(4\sin^2 \dfrac{x}{2}\right)^b} = \dfrac{1}{2}a_0(b) +\sum_{n=1}^{\infty}a_n(b) \cos{nx} \ ,
\end{equation} where the Fourier coefficients are given by \begin{equation}
	\begin{split}
		\label{eq:Fourier coefficients}
		a_n(b) &= \dfrac{1}{\pi} \int_0^{2\pi}  \dd{x} \dfrac{\cos{nx}}{(4\sin^2{\dfrac{x}{2}})^b} \\
		&= \dfrac{2\cos(n\pi)\Gamma(1-2b)}{\Gamma(1-b+n)\Gamma(1-b-n)}\\     
		& = \dfrac{\text{sec}(\pi b)\Gamma(n+b)}{\Gamma(2b)\Gamma(1-b+n)} \ .
	\end{split}
\end{equation} 
Another useful relation is the Fourier series of the function $\log \sin(\dfrac{x}{2})$. A straightforward calculation shows that \begin{equation}
	\label{eq:Fourier series log}
	\log \sin(\dfrac{x}{2})=-\log(2)-\sum_{n=1}^{\infty}\dfrac{\cos(nx)}{n} \ .   
\end{equation}  The first integral we consider is  \begin{equation}
	\label{eq:first integral}
	\begin{split}
		I_1(\alpha) &=\oint\dd^2{s}  \dfrac{ {R}^2-\dot{X}_1\cdot\dot{X}_2}{[X_{12}^2]^{\alpha+1}} \ \log \dfrac{4{R}^2}{X_{12}^2} 
	=-\dfrac{2\pi}{{R}^{2\alpha}}\int_0^{2\pi}\dd{x} \dfrac{\log \sin \dfrac{x}{2}  }{\left(4\sin^{2} \dfrac{x}{2} \right)^\alpha} \ ,
\end{split}
\end{equation} where to obtain the second equality we exploited the parametrization of the coordinates (\ref{eq:parametrization on the sphere}) and translation invariance to integrate over one of the coordinates. To proceed, we express the trigonometric functions in terms of their Fourier expansions (\ref{eq:Fourier series log}) and (\ref{eq:Fourier series sin}). As a result, we find 
\begin{equation}
\label{eq:first integral computation}
\begin{split}
		I_1(\alpha) &= \dfrac{2\pi^2}{{R}^{2\alpha}}\log 2  a_0(\alpha)+\dfrac{2\pi}{{R}^{2\alpha}}\int_0^{2\pi}\dd{x} \ \Big(\sum_{m=1}^{\infty}a_m(\alpha)\cos(mx)\Big) \Big( \sum_{n=1}^{\infty}\dfrac{\cos(nx)}{n}\Big)\\
	&=\dfrac{2\pi^2}{{R}^{2\alpha}}\log2 a_0(\alpha) + \dfrac{2\pi^2}{{R}^{2\alpha}}\sum_{n=1}^{\infty}\dfrac{\sec(\pi\alpha)\Gamma(n+\alpha)}{\Gamma(2\alpha)\Gamma(n-\alpha+1)n}\\
	&=\dfrac{2\pi^2}{{R}^{2\alpha}}\log2  a_0(\alpha) +\dfrac{4\pi^2\Gamma(1+\alpha)\sec(\pi\alpha)(\alpha-1)(\psi(1-2\alpha)-\psi(1-\alpha))}{{R}^{2\alpha}\Gamma(2\alpha+1)\Gamma(2-\alpha)} \ ,
\end{split}
\end{equation} where in the previous expression $a_0(\alpha)$ is given by (\ref{eq:Fourier coefficients}) with $n=0$. 
It is straightforward to show that in the limit $\alpha\to 0$ the second contribution in (\ref{eq:first integral computation}) vanishes and we have 
\begin{equation}
\lim_{\alpha \to 0}I_1(\alpha) = 4\pi^2\log 2 \ .
\end{equation}
The second integral we consider is    \begin{equation}
\begin{split}
	I_2(\alpha) &= \oint \dd^2{s} \ \dfrac{{R}^2-\dot{X}_1\cdot\dot{X}_2}{|X_1-X_2|^{2\alpha+2}} \dfrac{U(X_{12}^2)\log U(X_{12}^2)}{U(X_{12}^2)-1}
	=-\dfrac{2\pi}{{R}^{2\alpha}} \int_{0}^{2\pi} \dd{x}\dfrac{\log \sin \dfrac{x}{2}  }
	{\left(4\sin^{2} \dfrac{x}{2} \right)^\alpha  \cos^2 \dfrac{x}{2} }  \ ,
\end{split}
\end{equation} where to obtain the second line we employed again the (\ref{eq:parametrization on the sphere}) and integrated over one of the coordinates by translational invariance. The presence of the analytic continuation parameter $\alpha$ enables us to effectively integrate by parts neglecting  surface terms, i.e.  \begin{equation}
\begin{split}
	I_2(\alpha)	&=-\dfrac{4\pi}{{R}^{2\alpha}} \int_{0}^{2\pi} \dd{x} \dfrac{\dd }{\dd{x}}\tan(\dfrac{x}{2})\dfrac{\log \sin  \dfrac{x}{2}  }
	{\left(4\sin^{2} \dfrac{x}{2}\right)^\alpha}\\
	&=\dfrac{2\pi}{{R}^{2\alpha}} \Bigg[ \int_{0}^{2\pi} \dd{x}\dfrac{1}{\left(4\sin^{2} \dfrac{x}{2} \right)^\alpha}-2\alpha \int_{0}^{2\pi} \dd{x}\dfrac{\log \sin \dfrac{x}{2}}
	{\big(4 \sin^2\dfrac{x}{2}\big)^\alpha}
	\Bigg]\\
	&=\dfrac{2\pi^2}{{R}^{2\alpha}}\dfrac{\sec(\pi\alpha)\Gamma(\alpha)}{\Gamma(2\alpha)\Gamma(1-\alpha)}+2\alpha I_1(\alpha) \ .
\end{split}
\end{equation} To obtain the last line we employed the Fourier coefficients (\ref{eq:Fourier series sin}) and exploited the definition of  $I_1(\alpha) $, given by (\ref{eq:first integral computation}). Finally, in the limit $\alpha \to 0$ the result is \begin{equation}
\lim_{\alpha \to 0} I_2(\alpha) =4\pi^2 \ .
\end{equation}
 The third integral we analyse  is  \begin{equation}
 	\begin{split}
I_3(\alpha)&=\dfrac{1}{2} \oint\dd^2{s} \dfrac{{R}^2-\dot{X}_1\cdot\dot{X}_2}{|X_1-X_2|^{2\alpha +2}} \ \log^2 \dfrac{4{R}^2}{X_{12}^2}  =\dfrac{2\pi}{R^{2\alpha}} \int_0^{2\pi} \dd{x} \dfrac{ \log^2 \sin \dfrac{x}{2} }{ \left( 4\sin^2\dfrac{x}{2}\right)^\alpha}  \ .
\end{split} 
\end{equation} Obtaining a  closed-form expression for $I_3(\alpha)$ is slightly more complicated than the previous examples. Using the Fourier expansions (\ref{eq:Fourier series log}) and (\ref{eq:Fourier series sin}) it is extremely straightforward to show that the previous expression becomes \begin{equation}
\begin{split}
	I_3(\alpha)&=\dfrac{2\pi^2a_0(\alpha)}{R^{2\alpha}}\log^2 2  + \dfrac{\pi}{R^{2\alpha}} \sum^{\infty}_{m,l=1}\int_{0}^{2\pi}\dd{x} \dfrac{\cos m x \cos l x}{m} \left(\dfrac{a_0(\alpha)}{l}+ 4\log 2 \  a_m(\alpha)   \right) \\
	&= \dfrac{2\pi^2a_0(\alpha)}{R^{2\alpha}}\log^2 2 +\dfrac{\pi^2\zeta(2)}{R^{2\alpha}}a_0(\alpha) + \dfrac{4\pi^2\log 2 \sec\pi\alpha}{R^{2\alpha}\Gamma(2\alpha)}\sum_{m=1}^{\infty}\dfrac{\Gamma(m+\alpha)}{m\Gamma(1-\alpha+m)}  \  .
\end{split}
\end{equation} Let us note that the sum in the previous expression is finite and has been already computed in (\ref{eq:first integral computation}). Since it appears with a coefficient which vanishes in the limit $\alpha\to 0$, we find that  \begin{equation}
\begin{split}
\lim_{\alpha \to 0}I_3(\alpha) 
&=4\pi^2\log^2 2+ 2\pi^2\zeta(2) \ .
\end{split}
\end{equation} 
 Let us now turn our attention to the last integral we have to compute to determine the explicit expression of (\ref{eq:integral})  \begin{equation}
\label{eq:third integral}
I_4(\alpha)=\oint\dd^2{s}  \dfrac{{R}^2-\dot{X}_1\cdot\dot{X}_2}{[X_{12}^2]^{\alpha+1}} \ \text{Li}_2\Big( 1-\dfrac{4{R}^2}{X_{12}^2}\Big) \ ,
\end{equation} where in the previous expression the Dilogarithm is defined as \begin{equation}
\text{Li}_2(z)= \sum_{n=1}^{\infty}\dfrac{z^n}{n^2} \ ,
\end{equation} with $z$ being a complex number such that $|z|<1$. However, we can extend its definition to regions of the complex plane where $|z|>1$ via analytic continuation. Using the parametrization of the embedding coordinates (\ref{eq:parametrization on the sphere}) we find that \begin{equation}
	\label{eq:definition of g}
1-\dfrac{4R^2}{X_{12}^2}=-\dfrac{\cos^2 \dfrac{s_{12}}{2}}{\sin^2 \dfrac{s_{12}}{2}}\equiv g(s_{12})	  \  .
\end{equation} 
To proceed we manipulate the Dilogarithm as follows 
\begin{equation}
\begin{split}
	\text{Li}_2(g(s_{12}))&=\sum_{m=1}^{\infty}\dfrac{\cos(m\pi)}{m^2}\dfrac{\cos^{2m}\dfrac{s_{12}}{2} }{\big(\sin^2\dfrac{s_{12}}{2}\big)^m}\\	&=\sum_{m=1}^{\infty}\sum_{k=0}^{2m}\bigg(\dfrac{\Gamma(2m+1)}{\Gamma(2m-k+1)\Gamma(k+1)} \bigg)\frac{\cos(m\pi)}{m^2}\dfrac{\mathrm{e}^{\ii s_{12}(m-k)}}{\left(4\sin^2\dfrac{s_{12}}{2}\right)^m}.
\end{split}
\end{equation} 

Let us insert the previous expression in (\ref{eq:third integral}), and we find that  \begin{equation}
\begin{split}
	I_4(\alpha) &= \dfrac{\pi}{{R}^{2\alpha}}\int_{0}^{2\pi}\dd x\sum_{m=1}^{\infty}\sum_{k=0}^{2m}\bigg(\dfrac{\cos(m\pi)\Gamma(2m+1)}{m^2\Gamma(2m-k+1)\Gamma(k+1)} \bigg)\dfrac{\mathrm{e}^{\ii x(m-k)}}{\left(4\sin^2\dfrac{x}{2}\right)^{m+\alpha}}\\
	&=\dfrac{\pi}{{R}^{2\alpha}}\sum_{m=1}^{\infty}\sum_{k=0}^{2m}\bigg(\dfrac{\cos(m\pi)\Gamma(2m+1)}{m^2\Gamma(2m-k+1)\Gamma(k+1)} \bigg) \int_{0}^{2\pi} \dd x \dfrac{\mathrm{e}^{\ii x(m-k)}}{\left(4\sin^2\dfrac{x}{2}\right)^{m+\alpha}}\\
	&=I_4^{A}(\alpha) +I_4^B(\alpha),
\end{split}
\end{equation} 
where to exchange the integral with the infinite sum we perform an analytical continuation\footnote{To do so, we first have to integrate over a region where the image of (\ref{eq:definition of g}) is in $(-1,1)$. As a result, we can exchange the sum with the integral and perform the continuation to the interval $(0,2\pi)$. } and split the expression into two contributions arising from the Fourier expansion of the sine given by (\ref{eq:Fourier series sin}). The first one reads, \begin{equation}
	\label{eq:first contribution}
\begin{split}
I_4^A(\alpha)&=\dfrac{\pi}{2{R}^{2\alpha}}\int_{0}^{2\pi}\dd x\sum_{m=1}^{\infty}\sum_{k=0}^{2m}\bigg(\dfrac{\cos(m\pi)\Gamma(2m+1)}{m^2\Gamma(2m-k+1)\Gamma(k+1)} \bigg)\mathrm{e}^{\ii x(m-k)}a_0(m+\alpha)\\
	&=\dfrac{\pi^2}{{R}^{2\alpha}}\sec(\pi\alpha)\sum_{m=1}^{\infty}\dfrac{\Gamma(2m)\Gamma(m+\alpha)}{\Gamma^2(m+1)m\Gamma(1-(m+\alpha))\Gamma(2m+2\alpha)}\\
	&=\dfrac{2\pi^2\sec(\pi\alpha)\Gamma(\alpha+1)}{{R}^{2\alpha}\Gamma(-\alpha)\Gamma(2+2\alpha)}\prescript{}{5}{F}_4(\{1,1,1,\dfrac{3}{2},1+\alpha\},\{2,2,2,\dfrac{3}{2}+\alpha\},-1) \ .
\end{split}
\end{equation} The previous expression vanishes in the limit $\alpha\to 0$ and consequently, we can neglect it. The second contribution we have to consider is more involved and is given by 
\begin{equation}
\begin{split}
I^B_4(\alpha)&=\dfrac{\pi}{{R}^{2\alpha}}\sum_{m,n=1}^{\infty}\sum_{k=0}^{2m}\dfrac{\cos(m\pi)}{m^2}\binom{2m}{k}a_n(m+\alpha)\int_{0}^{2\pi} \dd x \ \mathrm{e}^{\ii x(m-k)}\cos(nx)\\
	&=\dfrac{\pi^2}{{R}^{2\alpha}}\sec[](\pi\alpha)\sum_{n,m=1}^{\infty}\sum_{k=0}^{2m}\binom{2m}{k}\dfrac{\Gamma(n+m+\alpha)}{\Gamma(2m+2\alpha) \Gamma(1-m-\alpha+n)m^2}(\delta_{k,m+n}+\delta_{k,m-n})\  .
\end{split}
\end{equation}  Employing the Kronecker symbol to perform the sum over $k$ we find that 
 eq. (\ref{eq:third integral}) can be cast as follows  \begin{equation}
\begin{split}
	I_4(\alpha)&=\oint\dd^2{s}  \dfrac{{R}^2-\dot{X}_1\cdot\dot{X}_2}{[X_{12}^2]^{\alpha +1}} \ \text{Li}_2\Big( 1-\dfrac{4{R}^2}{X_{12}^2}\Big)\\
	&=I_4^A(\alpha) + \dfrac{2\pi^2}{{R}^{2\alpha}}\sec[](\pi\alpha)\sum_{m=1}^{\infty}\sum_{n=1}^{m}\binom{2m}{m+n}\dfrac{\Gamma(n+m+\alpha)}{\Gamma(2m+2\alpha) \Gamma(1-m-\alpha+n)m^2} \ .
\end{split}
\end{equation} 
The previous expression is finite for $\alpha \to 0$ and gives us the following result 
\begin{equation}
\begin{split}
	\sum_{m=1}^{\infty}\sum_{n=1}^{m}\binom{2m}{m+n}\dfrac{2\Gamma(n+m)}{\Gamma(2m+1) \Gamma(1-m+n)m}=\dfrac{\pi^2}{6} \ .
\end{split}
\end{equation} Thus, we find the following result \begin{equation}
\lim_{\alpha \to 0} I_3(\alpha) = 2\pi^2 \zeta(2) \ .
\end{equation}
 Combining together the results we obtained in this section,  we can easily compute 
 	\begin{equation}
 		\label{eq:final result for I}
 		\begin{split}
 			\lim_{\alpha \to 0} \  \mathcal{I}(\alpha) = 4\pi^2\zeta(2) + 2\pi^2(\log(2)^2-\log 2 +1) \ .
 		\end{split}
 	\end{equation} 

\bibliographystyle{JHEP}
\providecommand{\href}[2]{#2}\begingroup\raggedright\endgroup

\end{document}